\documentclass[a4paper,11pt]{article}
\pdfoutput=1 

\usepackage{jcappub} 
\usepackage{graphicx}
\usepackage{dcolumn}
\usepackage{bm}
\usepackage{amssymb}
\usepackage{amsmath}
\usepackage{epsfig}
\usepackage{varwidth}
\usepackage{subfig}
\usepackage{hyperref}
\usepackage{hhline}
\usepackage{xcolor}
\usepackage{multirow}
\usepackage[normalem]{ulem}
\usepackage{slashed}
\usepackage[utf8x]{inputenc}
\usepackage[section]{placeins}
\usepackage[section]{placeins}
\definecolor{Zsug}{RGB}{0, 145, 33} 
\definecolor{Zcor}{RGB}{210, 0, 210}
\definecolor{Zque}{RGB}{0, 180, 190} 
 
\definecolor{jd}{rgb}{0.858, 0.188, 0.478}

\newcommand{\bea}{\begin{eqnarray}}
\newcommand{\eea}{\end{eqnarray}}
\newcommand{\beq}{\begin{equation}}
\newcommand{\eeq}{\end{equation}}
\newcommand{\ec}{\end{center}}
\newcommand{\bc}{\begin{center}}

\newcommand{\pdir}{p\kern -5.2pt\raise 0.2ex\hbox {/}}

\newcommand{\vdir}{v\kern -5.75pt\raise 0.15ex\hbox {/}}
\newcommand{\kdir}{k\kern -5.75pt\raise 0.15ex\hbox {/}}
\newcommand{\epsdir}{\epsilon\kern -5.0pt\raise 0.15ex\hbox {/}}
\newcommand{\bvdir}{\bar{v}\kern -5.75pt\raise 0.15ex\hbox {/}}
\newcommand{\Ddir}{D\kern -7.75pt\raise 0.20ex\hbox {/}}
\newcommand{\Adir}{A\kern -7.75pt\raise 0.20ex\hbox {/}}
\newcommand{\ldir}{l\kern -5.0pt\raise 0.2ex\hbox{/}}
\newcommand{\varepsdir}{\varepsilon\kern -5.5pt\raise 0.15ex\hbox{/}}

\makeatother

\def\be{\begin{equation}}
\def\ee{\end{equation}}

\definecolor{straw}{rgb}{1,1,0.50}
\definecolor{lstraw}{rgb}{1,1,0.70}
\definecolor{red}{rgb}{1.00,0.00,0.00}
\definecolor{teal}{rgb}{0.70,0.70,0.90}
\definecolor{violet}{rgb}{1,0,1}
\definecolor{lblue}{rgb}{0.8,0.8,1}

\author{Subhaditya Bhattacharya,}
\author{Purusottam Ghosh,}
\author{Shivam Verma}
\affiliation{Department of Physics, Indian Institute of Technology Guwahati\\Guwahati, Assam 781039, India}
\emailAdd{subhab@iitg.ac.in}
\emailAdd{pghoshiitg@gmail.com}
\emailAdd{shivam.59910103@gmail.com}

\abstract{
With growing agony of not finding a dark matter (DM) particle in direct search experiments so far (for example in XENON1T), 
frameworks where the freeze-out of DM is driven by number changing processes within the dark sector itself and do not contribute to direct search, 
like Strongly Interacting Massive Particle (SIMP) are gaining more attention. In this analysis, we ideate a simple 
scalar DM framework stabilised by $\mathcal{Z}_3$ symmetry to serve with a SIMP-like DM ($\chi$) with additional light 
scalar mediation ($\phi$) to enhance DM self interaction. We identify that a large parameter space for such DM 
is available from correct relic density and self interaction constraints coming from Bullet or Abell cluster data. 
We derive an approximate analytic solution for freeze-out of the SIMP like DM in Boltzmann equation describing $3_{\rm DM} \to 2_{\rm DM}$ number 
changing process within the dark sector. We also provide a comparative analysis of the SIMP like solution with the 
Weakly Interacting Massive Particle (WIMP) realisation of the same model framework here.}

\keywords{SIMP Dark Matter.}

\begin{document}

\renewcommand*{\thefootnote}{\fnsymbol{footnote}}
\title{\bf SIMPler realisation of Scalar Dark Matter\\} 

\maketitle
\flushbottom

\setcounter{footnote}{0}
\renewcommand*{\thefootnote}{\arabic{footnote}}

\section{Introduction}

Numerous experimental observations at wide range of length scales~\cite{Rubin:1967msa,Rubin:1970zza,Bertone:2004pz}, 
have indicated that about 80$\%$ of total matter density is dominated by dark matter (DM)~\cite{Hu:2001bc,Hinshaw:2012aka}, although we know very little about it. 
The absence of a particle of its kind within the Standard Model (SM), also provides a very strong motivation for the existence of 
physics beyond the Standard Model. Efforts are therefore being made to characterise the nature of DM and discover them in experiments.
We know of it's existence through gravitational interaction, but as it doesn't interact with the electromagnetic radiations, 
its quite hard to detect DM. Two popular ways to detect DM have so far been looked at; through Direct search, for example, XENON1T~\cite{Aprile:2018dbl,Akerib:2017kat}, 
and Collider search, for example, Large Hadron Collider (LHC)~\cite{Abercrombie:2015wmb}. One can also see an evidence of DM in excess 
of antiparticles, photon etc., however that serves as indirect search~\cite{Ahnen:2016qkx} of DM. 
After searching for more than a decade and not being able to find a DM so far, one has to evidently constrain DM properties, 
particularly on its coupling to the visible sector. 

Amongst theoretical efforts to construct a viable DM candidate, Weakly Interacting Massive Particle (WIMPs)~\cite{Kolb} in extensions of SM 
turns out to be simplest and hence most popular. In such a case, the DM is assumed to freeze-out from the equilibrium via $2_{\rm DM} \to 2_{\rm SM}$ annihilations to SM 
and easily satisfies the relic density $\Omega h^2 \simeq 0.12$ (as indicated by PLANCK data~\cite{Ade:2015xua}), if the DM-SM interaction is of the 
order of weak interaction strength. For WIMP like solutions, the same DM-SM interaction also provides direct search scattering and collider
production. Therefore it is difficult to explain the non-observation of the DM in these experiments while addressing correct relic density. Alternate 
possibilities within the WIMP paradigm is therefore to decouple the number changing processes for freeze-out from direct search graphs through 
co-annihilation, semi-annihilation or DM-DM conversion (see for example, in ~\cite{Bhattacharya:2017fid,Bhattacharya:2018cgx}).

Strongly Interacting Massive Particle (SIMP) predicts an interesting alternative to produce the freeze out through number changing process 
within the dark sector itself through for example, $3_{\rm DM}\to 2_{\rm DM}$ or $4_{\rm DM} \to 2_{\rm DM}$ processes. Evidently, for these processes to 
contribute significantly and govern the freeze-out, one requires very small $2_{\rm DM} \to 2_{\rm SM}$ annihilation, i.e. very small DM-SM interaction. 
Therefore SIMP models have a natural explanation for non-observation of DM in direct and collider searches. 
DM in such a framework typically has sub-GeV mass and a large self-scattering cross section, unlike
the WIMP case~\cite{yonit}. Then, although such a large self-scattering cross section is constrained
by Bullet cluster~\cite{bullet} and spherical halo shapes, it can lead to distinct signatures in
galaxies and galaxy clusters, such as the offset of the dark matter sub halo from the galaxy centre, as hinted in Abell 3827~\cite{abell}.
Recently in~\cite{yonit}, it was shown that if we consider a paradigm where DM particles have a strong number changing self interaction, 
then the required thermal relic density can be obtained along with addressing the problems like core vs cusp \cite{ccp} 
and too big to fail \cite{tbtf} that poses a conundrum to face. 


The aim of the paper is to ideate a simple dark sector that inherits the above SIMP-like credentials. The models studied with a scalar 
DM so far had an additional $U(1)$ gauge symmetry to aid self interaction through additional vector boson mediation 
and the remnant symmetry (after symmetry breaking) stabilizes the DM ~\cite{Ko:2014nha,Choi,Choi:2016hid,Choi:2016tkj,Choi:2017mkk,Bernal:2015bla,Bernal:2015ova}. Some other attempts to model a SIMP like DM can be seen in  
~\citep{Lee:2015gsa,Yamanaka:2015tba,Hochberg:2015vrg,Hochberg:2014kqa,lightDM,Choi:2017zww,crossection,Choi:2018jsb,Herms:2018ajr,
Ma:2015mjd,Bernal:2015xba,Bernal:2017mqb}.
We propose a dark sector consisting of one complex scalar singlet 
field $\chi$ and a real scalar singlet $\phi$, where $\chi$ transforms under an unbroken $\mathcal{Z}_3$ symmetry and serves as DM. 
The scalar field $\phi$ (even under $\mathcal{Z}_3$), acquires a vacuum expectation value (vev) during
spontaneous symmetry breaking (SSB) and mixes with the SM scalar doublet to predict an additional light physical scalar apart from Higgs boson, 
and aid DM self interaction. We perform a detailed analysis of the relic density of the DM for freeze-out through $3_{\rm DM} \to 2_{\rm DM}$ number changing 
process in the dark sector, with a brief sketch of $4_{\rm DM} \to 2_{\rm DM}$ process. As emphasised before, for these processes to dictate freeze-out, 
the Higgs portal DM-SM coupling has to be small. In this limit, we also find out that the relic density allowed parameter space is highly constrained by 
the DM self scattering cross-section from Bullet and Abell cluster data. The same model can also serves as WIMP DM with non vanishing Higgs portal coupling, 
which leads us to compare the outcome of SIMP solution to WIMP paradigm of the model. 

We also make a thorough review of the Boltzmann Equation (BEQ) describing a SIMP DM (in a model independent way) and obtain an approximate analytical 
solution. The approximate analytical solution turns out to match closely to the numerical solution of BEQ in a wide range of DM mass.  
 
The paper is organised as follows: Thermal freeze out for SIMP is discussed first in Section \ref{sec2}; the model under consideration and its relic density outcome 
together with self scattering cross-section constraints are discussed in Section \ref{sec3}; brief sketch of WIMP like solution of the model is discussed in Section \ref{sec4}. 
We finally conclude in Section \ref{sec5}. The Appendix of the paper is quite elaborate: DM annihilation cross-section to both
DM and SM ($3_{\rm DM} \to 2_{\rm DM}$, $2_{\rm DM} \to 2_{\rm SM}$, $4_{\rm DM}\to 2_{\rm DM}$) and scattering cross-section of DM with DM and SM are explicitly 
demonstrated. Freeze-out temperature of MeV order SIMP DM in the model also demonstrate in the appendix.  
\section{Thermal freeze out of Dark Matter in SIMP framework}
\label{sec2}
In this section, we review the thermal freeze out of DM governed by BEQ. The equation 
can only be solved numerically. However, for a better understanding of 
relic density of DM governed by the number changing process within the dark sector itself 
(for example, $3_{\rm DM} \to 2_{\rm DM}$ process as elaborated in this paper), 
we will try to identify an approximate analytical solution for the corresponding BEQ. 
We start with a quick recap of thermal freeze-out of DM governed by $2_{\rm DM} \to 2_{\rm SM}$ annihilation, well known to yield a WIMP like solution. 
This will help us to construct and solve SIMP like BEQ and eventually obtain an approximate analytical solution.
\subsection{ A quick recap of thermal freeze-out in WIMP scenario}
\begin{figure}[htb!]
$$
 \includegraphics[height=4.5cm]{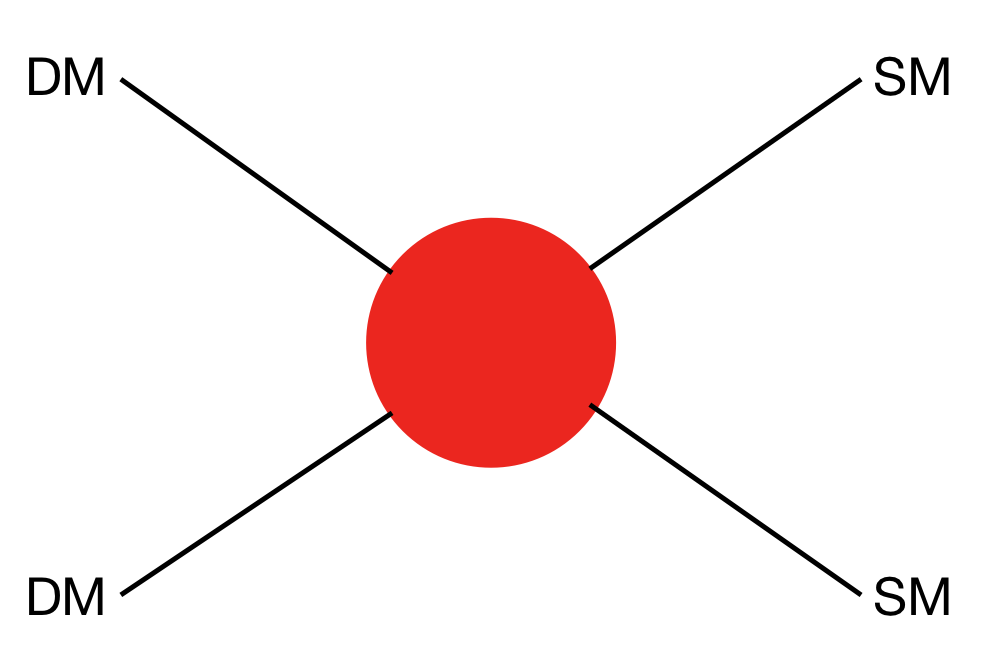}
 $$
 \caption{A cartoon of two DM particles annihilating to two SM particles to yield a WIMP-like scenario.}
 \label{fig:2-2-WIMP}
\end{figure}

The very idea of thermal freeze-out of DM is based on the assumption that the DM was in thermal and chemical equilibrium in early universe. 
As the universe expands with Hubble rate ($\mathcal{H}$), at a particular epoch the interaction rate of the DM ($\Gamma$) 
falls below the rate of expansion ($\mathcal{H}$)~\cite{Kolb} i.e. 
\begin{equation}
 \mathcal{H} \:(\textrm{Hubble expansion rate})\:>\:\Gamma\:(\textrm{particle interaction rate}), 
\end{equation}
and the DM freezes out from equilibrium, to yield a constant DM number density in co moving volume, known as relic density. 
A successful DM model must yield correct relic density as observed in Cosmic Microwave Background (CMB) data for example, given by PLANCK ~\cite{Ade:2015xua}:
\bea
0.1177~\leq \Omega_{{\textrm{DM}}} h^2 \leq ~0.1221, 
\eea
where $\Omega_{{\textrm{DM}}}=\rho_{\rm DM}/\rho_c$ is the cosmological DM density scaled with respect to critical density 
$\rho_c=3\mathcal{H}^2/(8\pi G_N)$, with $G_N$ denoting Newton's gravitational constant~\cite{Kolb}. 
The phenomena of freeze-out or thermal decoupling happens when the temperature of the thermal bath falls (roughly) 
below the mass of the DM particle. The number density of the DM after freeze-out depends 
on its interaction rate ($\Gamma$), which in turn depends on DM mass and coupling(s) to the visible sector.  
The BEQ that governs the thermal freeze-out of DM species, is described as time evolution of the DM phase space distribution function
$f(\textbf{r},\textbf{p},t)$ through ~\cite{Kolb}:
\begin{equation}
\hat{\mathcal{L}}[f]=\hat{\mathcal{C}}(f),
\end{equation}
where $\hat{\mathcal{L}}[f]$ is the Liouville operator describing the change in $f$ with time, while $\hat{\mathcal{C}}(f)$ denotes the change in $f$ through collision. 
Left hand side of the above equation remains unchanged in a homogeneous and isotropic universe (governed by Friedman-Robertson-Walker metric)
\footnote {~which also dictates $f(\textbf{r},P,t) \to f(E,t)$.}, 
while different possibilities of DM collision term $\hat{\mathcal{C}}(f)$ can yield different possibilities of DM freeze-out and relic density, as we elaborate here. 
The simplest realisation for the collision term $\hat{\mathcal{C}}(f)$ is obtained when two DM particles annihilate to two SM particles 
following the cartoon in Fig.~\ref{fig:2-2-WIMP}.
This is a standard number changing process for DM to yield WIMP like solution, which dictates that DM have annihilation 
cross-section of weak interactions strength to justify the observed relic density. 
The BEQ describing $2_{\rm DM} \to 2_{\rm SM}$ process can be written in terms of DM number density 
$n=(g_{\rm DM}/(2\pi)^{3})\int d^{3}P~f_{\rm DM}(E,t)$ as~\cite{Kolb}:

\bea
\frac{dn}{dt} + 3\mathcal{H}n &=& \int \frac{g_{\rm DM} ~d^{3}P_{1}}{(2\pi)^{3}2E_{1}}~\frac{g_{\rm DM}\:d^{3}P_{2}}{(2\pi)^{3}2E_{2}}~\frac{g_{\rm SM}~d^{3}P_{3}}{(2\pi)^{3}2E_{3}}~\frac{g_{\rm SM}~d^{3}P_{4}}{(2\pi)^{3}2E_{4}}(2\pi)^{4}\delta^{4}(P_{1}+P_{2}-P_{3}-P_{4})\nonumber \\
&& ~~~~~~~~~~~~~~~~~~~~~~~~~~~~~~\times |\mathcal{M}_{1+2\to 3+ 4}|^2 ~~(f_{\rm DM}f_{\rm DM}-f_{\rm DM}^{eq}f_{\rm DM}^{eq}) \nonumber \\
 &=& {-\langle\sigma v \rangle_{2_{\rm DM} \to 2_{\rm SM}}} \Big[n^2-{n^{eq}}^2\Big]~,
\eea
where $P_i$ stands for three momentum of $i^{ih}$ particle, $f_{\rm DM}^{eq}\sim{e^{-{E_{\rm DM}/T}}}$ denotes Maxwell's distribution, $g_{\rm DM} $ denotes internal degrees of freedom of DM particles, $g_{\rm SM}$ denotes internal degrees of freedom of SM particles 
and $\langle\sigma v \rangle_{2_{\rm DM} \to 2_{\rm SM}}$ is the thermal average annihilation cross-section given by~\citep{Kolb,Feng:2014vea,Bhattacharya:2016ysw}, 
\bea
\langle\sigma v \rangle_{2_{\rm DM} \to 2_{\rm SM}}&=&\frac{1}{n_1^{eq}~n_2^{eq}}\int~\frac{g_{\rm DM}\:d^{3}P_{1}}{(2\pi)^{3}2E_{1}}~\frac{g_{\rm DM}\:d^{3}P_{2}}{(2\pi)^{3}2E_{2}}~\frac{g_{\rm SM}~d^{3}P_{3}}{(2\pi)^{3}2E_{3}}~\frac{g_{\rm SM}~d^{3}P_{4}}{(2\pi)^{3}2E_{4}}(2\pi)^{4} \nonumber \\&&~~~~~~~~~~~~~~~~~~~~~~~~
\times\delta^{4}(P_{1}+P_{2}-P_{3}-P_{4}) |\mathcal{M}_{1+2\to 3+ 4}|^2~f_1^{eq}~f_2^{eq}~.\nonumber \\
&=& \int_{4m_{\rm DM}^{2}}^{\infty}ds \frac{s\sqrt{(s-4m_{\rm DM}^{2})}~K_{1}(\sqrt{s}/T)~(\sigma v)_{2_{\rm DM}\to 2_{\rm SM}}}{16~T~m_{\rm DM}^{4}~[K_{2}(m_{\rm DM}/T)]^{2}}.
\eea
One can further parameterize this equation by substituting the 
number density per co-moving volume: $Y = n/s$, where $s$ is the entropy density and $x= m_{DM}/T$ to yield~\cite{Kolb}:
\bea
\frac{d Y}{d x} = -0.264~ \frac{g_{*s}}{\sqrt{g_{*}}}~M_{pl}~\frac{m_{\rm DM}}{x^2}~\langle{\sigma v}\rangle_{2_{\rm DM}\to 2_{\rm SM}} ~\Big(Y^{2}-Y_{eq}^{2}\Big)~.
\label{eq:BEQ2to2pre}
\eea
In above equation, 
\bea
\nonumber
g_{*s}&=&\sum_{i=bosons}g_{i}\Big(\frac{T_{i}}{T}\Big)^{3}+\frac{7}{8}\sum_{i=fermions}g_{i}\Big(\frac{T_{i}}{T}\Big)^{3}~, \\
g_{*}&=&\sum_{i=bosons}g_{i}\Big(\frac{T_{i}}{T}\Big)^{4}+\frac{7}{8}\sum_{i=fermions}g_{i}\Big(\frac{T_{i}}{T}\Big)^{4}~,
\label{eq:DOFeqn}
\eea
denote effective degrees of freedom associated with entropy and energy density respectively. $g_{i}$ is the degrees of freedom for the $i^{th}$ species.
Since, for most of the history of the universe, all particles species shared a common temperature, it can be approximated as $g_{*s} \simeq g_{*}$ ~\cite{Kolb}. Thus, we can write \ref{eq:BEQ2to2pre} as:
\bea
\frac{d Y}{d x} = -0.264~ \sqrt{g_{*}}~M_{pl}~ \frac{m_{\rm DM}}{x^2}~\langle{\sigma v}\rangle_{2_{\rm DM}\to 2_{\rm SM}} ~\Big(Y^{2}-Y_{eq}^{2} \Big)~.
\label{eq:BEQ2to2}
\eea

Using Maxwell-Boltzmann statistics for both fermions and Bosons in non-relativistic regime, the equilibrium number density per co-moving volume turns out ~\cite{Kolb}:
 \bea
 Y_{eq}(x)=0.145~\frac{g_{\rm DM}}{g_{*s}} x^{3/2}e^{-x}~.
\eea
For $m_{\rm DM}\sim \mathcal{O}$(GeV), 
$g_{*s} \simeq g_{*}=106.75$. With all these inputs, one can now solve the BEQ \ref{eq:BEQ2to2} numerically to obtain freeze out and 
present yield $Y(x \to \infty)$. Using $n = s~Y(x \to \infty)$, one can find relic density of DM as ~\cite{Kolb}: 
\bea
\Omega h^2 = 2.752 \times 10^8~ \Big( \frac{m_{\rm DM}}{\textrm{GeV}}\Big)~Y(x \to \infty)~.
\eea
 
One can also estimate $Y(x \to \infty)$ approximately without solving BEQ numerically (Eqn.\ref{eq:BEQ2to2}) and relic density of DM can be expressed in terms of 
annihilation cross-section $\langle{\sigma v}\rangle_{2_{\rm DM} \to 2_{\rm SM}}$ (see for example, \cite{Kolb}): 
\bea
\Omega h^2 \approx \frac{854.45 \times 10^{-13}}{\sqrt{g_*}}~x_f~\Big(\frac{{\textrm{GeV}}^{-2}}{\langle{\sigma v}\rangle_{2_{\rm DM}\to 2_{\rm SM}}}\Big)~,
\eea
where $x_{f}$ correspond to freeze-out temperature of DM that is given by~\cite{Kolb}: 
\bea
x_f &\approx& \ln\Big[0.038~\frac{g_{\rm DM}}{\sqrt{g_*}}~ M_{Pl}~ m_{\rm DM}~ (c+2)c ~\langle{\sigma v}\rangle_{2_{\rm DM}\to 2_{\rm SM}} \Big] \nonumber \\
&& ~~~~~~~~~~~~~~ -\frac{1}{2} \ln \ln\Big[0.038~\frac{g_{\rm DM}}{\sqrt{g_*}}~ M_{Pl}~ m_{\rm DM}~ (c+2)c ~\langle{\sigma v}\rangle_{2_{\rm DM}\to 2_{\rm SM}} \Big].
\eea
In the above equation, at $x=x_{f}$, $\Delta(x_{f})=c Y_{eq}(x_{f})$ where $c$ is an unknown constant and 
$\Delta=Y-Y_{eq}$. An example of DM freeze-out in WIMP-like scenario is 
shown in the right hand side (RHS) of Fig.~\ref{fig:DBF} for a DM mass of 100 GeV with different values of annihilation 
cross-section $\langle{\sigma v}\rangle_{2_{\rm DM} \to 2_{\rm SM}}$ in $Y-x$ plane. The correct relic density $\Omega_{\rm DM} h^2 \sim 0.12$ line 
is also shown, which corresponds to $\langle{\sigma v}\rangle_{2_{\rm DM} \to 2_{\rm SM}} \sim 1.5\times10^{-9}~\rm{GeV}^{-2}$, typical cross-section of 
weak interaction strength. We will now follow the same procedure to find out the freeze-out in SIMP mechanism. 
\subsection{SIMP scenario}
\begin{figure}
\centering
\subfloat[\label{3to2_1}]
{\includegraphics[scale=0.2]{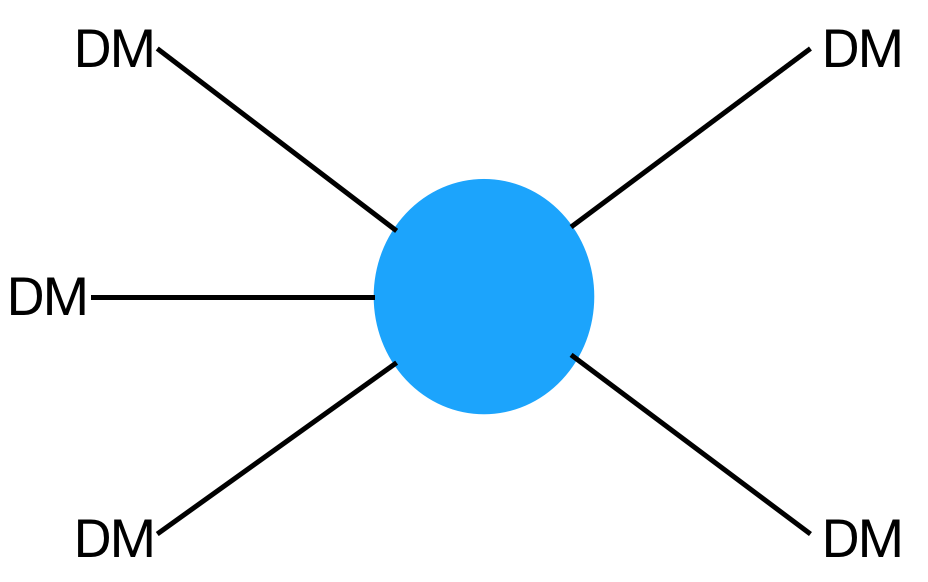}}\hspace{1cm}
\subfloat[\label{3to2_2}]{\includegraphics[scale=0.25]{two-two.png}}
 \caption{A cartoon of annihilation process of three DM particles to two DM particles in SIMP scenario assisted with $2_{\rm DM} \to 2_{\rm SM}$ annihilation to SM 
 particles. The sizes of the diagrams roughly indicate the strengths of the processes (not in exact scale).}
 \label{fig:3-2-SIMP}
\end{figure}

SIMP mechanism can be achieved when $2_{\rm DM} \to 2_{\rm SM}$ annihilation to SM is suppressed and change in 
DM number density is mainly dictated within dark sector for example, by $3_{\rm DM}\rightarrow 2_{\rm DM}$ process. 
Given the fact that the DM still has to be in equilibrium with visible sector particles (SM particles) in thermal bath in the early 
universe for thermal freeze-out to provide correct relic\footnote{One can also achieve correct DM relic density, 
when the DM is out of equilibrium and is produced via decay or annihilation of particles in equilibrium catering to the possibility 
of freeze-in, see for example~\cite{Hall:2009bx}}, and since DM-SM interaction is responsible for maintaining the equilibrium, it can not be completely neglected. 
The scattering of DM with the SM via the same interaction can still be sizeable enough even if the annihilation cross section 
$2_{\rm DM} \to 2_{\rm SM}$ is low due to the large SM number density compared to equilibrium DM number density 
(A numerical estimate is presented later in Sec.~\ref{sec:equilibrium}). 
This helps DM to keep up with equilibrium while not heating up the dark sector until the DM freezes out, 
following the inequality condition ~\cite{yonit}:
\begin{equation}
\Gamma_{{\rm DM} + {\rm SM} \to {\rm DM} +{\rm SM} ~\textrm{scattering}}\gtrsim~\Gamma_{3_{\rm DM}\to 2_{\rm DM} ~\textrm{annihilation}} ~\gg ~\Gamma_{2_{\rm DM}\to 2_{\rm SM} ~\textrm{annihilation}}~.
\label{eq:condition}
\end{equation}
In above equation, $\Gamma_{{\rm DM} + {\rm SM} \to {\rm DM} +{\rm SM}}=n^{eq}~\langle \sigma v\rangle$, 
$\Gamma_{2_{\rm DM}\to 2_{\rm SM}}= n\times{\langle \sigma v\rangle}_{2_{\rm DM}\to 2_{\rm SM}}$ and  
 $\Gamma_{3_{\rm DM}\to 2_{\rm DM}}$ $= n^{2}~{\langle \sigma v^2\rangle}_{3_{\rm DM}\to 2_{\rm DM}}$ define the rate of the corresponding interactions, 
 where $n$ denotes DM number density following our earlier convention. 
We will put up an explicit demonstration of the inequality Eq.~\ref{eq:condition} in context of the model 
described here later. The scattering does not contribute to the relic density of the DM caveat to a kinetic decoupling (see for example, the discussion on 
ELDER DM as in ~\citep{elder}); therefore the number changing processes that govern the freeze-out for SIMP can be described by the cartoon diagram of 
Fig. \ref{fig:3-2-SIMP}, where the sizes of the diagrams ($3_{\rm DM} \to 2_{\rm DM}$ versus $2_{\rm DM} \to 2_{\rm SM}$ annihilation) roughly indicate the 
dominant and sub-dominant contributions.  

Thermally averaged cross section for $n\rightarrow 2$ annihilation processes, where n is the initial number of DM particle and 2 
correspond to the number of particles in the final state can be expressed in terms of the characteristic mass scale $M$ as \cite{lightDM}:
\begin{equation}
\label{dimension}
[<\sigma_{n\rightarrow 2} v^{n-1}>]=[M^{-3n+4}].
\end{equation}
 Eq.~\ref{dimension} can simply be derived from equating the Hubble constant ($\mathcal{H}$) to the rate of interaction ($\Gamma$) for $n\rightarrow 2$ annihilation process.
According to Eq.~\ref{dimension}, a $2_{\rm DM} \to 2_{\rm SM}$ process is: $[\langle\sigma v\rangle] = [M]^{-2}$, with unit $\rm{GeV}^{-2}$ (assuming the 
mass of the DM $\sim$ GeV and 'v' to be dimensionless in natural units). {Similarly for a $3_{\rm DM} \to 2_{\rm DM}$ process, $[\langle\sigma v^{2}\rangle] = [M]^{-5}$, 
so it has unit $\rm{GeV}^{-5}$ and for $4_{\rm DM} \to 2_{\rm DM}$ process, $[\langle\sigma v^{3}\rangle] = [M]^{-8}$, with unit $\rm{GeV}^{-8}$.}
Next we discuss BEQ for $3_{\rm DM} \to 2_{\rm DM}$ process and its possible analytical solutions for freeze-out.  
\subsubsection{Boltzmann Equation and numerical solution to freeze-out} 
The BEQ that dictates the freeze-out through $3_{\rm DM} \to 2_{\rm DM}$ number changing process in dark sector (see Fig. \ref{3to2_1} only),
in terms of DM number density, n~\citep{Kolb,crossection} is given by~\footnote {As argued before, DM-SM interaction can not be neglected for the DM to be in thermal bath, however contribution of $2_{\rm DM} \to 2_{\rm SM}$ for the DM
freeze-out can be neglected in SIMP paradigm.}:
\bea
\frac{dn}{dt} + 3\mathcal{H}n &=& \int \frac{g_{\rm DM} ~d^{3}P_{1}}{(2\pi)^{3}2E_{1}}~\frac{g_{DM}\:d^{3}P_{2}}{(2\pi)^{3}2E_{2}}~\frac{g_{\rm DM}~d^{3}P_{3}}{(2\pi)^{3}2E_{3}}~\frac{g_{\rm DM}~d^{3}P_{4}}{(2\pi)^{3}2E_{4}}~\frac{g_{DM}~d^{3}P_{5}}{(2\pi)^{3}2E_{5}}~(2\pi)^{4}\nonumber \\
&&~\delta^{4}(P_{1}+P_{2}+P_{3}-P_{4}-P_{5})\times \overline{|\mathcal{M}_{1+2+3 \to 4+ 5}|^2}\times(f_{\rm DM}~f_{\rm DM}~f_{\rm DM}-f_{\rm DM}^{eq}~f_{\rm DM}^{eq}) \nonumber \\
 &=& -\langle\sigma v^2 \rangle_{3_{\rm DM} \to 2_{\rm DM}} \Big(n^3- n^2 {n^{eq}}\Big),
\eea
where again $g_{\rm DM}$ denotes the internal degrees of freedom in the DM sector. The thermal average of annihilation cross section 
$\langle\sigma v^2 \rangle_{3_{\rm DM} \to 2_{\rm DM}}$ in this case is given by~\citep{crossection}:
\bea
\langle\sigma v^{2} \rangle_{3_{\rm DM} \to 2_{\rm DM}}&=&\frac{1}{n_{1}^{eq}~n_{2}^{eq}~n_{3}^{eq}}\int\frac{g_{\rm DM} ~d^{3}P_{1}}{(2\pi)^{3}2E_{1}}~\frac{g_{\rm DM}\:d^{3}P_{2}}{(2\pi)^{3}2E_{2}}~\frac{g_{\rm DM}~d^{3}P_{3}}{(2\pi)^{3}2E_{3}}~\frac{g_{\rm DM}~d^{3}P_{4}}{(2\pi)^{3}2E_{4}}~\frac{g_{\rm DM}~d^{3}P_{5}}{(2\pi)^{3}2E_{5}}\nonumber \\
&&~~~~~(2\pi)^{4}\delta^{4}(P_{1}+P_{2}+P_{3}-P_{4}-P_{5})
\times \overline{|\mathcal{M}_{1+2+3\to 4+5}|^2}f_{1}^{eq}f_{2}^{eq}f_{3}^{eq}
\eea

\begin{figure}[htb!]
$$
\hspace{-0.75cm}
 \includegraphics[scale=0.275]{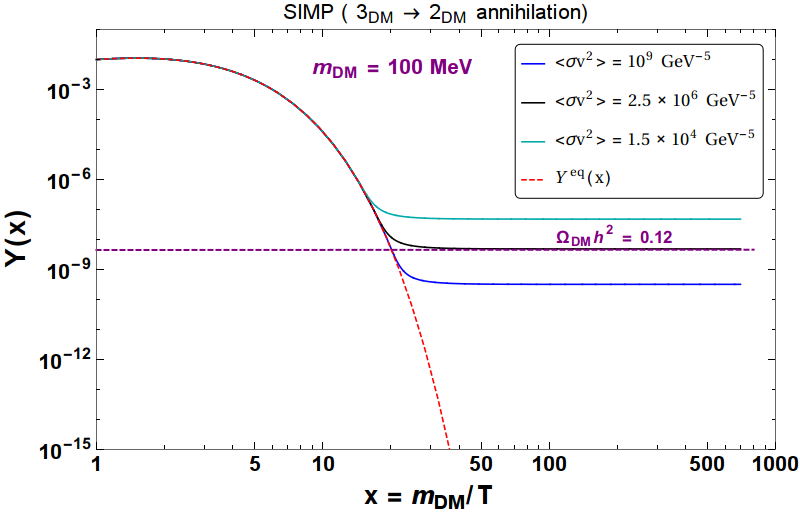}
 ~~
  \includegraphics[scale=0.275]{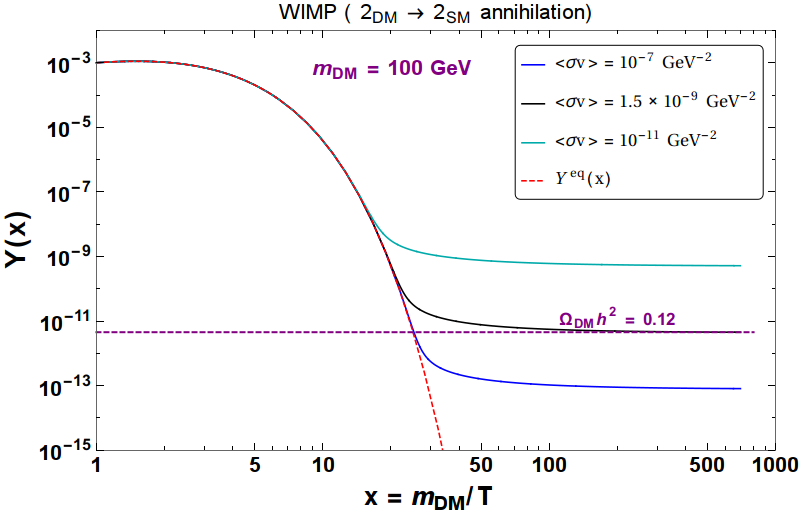}
 $$
 \caption{Freeze out of SIMP like DM ($3_{\rm DM} \to 2_{\rm DM}$) (left panel) and WIMP like DM ($2_{\rm DM} \to 2_{\rm SM}$)(right panel) from equilibrium 
 $Y_{eq}(x)$ (red dashed line) in $Y(x)-x$ plane obtained from the numerical solution of the corresponding BEQ (Eq.~\ref{eq:BEQ3to2} and 
 Eq.~\ref{eq:BEQ2to2} respectively for SIMP and WIMP case). DM mass and annihilation cross-sections have been chosen in a model independent way and mentioned 
 in figure inset.}
 \label{fig:DBF}
\end{figure}

In terms of co-moving number density, i.e. $Y = n/s$ and $x=m_{\rm DM}/T$, the BEQ turns out to be~\citep{Kolb}: 
\bea
\frac{d Y}{d x} = -0.116~\frac{g_{*s}^{2}}{\sqrt{g_{*}}}~ M_{Pl} \frac{{m_{\rm DM}}^4}{x^5}~\langle{\sigma v^2}\rangle_{3_{\rm DM} \to 2_{\rm DM}} ~\Big(Y^{3}-Y^2 ~Y_{eq}\Big).
\label{eq:BEQ3to2}
\eea
Since the temperature scale considered here allows us to take $g_{*s} \simeq g_{*}$, we can rewrite the above BEQ as,
\bea
\frac{d Y}{d x} = -0.116~g_{*}^{3/2}~ M_{Pl} \frac{{m_{\rm DM}}^4}{x^5}~\langle{\sigma v^2}\rangle_{3_{\rm DM} \to 2_{\rm DM}} ~\Big(Y^{3}-Y^2 ~Y_{eq}\Big) .\eea
The equilibrium yield is $Y_{eq}(x)=0.145~(g_{\rm DM}/g_{*s}) x^{3/2}e^{-x}$, with $g_{*s} \simeq g_{*}=10.75$ for MeV order DM. Again, one can solve 
the BEQ (Eq.~\ref{eq:BEQ3to2}) numerically to find the yield after freeze out: $Y (x\to\infty)$. One such numerical solution is demonstrated in the 
left panel of Fig.~\ref{fig:DBF}.  
For illustration, we have chosen mass of the DM to be 100 MeV and different magnitudes of annihilation cross-section to lie within:
$\langle{\sigma v^2}\rangle_{3_{\rm DM} \to 2_{\rm DM}} \sim \{10^4-10^9\} \rm{GeV}^{-5}$. The one corresponding to 
correct relic density (horizontal black dashed line in left panel of Fig.~\ref{fig:DBF}) is 
$\langle{\sigma v^2}\rangle_{3_{\rm DM} \to 2_{\rm DM}} \sim 2.5\times10^6~ \rm{GeV}^{-5}$, that lies in the strong interaction range. 
This can now be contrasted to WIMP case ($2_{\rm DM} \to 2_{\rm SM}$) on the right panel graph, where correct relic density is obtained for 100 GeV DM with 
$\langle{\sigma v}\rangle_{2_{\rm DM} \to 2_{\rm SM}} \sim 1.5\times10^{-9}~ \rm{GeV}^{-2}$. As stated earlier, relic density of DM in terms of yield after freeze out reads as~\citep{Kolb,Bhattacharya:2016ysw}:
\bea \label{eq:OmegaExp1}
\nonumber \Omega h^2 &=& 2.752 \times 10^8~ \Big( \frac{m_{\rm DM}}{\textrm{GeV}}\Big)~Y(x \to \infty)~, \\
&=&  2.752 \times 10^5~ \Big( \frac{m_{\rm DM}}{\textrm{MeV}}\Big)~Y(x \to \infty)~,
\eea
where the numerical pre factor depends on the choice of DM mass to be in MeV or in GeV order. 
 
\subsubsection{Approximate analytical solution to Boltzmann Equation} 
The main idea of this section is to find an approximate analytical solution for BEQ governed by $3_{\rm DM} \to 2_{\rm DM}$ process as in Eq.~\ref{eq:BEQ3to2}. 
Such an exercise is already standardised for $2_{\rm DM} \to 2_{\rm SM}$ case and we will follow a similar path. We first rewrite the BEQ (Eq. \ref{eq:BEQ3to2}) 
in terms of $\Delta=Y-Y_{eq}$, that marks the difference of DM yield from the corresponding equilibrium 
yield. When $\Delta$ is small, the DM follows equilibrium distribution, when $\Delta$ turns large, the DM freezes out. 
The BEQ in terms of $\Delta$ reads as~\citep{Kolb}:
\bea\label{eq:BEQprx1}
\frac{d\Delta}{dx}+\frac{dY_{eq}}{dx} &=&-\frac{A}{x^5} \Delta \Big(Y_{eq}^{2}+2\Delta Y_{eq}+\Delta^{2}\Big) ~, 
\eea
where we have dumped everything else into 
$A = 0.116~{g_*}^\frac{3}{2} ~M_{Pl} ~{m_{\rm DM}}^4~\langle{\sigma v^{2}}\rangle_{3_{\rm DM} \to 2_{\rm DM}}$.
Before freeze-out, i.e. for $1<x \leq x_{f}$ ($x_f$ denotes freeze out of DM), $\Delta<<Y_{eq}$ and $d\Delta/dx \to 0$. Then BEQ simplifies to:
\bea
\label{deltaxf}
\frac{dY_{eq}}{dx} &=&-\frac{A}{x^5} \Delta \Big((Y_{eq})^{2}+2\Delta Y_{eq}+\Delta^{2}\Big) ~. 
\eea

\begin{figure}[htb!]
$$
\includegraphics[scale=0.25]{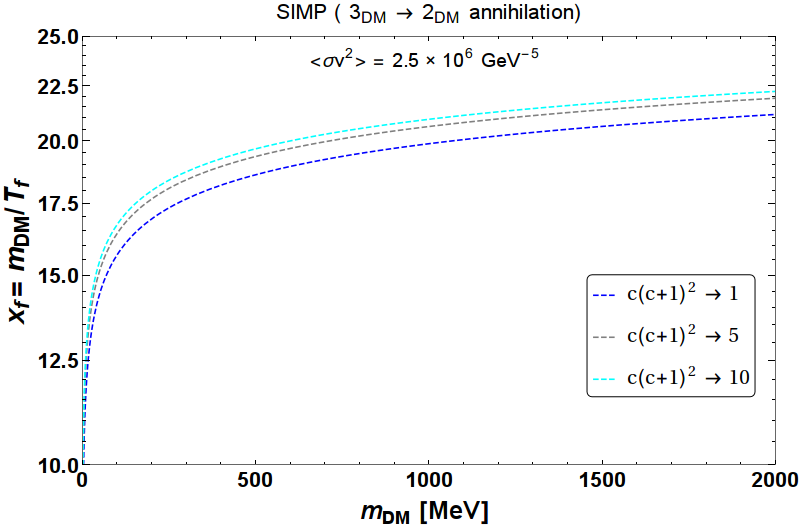}
$$
\caption{Variation in analytical solution of $x_{f}~(=\frac{m_{\rm DM}}{T_f})$ as in Eq.~\ref{analyticformxf} by choosing different values of $c$, where $c=\Delta(x_{f})/Y_{eq}(x_{f})$. }
\label{fig:comp_BEQ_Apprx1}
\end{figure}

Near freeze-out, i.e. for $x \sim x_{f}$, one can assume $\Delta(x_f) =c~Y_{eq}(x_f)$~\citep{Kolb} where $c$ is an unknown constant. The BEQ 
in such a case turns out to be:
\bea
\frac{dY_{eq}}{dx}|_{x=x_f} &=&-\frac{A}{x_f^5} \Delta(x_f)~ \Big(Y_{eq}^2(x_f)+2\Delta(x_f)~ Y_{eq}(x_f)+{\Delta^2(x_f)}\Big) ~, \nonumber \\
\Rightarrow \Big(1-\frac{3}{2 x_f} \Big) &=& \frac{A}{x_f^5} c (c+1)^2 Y_{eq}^2(x_f) \hspace{2.5cm}~~~~~~~~~~~~\bigg[{\rm using}\;\Delta(x_f) =c~Y_{eq}(x_f)\bigg] \nonumber \\
\Rightarrow \Big(1-\frac{3}{2 x_f} \Big) &=& \frac{A}{x_f^5} c (c+1)^2 \Big(0.145~\frac{g_{\rm DM}}{g_*} x_f^{3/2}e^{-x_f} \Big)^2 ~~ \bigg[{\rm using} ~Y_{eq}=0.145~\frac{g_{\rm DM}}{g_*} x^{3/2}e^{-x}\bigg] \nonumber \\
\Rightarrow  \Big(x_f^2 - \frac{3}{2} x_f \Big) &=& 0.0024 \frac{g_{\rm DM}^2}{\sqrt g_*}~c (c+1)^2~M_{Pl} ~m_{\rm DM}^4~\langle{\sigma v^2}\rangle_{3_{\rm DM} \to 2_{\rm DM}}~ e^{-2 x_f} ~~~~~~~~~\:\bigg[{\rm using}~ A\bigg] \nonumber \\
\Rightarrow x_f^2 &=& 0.0024 \frac{g_{\rm DM}^2}{\sqrt g_*}~c (c+1)^2~M_{Pl} ~m_{\rm DM}^4~\langle{\sigma v^2}\rangle_{3_{\rm DM} \to 2_{\rm DM}}~ e^{-2 x_f} 
\eea
One can solve for $x_f$ iteratively from above equation to obtain:
\bea\label{analyticformxf}
x_f && \approx \frac{1}{2} \ln \Big[ 0.0024 \frac{g_{\rm DM}^2}{\sqrt g_*}~c (c+1)^2~M_{Pl} ~m_{\rm DM}^4~\langle{\sigma v^2}\rangle_{3_{\rm DM} \to 2_{\rm DM}} \Big] \nonumber \\
&& ~~~ -2 \ln\Big[\frac{1}{2} \ln \Big[ 0.0024 \frac{g_{\rm DM}^2}{\sqrt g_*}~c (c+1)^2~M_{Pl} ~m_{\rm DM}^4~\langle{\sigma v^2}\rangle_{3_{\rm DM} \to 2_{\rm DM}} \Big]\Big]
\eea
Therefore, given the knowledge of DM mass and annihilation cross-section $\langle{\sigma v^2}\rangle_{3_{\rm DM} \to 2_{\rm DM}}$, one can find the decoupling or freeze-out temperature 
$x_f$. It is straightforward to show that for correct relic density (for example, with 
$m_{\rm DM} \sim 100$ MeV and $\langle{\sigma v^2}\rangle_{3_{\rm DM} \to 2_{\rm DM}} \sim 2.5 \times 10^6~ \rm{GeV}^{-5}$
as shown in the left panel of Fig.~\ref{fig:DBF}), $x_f \sim 20$, which is similar to WIMP like scenarios. This is shown in Fig.~\ref{fig:comp_BEQ_Apprx1} for different values of 
the unknown constant $c$ as a function of DM mass. We see that a large variation in $c$ produces only a small change in $x_f$ and indicate the stability of the solution. 

To evaluate relic density of DM, one needs to find out the yield after freeze out. We therefore need to focus at $x >> x_{f}$, where $Y_{eq} \to 0$. 
The Eq.~\ref{eq:BEQprx1} simplifies to a great extent to take the following form:
\bea
\frac{d\Delta}{dx} &=&-\frac{A}{x^5} \Delta^3 ~  \\
\int_{\Delta(x_f)}^{\Delta(x\to\infty)}-\frac{d\Delta}{\Delta^3}&=& A \int_{x_f}^{\infty} \frac{dx}{x^5} \nonumber \\
\Rightarrow \frac{1}{\Delta(x\to\infty)^{2}} &=&\frac{A^{2}Y_{eq}^{2}}{x_{f}^{10}}+\frac{A}{2x_{f}^{4}}=\frac{A(2AY_{eq}^{2}+x_{f}^{6})}{2x_{f}^{10}}~~\bigg[\rm{from} \: Eq.\ref{deltaxf},\Delta(x_{f})=\frac{x_{f}^{5}}{A\: Y_{eq}(x_{f})}\bigg]\nonumber \\  \Rightarrow \Delta(x\to \infty) &=& \sqrt{\frac{2}{A}}~x_f^2 \hspace{3cm}~~~~~~~~~~~~~~~~~~~~~~~~~~~~~~~~~~~\Bigg[A\: Y_{eq}^{2}<< x_{f}^{6}\bigg]\nonumber \\
\Rightarrow Y(x\to\infty) &=&  x_f^2 ~~\sqrt{\frac{2}{0.12 g_*^{\frac{3}{2}}~M_{Pl}~ m_{DM}^4~\langle{\sigma v^2}\rangle}_{3_{\rm DM} \to 2_{\rm DM}}}.
\label{eq:Yapprox}
\eea

Now, From Eq.~\ref{eq:OmegaExp1} and Eq.~\ref{eq:Yapprox}, one can write the expression of relic density as follows:
\bea\label{eq:Omegafinal}
\Omega h^2 &=& 2.752 \times 10^8 \left(\frac{m_{\rm DM}}{{\textrm {MeV}\times 10^{3}}}\right) \times \sqrt{\frac{2}{0.12 g_*^{\frac{3}{2}}~M_{Pl}~ m_{\rm DM}^4~\langle{\sigma v^2}\rangle_{3_{\rm DM} \to 2_{\rm DM}}}}~x_{f}^{2} \nonumber \\
&=& \frac{0.33}{{g_*}^{\frac{3}{4}}} \Big(\frac{\textrm {MeV} \times 10^{3}}{m_{\rm DM}}\Big)~x_f^2~\sqrt{\Big(\frac{{\textrm {GeV}}^{-5}}{\langle{\sigma v^2}\rangle_{3_{\rm DM} \to 2_{\rm DM}}}\Big)}~.
\eea

\begin{figure}[htb!]
$$
\includegraphics[scale=0.26]{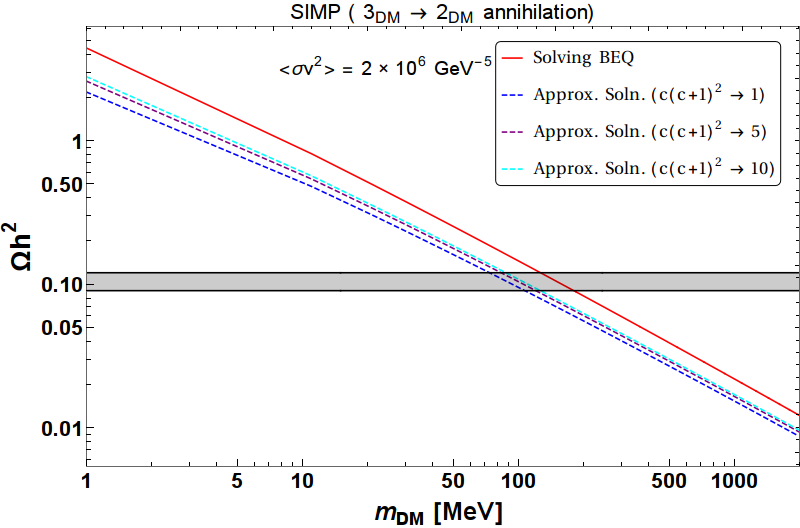}
\includegraphics[scale=0.26]{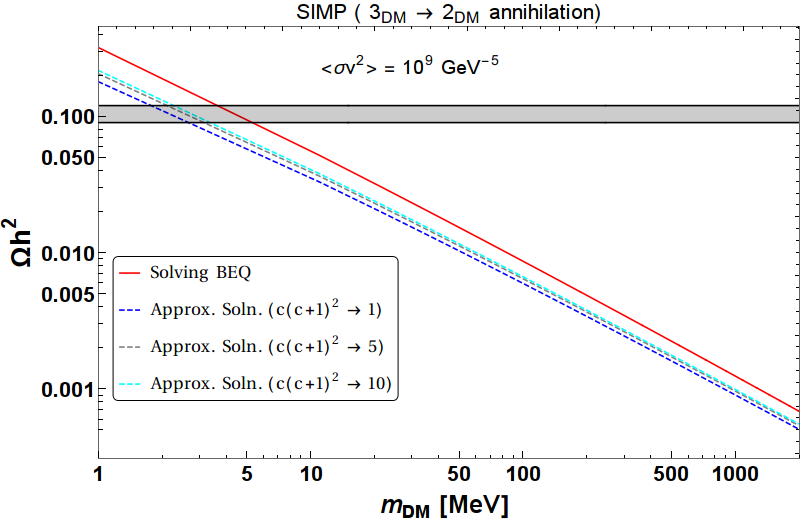}
$$
\caption{Comparison of relic density obtained by numerical solution to BEQ in Eq.~\ref{eq:BEQ3to2} and that from approximate analytical solution 
obtained in Eq.~\ref{eq:Omegafinal} as a function of DM mass for different choices of $c$. We choose two different annihilation cross-section 
$\langle\sigma v^{2}\rangle_{3_{\rm DM} \to 2_{\rm DM}}=\{2\times10^6,10^9\} ~{\rm GeV^{-5}}$ in left and right panel respectively. 
Correct relic density ($\Omega_{\rm DM}h^{2}=0.1199\pm 0.0022$) is indicated by the grey shaded band.}
\label{fig:comp_BEQ_Apprx2}
\end{figure}

Now, we are in a position to check the reliability of the analytical solution for DM relic density obtained for the SIMP like case (Eq.~\ref{eq:Omegafinal}) 
to that of the numerical solution obtained from the BEQ~\ref{eq:BEQ3to2}. This is shown in Fig.~\ref{fig:comp_BEQ_Apprx2}, where
we plot relic density obtained from both numerical solution and approximate analytical solution together for different 
values of $c$. Two different annihilation cross-sections $\langle\sigma v^{2}\rangle_{3_{\rm DM} \to 2_{\rm DM}}=\{2\times10^6,10^9\} ~{\rm GeV^{-5}}$ are shown
in left and right panel respectively. We see from Fig.\ref{fig:comp_BEQ_Apprx2}, that the analytical solution closely 
mimic the numerical solution for higher values of DM mass ($\sim$ GeV).
Actually, the cause of this discrepancy in relic density obtained between numerical and analytical solution occurs 
when we simplify the Eq.\ref{eq:BEQprx1} to Eq.\ref{eq:Yapprox} to only retain terms of the order $\sim \Delta^3$. 
If we consider second order term in $\Delta(x)$, the equation looks like that of Abel equation of first kind~\citep{eq_abel}, 
solution of that will mimic the numerical solution even more closely. 
\section{Model specific analysis of a SIMP Framework}
\label{sec3}
\subsection{The Model}

If simplicity is the guiding principle to realise a SIMP paradigm, one should focus on scalar DM ($\chi$). 
The DM also need to possess an additional symmetry for stability (call it a
dark symmetry) distinct from that of the SM. If we require a vertex consisting of three DM fields ($\chi^3$) for the DM to enable a $3_{\rm DM} \to 2_{\rm DM}$ interaction, 
the minimal choice for the symmetry under which $\chi$ transforms non trivially is $\mathcal{Z}_3$. 
As the roots of $\mathcal{Z}_3$ are complex ($1,\omega,\omega^2$), the scalar DM $\chi$ needs to be complex. 
In principle, this is enough to ideate $3_{\rm DM} \to 2_{\rm DM}$ interactions through $\chi$ mediation itself. However, it turns out that relic density allowed parameter space 
for this simplest possibility is quite restrictive and even more so when we impose the self scattering 
(we will have explicit demonstration later) and unitarity bound. We can enlarge the available parameter space by connecting the graph for 
 $3_{\rm DM} \to 2_{\rm DM}$ process to the other end in presence of a mediator, which doesn't have $\mathcal{Z}_3$ charge. 
But, this can not be realised with a SM particle (even if Higgs has a portal interaction with our DM) 
unless we augment the SM with another additional field. Again, the minimal choice of such mediator 
will be another scalar $\phi$ (real scalar for simplicity) which is singlet under SM. 

Therefore, in this model, we consider a complex scalar singlet field $\chi$ which transforms under $\mathcal{Z}_3$ 
and acts as DM, while the real scalar singlet $\phi$ do not transform under $\mathcal{Z}_3$. 
The $\mathcal{Z}_3$ transformation properties of the fields is mentioned in Table \ref{table:1}.
In SIMP paradigm, the freeze out is mainly driven by $3_{\rm DM} \to 2_{\rm DM}$ number changing process, so the $2_{\rm DM} \to 2_{\rm SM}$ interaction can be 
killed by choosing a negligible value of the Higgs portal coupling. Now, if we provide VEV to $\phi$, then it will mix with SM Higgs 
after spontaneous symmetry breaking and will mediate the number changing process in the dark sector. The mass of the additional scalar can be 
fairly light (being singlet) and will aid to annihilation cross-section providing cushion to the DM coupling to remain within perturbative limit.  

\begin{table}[htb!]
\centering
\begin{tabular}{|c|c|c|}
\hline
\textbf{ Particle } & \textbf{ Nature } &\textbf{ $\mathcal{Z}_3$ transformation}\hspace{0.35cm}\\
\hline 
$\chi$ \,\, & Complex Scalar Singlet \,\,&$\omega$ \hspace{0.02cm}\\
$\phi$\,\,& Real Scalar Singlet \,\,& 1 \hspace{0.05cm}\\
$H$ \,\, & SM Higgs Doublet \,\,&$1$ \hspace{0.02cm}\\
 \hline
\end{tabular}
\caption{$\mathcal{Z}_3$ charges of the additional scalar fields assumed in the model ($\chi,\phi$).}
\label{table:1}
\end{table}

The relevant Lagrangian for this model can be mainly segregated into two parts :
\begin{equation}
\mathcal{L} = \mathcal{L}_{\textrm{SM}} + \mathcal{L}_{\textrm{BSM}}. 
\end{equation}
Here, we are interested in the part describing the dark sector:
\begin{equation}
\label{test}
\mathcal{L}_{\textrm{BSM}} = \frac{1}{2}(\partial^ \mu \phi) (\partial _\mu \phi)+(\partial ^\mu \chi)^*(\partial _\mu \chi) - V(H,\phi,\chi).
\end{equation}
The scalar potential involving the additional scalars and SM Higgs ($H$) reads as~\cite{yonit,Beniwal:2018hyi}:
\bea
\label{eq:potential}
V(H,\phi,\chi)&=&
- \mu_H^2 H^{\dagger }H+ \lambda _H (H^{\dagger} H)^2-\frac{1}{2}\mu_{\phi }^2 \phi^2 + \frac{1}{4} \lambda _{\phi }~\phi ^4 \nonumber \\
&& + \frac{\mu_{3}}{3} \phi^3 +\frac{1}{2}\lambda _{\phi h} ~\phi^2~ H^{\dagger}H + \mu_{\phi h}~\phi~ (H^{\dagger} H) \nonumber \\
&& + \mu^2 |\chi|^2 + \lambda _{\chi } |\chi|^4 +\frac{1}{3!} \mu _{\chi }(\chi ^3+\chi^{* 3})+\lambda_{\chi h}~ |\chi|^2  H^{\dagger } H \nonumber\\
&& +\frac{1}{2}\lambda _{\chi \phi} |\chi|^2  \phi ^2 +\mu_{\chi \phi} ~\phi~ |\chi|^2+\frac{1}{3!} Y_{\chi \phi }~\phi ~ (\chi^{3}+\chi^{* 3}).
\eea

As has already been mentioned, $3_{\rm DM}\to 2_{\rm DM}$ interactions are mediated by the self couplings of $\chi$, namely involving $|\chi|^4$ and $ \chi ^3$ terms.
$\phi$ mediates additional channels through the two terms $\chi^3\phi$ and $|\chi|^2  \phi ^2$, when $\phi$ acquires a VEV. 
After spontaneous symmetry breaking (SSB), $\phi$ and $H$ mixes through their VEVs ($v_\phi$ and $v_h$) as follows:

\begin{align}
\phi \to \Phi+v_{\phi},\\
H \to \begin{pmatrix}0\\\frac{h+v_{h}}{\sqrt{2}} \end{pmatrix}.
\end{align}
The squared mass matrix for the interaction basis, $(h~\Phi)^{T}$ is given as,
\begin{align}
\label{nondiag}
M^{2}_{h \Phi}=
\begin{pmatrix}
2 v_{h}^{2}\lambda_{h}&\:v_{h}v_{\phi}\lambda_{\phi h}+v_{h}\mu_{h\phi}\\v_{h}v_{\phi}\lambda_{\phi h}+v_{h}\mu_{h\phi}&\quad\mu_{3} v_{\phi}+2v_{\phi}^{2}\lambda_{\phi}-\mu_{h\phi}(v_{h}^{2}/2v_{\phi}),
\end{pmatrix}=
\begin{pmatrix}
A&B
\\
B&C
\end{pmatrix}.
\end{align}
 
The physical scalars ($h_1$ and $h_2$) are obtained from $h,\Phi$ by choosing the following transformation,
\begin{align}
\begin{pmatrix}
h_{1}\\ h_{2}
\end{pmatrix}
=\begin{pmatrix}
\cos\theta &-\sin\theta \\ \sin\theta & \cos\theta
\end{pmatrix}
\begin{pmatrix}
h\\ \Phi
\end{pmatrix}
\end{align}

The mass eigenvalues are therefore obtained by diagonalising the above mass matrix ($M^{2}_{h \Phi}$) and are given by:
\begin{equation}
\begin{aligned}
m_{h_1}^{2}=&\: A\: \cos^{2} \theta+C\:\sin^{2}\theta-B\:\sin 2\theta\\
m_{h_2}^{2}=& \: A\: \sin^{2} \theta+C\:\cos^{2}\theta+B\:\sin 2\theta.
\end{aligned}
\end{equation}
The physical states are related to the flavour states through the mixing angle $\theta$ as:
\bea
\label{angle}
\tan\:2\theta = \frac{2B}{C-A}
\eea

Now, we are all set to address the phenomenology of the scalar sector. Let $h_{2}$ be the SM like Higgs ($m_{h_{2}}=125 ~{\rm GeV}$ and $v_{h}=246 ~{\rm GeV}$)
and $h_{1}$ be the additional scalar boson. The additional scalar being a singlet predominantly, can be heavier or lighter than the SM Higgs, 
because it can't be produced at colliders easily. We will be interested in the light Higgs mass region, where we will have $\sin\theta \to 1$, for above mixing assignment. 
Finally, we point out that we can easily rewrite some of the coupling parameters as a function of the physical masses after SSB as follows
~\citep{yonit,subhaditya,Ghosh:2017fmr}:
\bea
\label{eq:derived-para}
\nonumber \mu _{\text{$\phi $h}}&=&-\frac{2 v_{\phi } }{v_h^2}\bigg(\sin ^2 \theta m_{h_1}^2+ \cos ^2\theta m_{h_2}^2+v_{\phi } \left(-2 \lambda
   _{\phi } v_{\phi }+\mu _3\right)\bigg)~, \\
\nonumber \lambda _{\text{$\phi $h}}&=&\frac{1}{v_h v_{\phi }}\bigg(\text{sin$\theta $}\: \text{cos$\theta $}  \left(m_{h_2}^2-m_{h_1}^2\right)-v_h \mu _{\text{$\phi $h}}\bigg)~,\\
\lambda _h&=&\frac{1}{2 v_h^2}\bigg(\sin ^2 \theta m_{h_2}^2+ \cos ^2\theta m_{h_1}^2\bigg).
\eea

The freedom of choosing other parameters will help us to get a correct Higgs mass even if we vary the following 
parameters to address correct relic density for DM in this model:
\begin{equation}
\label{parameters}
\{m_{\chi}(=m_{\rm DM}),Y_{\chi \phi},\sin \theta, \lambda_{\chi \phi},\lambda_{\chi h},m_{h_{1}},v_{\phi},\mu_{\chi},\mu_{\chi\phi},\lambda_{\chi}\} ~.
\end{equation} 
After SSB the DM mass turns out to be : $m_\chi^2=\mu^2+\frac{1}{2}~\lambda_{\chi\phi}~v_\phi^2+\mu_{\chi\phi}~v_\phi+\frac{1}{2}\lambda_{\chi h}~v_h^2$. 
Again, due to large number of parameters dictating DM mass, we will vary DM mass ($m_\chi$), along with ($\mu_\chi, v_\phi, Y_{\chi \phi}$)
 independently to search for available parameter space of the model.
\subsection{Relic density outcome}
The model at hand offers both SIMP like and WIMP like solution as it has both self coupling and coupling to SM. 
For SIMP framework to be operative, a very tiny coupling with SM is realised by taming $\lambda_{\chi h}$ and $\lambda_{\chi \phi}$. 
The Feynman diagrams that leads to $3_{\rm DM} \to 2_{\rm DM}$ number changing processes in this framework are shown in Appendix \ref{feyn3to2dm1}. 
There are four annihilation processes that dictate relic density of the DM, they are 
$\chi\chi\chi\to\chi\chi^{*}$, $\chi\chi^{*}\chi^{*}\to\chi\chi$ and their complex conjugate processes 
i.e. $\chi^{*}\chi^{*}\chi^{*}\to\chi^{*}\chi$ and $\chi^{*}\chi\chi\to\chi^{*}\chi^{*}$ respectively \footnote{One may note that in presence of $\mathcal{Z}_3$ symmetry, 
one may also have semi annihilations like $\chi \chi \to \chi^{*} h_{1}$ or $\chi \chi \chi^{*} \to \chi^{*} \chi^{*} h_{1}$. However, their contributions will be small due to small 
$\lambda_{\chi h}$ and $\lambda_{\chi \phi}$ couplings assumed for SIMP realisation to work.} 
The diagrams in each cases can be categorized into two classes, (i) mediated by self interaction of $\chi$, (ii) mediated by the scalars $h_1~\&~h_2$.
We implemented this model using LanHEP~\cite{Semenov:2008jy}. To check the consistency with our numerical calculations, 
we have used CalcHEP \cite{calchep}, for drawing the Feynman diagrams we have used Tikz-Feynhand \cite{feyndiag} and in order to 
calculate the matrix amplitude and relic density, we have used Mathematica \citep{ram2010}. 
Vertex factors used in the calculation of each matrix amplitudes are also detailed in Appendix \ref{feyn3to2dm1}. 
Here we note that the numerical solution to the SIMP like BEQ
have been used to scan the parameter space to yield relic density, 
instead of the approximate analytical solution advocated before. 

It is straightforward to see that the matrix element squared for the complex conjugate processes are same: 
$$
|\mathcal{M}_{\chi\chi\chi\to\chi\chi^{*}}|^{2}=|\mathcal{M}_{\chi^{*}\chi^{*}\chi^{*}\to\chi^{*}\chi}|^{2}, ~ |\mathcal{M}_{\chi\chi^{*}\chi^{*}\to\chi\chi}|^{2}=|\mathcal{M}_{\chi^{*}\chi\chi\to\chi^{*}\chi^{*}}|^{2}.
$$
Therefore, the total $3_{\rm DM} \to 2_{\rm DM}$ annihilation cross section in this model is given by:
\bea
\nonumber \langle\sigma v^2\rangle_{3_{\rm DM}\to2_{\rm DM}}&=&2[\langle\sigma_{\chi\chi\chi\to\chi\chi^{*}} v^{2}\rangle+\langle\sigma_{\chi\chi^{*}\chi^{*}\to\chi\chi} v^{2}\rangle]~,
\\
&=&\frac{2\sqrt{5}}{192\pi m_{\chi}^{3}}\bigg(|\mathcal{M}_{\chi\chi\chi\to\chi\chi^{*}}|^{2}+|\mathcal{M}_{\chi\chi^{*}\chi^{*}\to\chi\chi}|^{2}\bigg),
\eea
where the last line corresponds to $s$-wave computation of the annihilation cross section, also detailed in appendix \ref{feyn3to2dm1}.  
For SIMP realization, we choose 
$\lambda_{\chi\phi}$ and $\lambda_{\chi h}$ very tiny $ \sim 0.001$. Since we are also interested in exploring the light Higgs mediation
to expedite the annihilation processes, we have kept the value of mixing angle $\sin\theta = 0.999(\to1)$. Keeping above parameters as quoted, 
we are now left with the following free parameters:
\begin{equation}
\{m_{\chi},Y_{\chi \phi},m_{h_{1}},v_{\phi},\mu_{\chi},\mu_{\chi\phi},\lambda_{\chi}\}.
\end{equation}

\begin{figure}[htb!]
$$
\hspace{-0.75cm}
\includegraphics[scale=0.26]{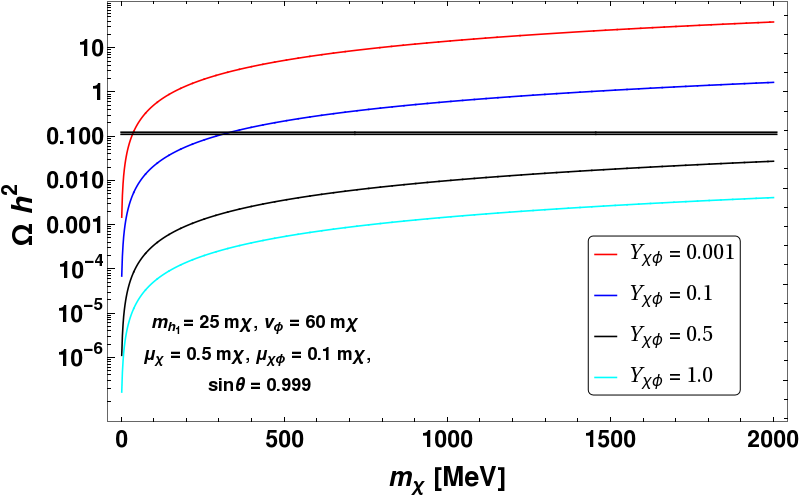}
\includegraphics[scale=0.26]{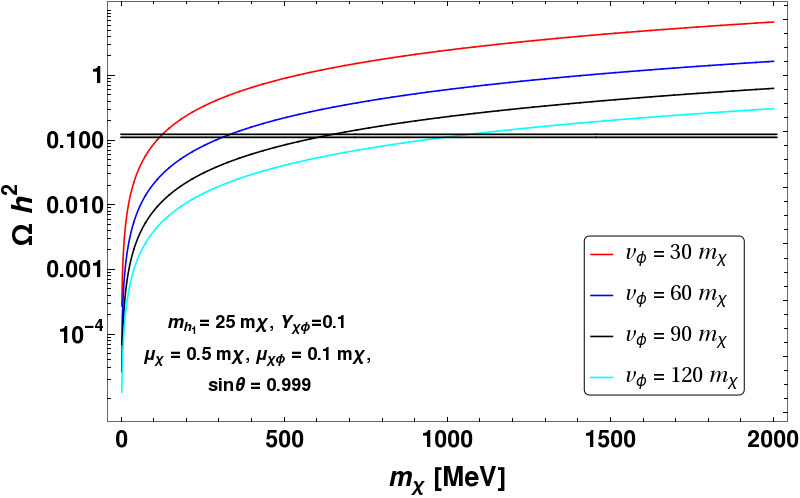}
$$
\caption{Variation of Relic density with DM mass  for different values of $Y_{\chi \phi}$ [Left Panel] and $v_{\phi}$ [Right panel]. We kept the self coupling large ($\lambda_\chi=1$) for both the plots. The correct relic density ($0.1177 \leq \Omega h^{2} \leq 0.1221$) is 
also indicated here by the horizontal grey band.}
\label{fig:sigma-v1}
\end{figure}

\begin{figure}[htb!]
$$
\includegraphics[scale=0.26]{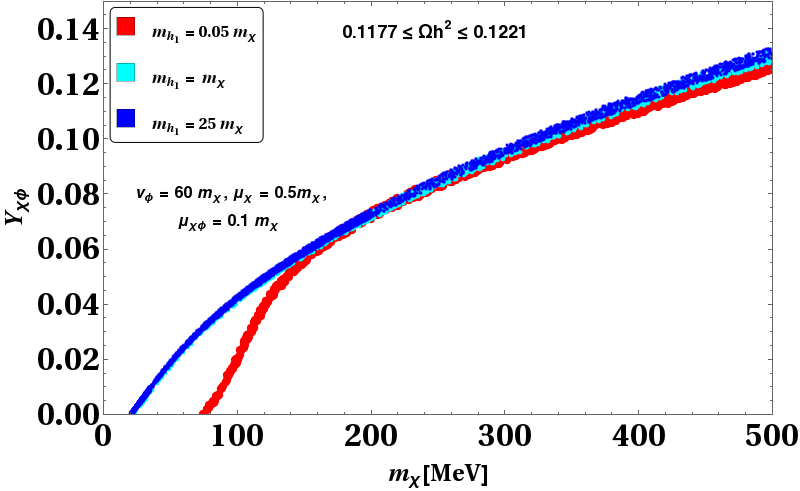}~
\includegraphics[scale=0.26]{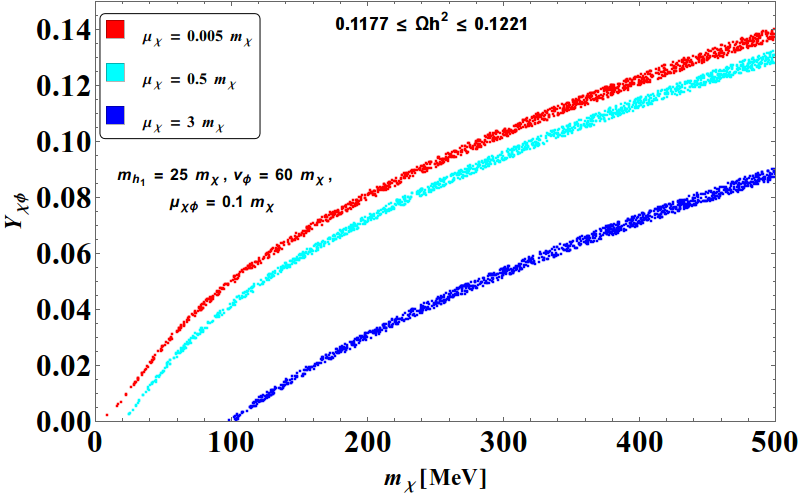}
$$
$$
\includegraphics[scale=0.26]{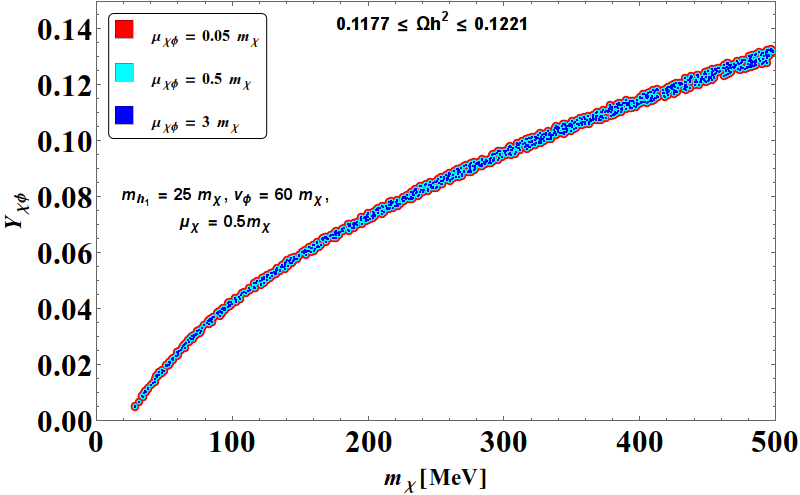}~~
\includegraphics[scale=0.26]{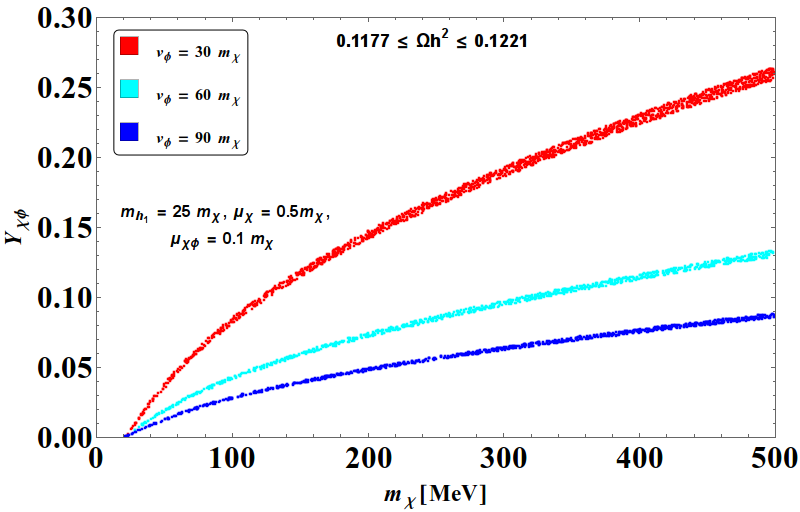}
$$
\caption{Relic density allowed ($0.1177 \leq \Omega h^{2} \leq 0.1221$) parameter space in $m_{\chi}-Y_{\chi\phi}$ plane with variation of 
$m_{h_{1}}$ (Top Left), $\mu_{\chi}$ (Top Right), $\mu_{\chi\phi}$ (Bottom Left) and $v_{\phi}$ (Bottom Right). Other parameters kept fixed, 
and the range of variation are mentioned in the respective figure inset. We choose $\lambda_\chi=1$ for illustration.}
\label{fig:Y-m}
\end{figure}

Now we will study the variation of relic density with DM mass, keeping most of the other parameters steady. 
In Fig.~\ref{fig:sigma-v1}, we show such a variation with respect to different choices of $Y_{\chi\phi} \sim\{0.001 \to 1\}$ in the left panel and for different choices of 
$v_\phi\sim \{30m_\chi \to 120m_\chi\}$ in the right panel (the parameters kept constant are mentioned in the figure inset). We have kept $\lambda_\chi=1$ for both the plots. The outcome from the left panel is 
understood easily, with larger $Y_{\chi\phi}$, the $3_{\rm DM} \to 2_{\rm DM}$ annihilation gets larger and that diminishes the relic density significantly.  Therefore, $Y_{\chi\phi}$ serves as one of the key parameters to find 
correct relic density in this model, and is used for the numerical scan performed later. Similarly, from the right panel, we see that $v_\phi$ turns out to be an important parameter to find the correct relic of this 
DM, as with larger $v_\phi$, the annihilation cross-section increases and subsequently the relic density drops. The effects of $Y_{\chi\phi}$ and $v_\phi$ can also be validated from the expressions 
of annihilation cross-sections detailed in Appendix \ref{feyn3to2dm1}.  As stated before, we use the numerical solution obtained from the BEQ. 

Next in Fig.~\ref{fig:Y-m}, we show the relic density allowed parameter space in $m_\chi-Y_{\chi\phi}$ plane by varying $m_{h_{1}}$ (Top Left), 
$\mu_{\chi}$ (Top Right), $\mu_{\chi\phi}$ (Bottom Left) and $v_{\phi}$ (Bottom Right) with other parameters fixed as mentioned in figure inset. We again choose 
$\lambda_\chi=1$ for this plot. The available parameter space has a large DM mass range upto GeV with larger $Y_{\chi\phi}$ (going upto 0.4). We also see that variation in $\mu_\chi$
and $v_\phi$ affect relic density quite significantly (top right and bottom right respectively) allowing a wide span of relic density allowed parameter space. 
This is easily seen from the vertex factors in Appendix \ref{feyn3to2dm1}, that the three point vertex is directly proportional to $\mu_\chi$ and also on $v_\phi$ thanks to 
$\phi \chi^3$ term, which crucially controls the annihilation cross-section through self mediation.  From the top left figure in Fig.~\ref{fig:Y-m}, we also see 
that a light scalar (red points depicted by choosing $m_{h_1}=0.05 m_{\chi}$) show a departure from the choices of heavy scalar ($m_{h_1}= m_{\chi}, 25m_{\chi}$ shown by 
cyan and dark blue points) for sufficiently small DM mass $\le 150$ MeV. Again, note here that due to the freedom of having a large 
number of parameters contributing to $m_{h_{1,2}}$, we can fix Higgs mass ($m_{h_2}$) to 125 GeV and still vary $m_{h_1}$ keeping $v_\phi=60 m_\chi$ as in the top left panel. 
Also note here, that stability of the scalar potential constrains the dimensionful cubic couplings $\mu_\chi$ and $\mu_{\chi\phi}$ to lie within $3 m_\chi$ in a conservative limit as adopted for the scans.

To summarise this section, we see that a large parameter space is available from relic density 
constraint, particularly the DM mass can vary in a large range even upto GeV, while the relevant couplings $Y_{\chi\phi}, \lambda_{\chi}$ 
do not require to be very large. These are all in contrary to the naive SIMP realisation of DM ideally having one self coupling and one 
mass parameter dictating them to be in the strong interaction range. However, we need to consider other constraints like unitarity, 
vacuum stability and self scattering cross section, which will constrain the relic density allowed parameter space as we discuss below.  
\subsection{Additional Constraints on dark matter parameter space}
In this section, we discuss three important constraints on the model parameter space coming from vacuum stability, 
unitarity and DM self interaction cross-section limit. All the couplings are assumed positive to cope up with the vacuum stability of the scalar potential.
\subsubsection{Self scattering cross section}
DM self scatters through $2_{\rm DM} \to 2_{\rm DM}$ scattering process like $\chi \chi \to \chi\chi$ and $\chi \chi^* \to \chi \chi^*$ and their complex conjugate processes. 
Feynman graphs and the matrix elements are detailed in Appendix \ref{selfDMann}. The self scattering cross-section is then obtained as:
\begin{align*}
\sigma_{self}&=2[\sigma_{\chi\chi\to\chi\chi}+\sigma_{\chi\chi^{*}\to\chi\chi^{*}}]\\&=\frac{2}{64\pi m_{\chi}^{2}}\bigg(|\mathcal{M}_{\chi\chi\to\chi\chi}|^{2}+|\mathcal{M}_{\chi\chi^{*}\to\chi\chi^{*}}|^{2}\bigg).
\end{align*}
Again, we have used the fact that the matrix element for $\chi \chi \to \chi \chi$ and $\chi^* \chi^* \to \chi^* \chi^*$ are same. There are two important bounds on the 
self scattering cross-section for DM coming from Bullet cluster and Abell cluster data as follows:
\begin{itemize}
\item Bullet cluster bound~\cite{bullet}:
\begin{align}
\label{bullet}
\sigma_{self}/m_{\chi}\;\lesssim 1~ {\rm cm}^{2}/{gm}\hspace{0.125cm}~(= 4555.8 ~{\rm GeV}^{-3})
\end{align}
\item Abell cluster bound~\cite{abell}:
\begin{align}
\label{abell}
1~{\rm cm}^{2}/{gm}\hspace{0.125cm}\lesssim \sigma_{self}/m_{\chi}\; \lesssim 3~{\rm cm}^{2}/{gm}\hspace{0.125cm}
\end{align}
\end{itemize}
As one can see that the bounds above do not have an overlap to each other. We will use one or the other to see the constraints on the model parameter space. 

\begin{figure}[htb!]
$$
 \includegraphics[scale=0.25]{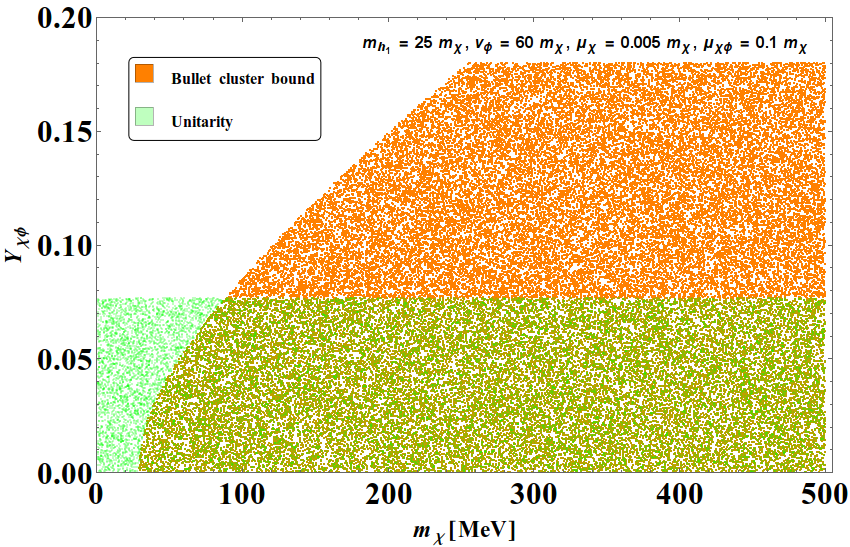}~~
 \includegraphics[scale=0.25]{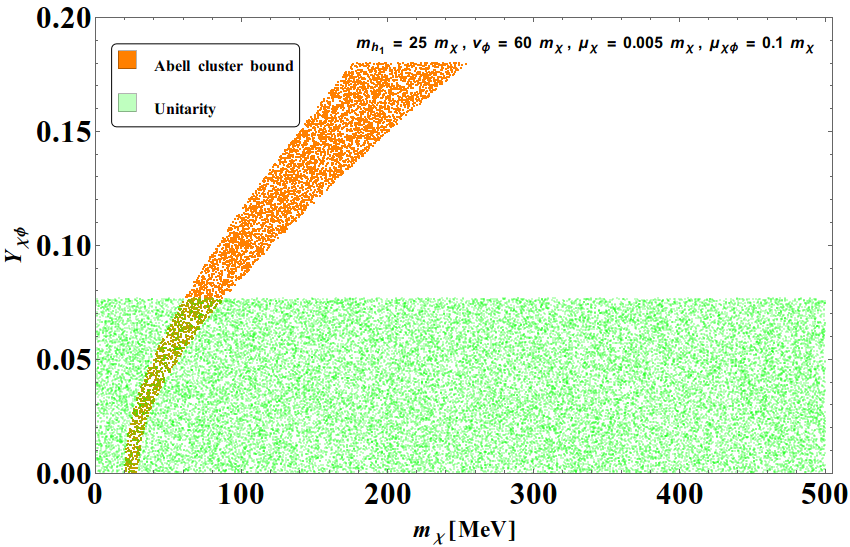}
$$	
\caption{Unitarity Bound (Green) and Self scattering cross-section limit (Orange) in $m_{\chi}-Y_{\chi\phi}$ plane of our model. 
We have kept other parameters fixed as mentioned in the figure inset. Bullet Cluster bound (Eq.~\ref{bullet}) is shown in the left panel 
and Abell Cluster bound (Eq.~\ref{abell}) is shown in the right panel. We have kept $\lambda_\chi=1$ for this plot.}
\label{unitarity-self}
\end{figure}
\subsubsection{Unitarity Bound}
Unitarity of $S$ matrix constrains the matrix element of the $2_{\rm DM} \to 2_{\rm DM}$ scattering process via\footnote{This can be derived from optical theorem using partial wave analysis~\cite{Peskin:1995ev}.}
\begin{align}
|\mathcal{M}_{\chi \chi\to\chi\chi}|\leq 8 \pi,~
|\mathcal{M}_{\chi \chi^{*}\to\chi\chi^{*}}|\leq 8 \pi.
\end{align}
$|\mathcal{M}_{\chi \chi\to\chi\chi}|$ and $|\mathcal{M}_{\chi \chi^{*}\to\chi\chi^{*}}|$ are mentioned in details of the model in Appendix \ref{selfDMann}.  It turns out to be one of the most stringent bounds on the model parameter space as we demonstrate below. In addition, we also obey the perturbative limit on each of the 
couplings as assumed in the model $|\lambda_i|<4 \pi$.

In Fig.~\ref{unitarity-self}, we have plotted the available parameter space in $m_{\chi}-Y_{\chi\phi}$ plane of our model coming from 
self scattering cross-section limits from Bullet cluster data (Eq.~\ref{bullet}) in the left panel 
and Abell cluster data (Eq.~\ref{abell}) in the right panel by green shaded region together with unitarity bound by 
orange shaded region. The plot is obtained by keeping $\lambda_\chi=1$, while other choices of parameters 
are mentioned in the figure inset. Unitarity bound strongly constrains $Y_\chi\phi \lesssim 0.07$.
\subsection{Summary of available parameter space from all constraints}
In this section we will address the available parameter space of the model which satisfy all the bounds together.
\begin{figure}[htb!]
$$
 \includegraphics[scale=0.25]{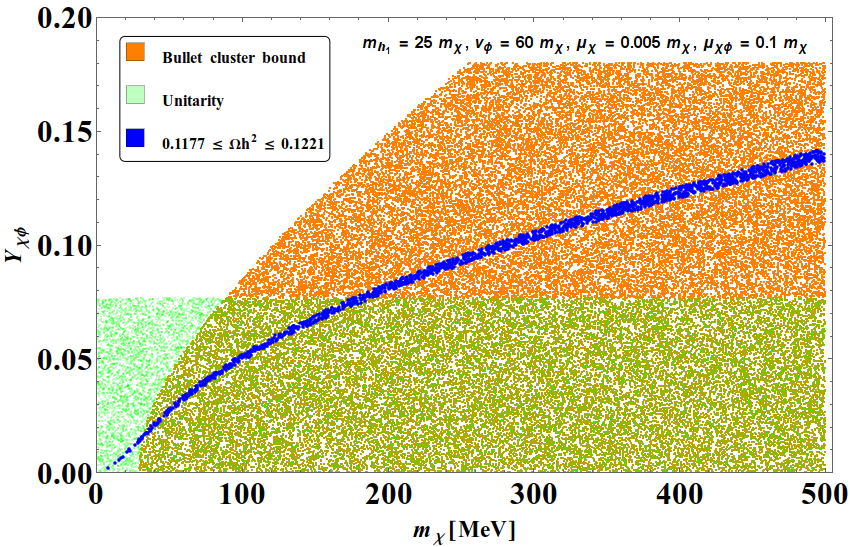}~
 \includegraphics[scale=0.27]{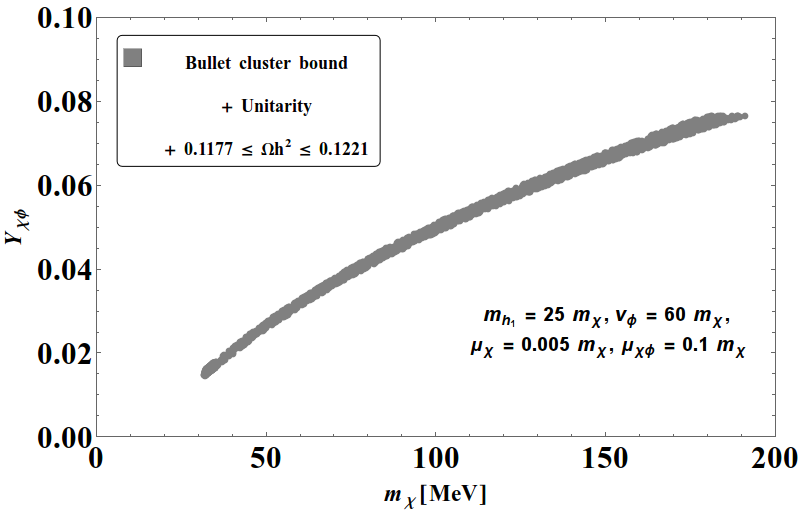}
 $$
 \caption{[Left Panel] Self scattering bound for Bullet cluster (Orange), Unitarity (Green) and Relic density (Blue) allowed regions are plotted in $m_{\chi}-Y_{\chi\phi}$ plane where the other parameters are mentioned inside the figure. 
 [Right Panel] Combined parameter space allowed from all the bounds.}   
\label{fig:bullet}
\end{figure}

\begin{figure}[htb!]
$$
\includegraphics[scale=0.25]{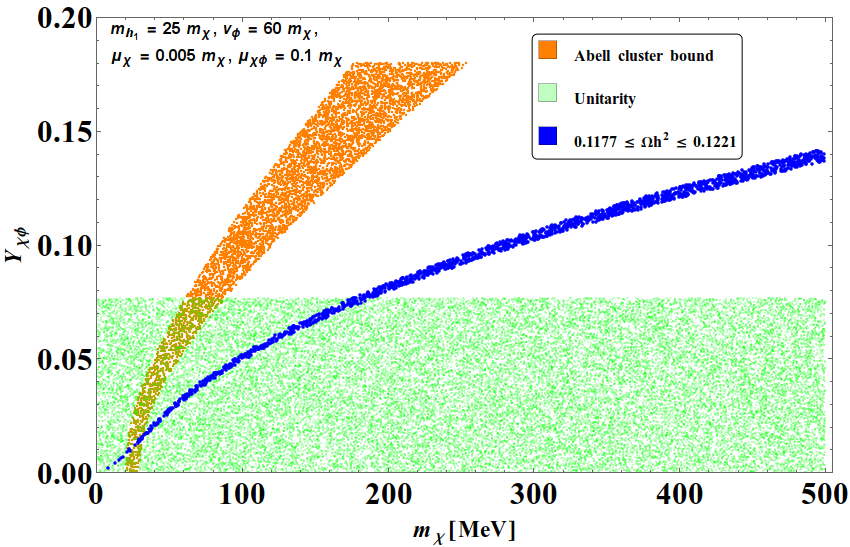}~
\includegraphics[scale=0.27]{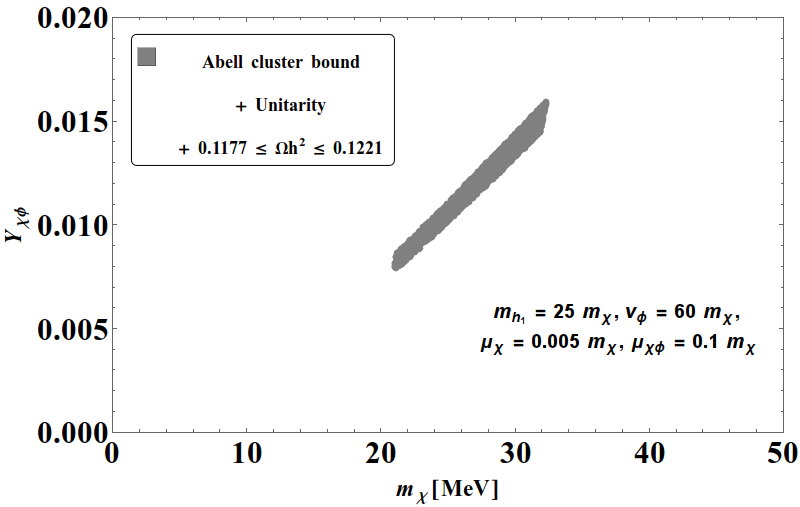}
$$
 \caption{[Left Panel] Self scattering bound for Abell cluster (Orange), Unitarity (Green) and Relic density (Blue) allowed regions are plotted in $m_{\chi}-Y_{\chi\phi}$ plane where the other parameters are mentioned inside the figure. [Right Panel] Combined parameter space for all Bounds.}
\label{fig:abell}
\end{figure}

\begin{figure}[htb!]
$$ \includegraphics[scale=0.25]{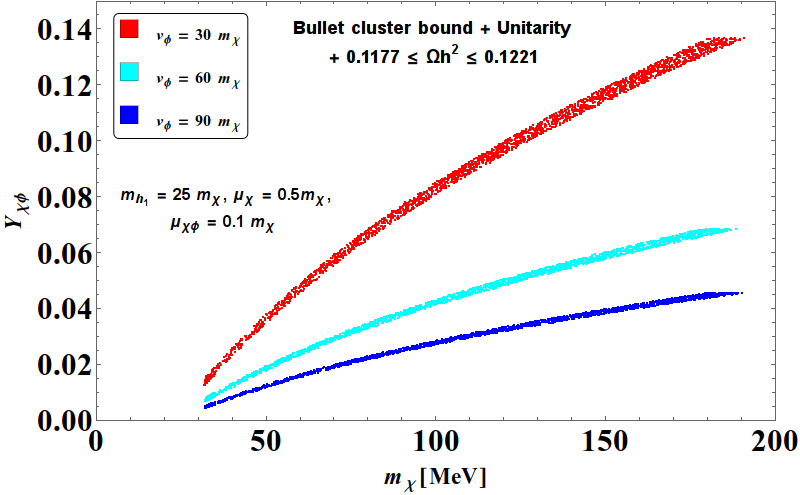}~~
\includegraphics[scale=0.25]{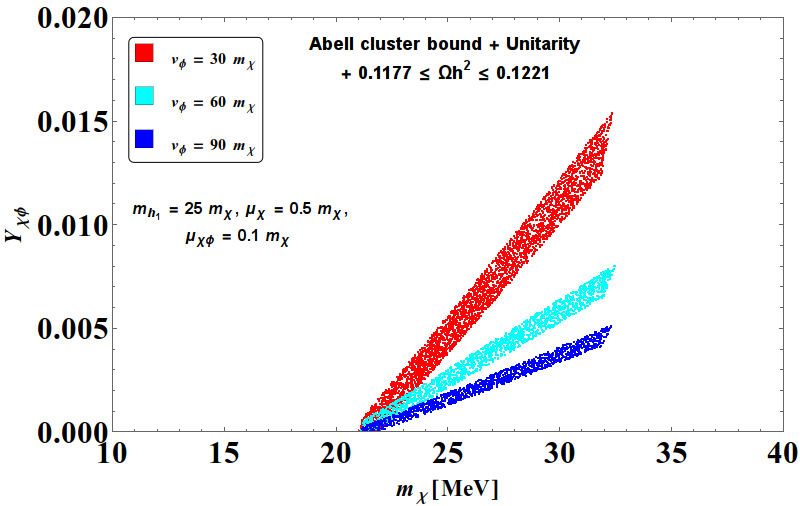}
$$
 \caption{Allowed parameter space in $m_{\chi}-Y_{\chi\phi}$ plane for different choices of $v_{\phi}$ from relic density, unitarity and self scattering cross-section coming from [Left Panel] Bullet Cluster, 
 [Right Panel] Abell Cluster constraints.}
 \label{fig:v-phi}
\end{figure}

\begin{figure}[htb!]
$$ \includegraphics[scale=0.25]{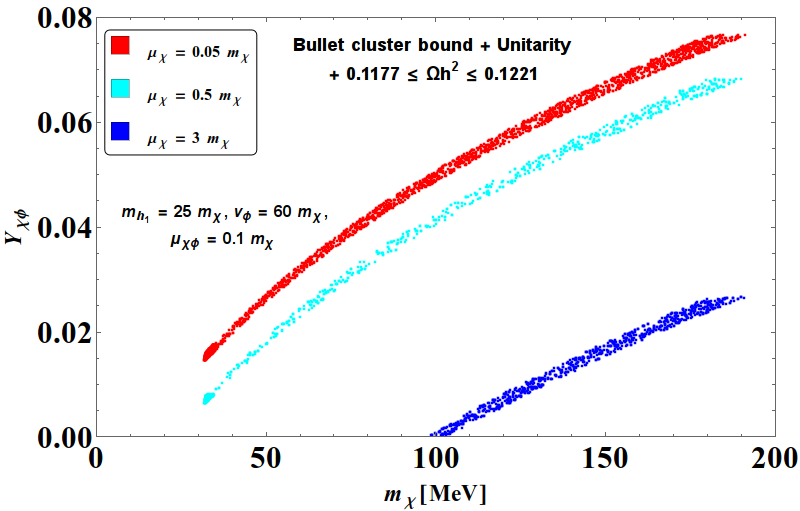}~~
\includegraphics[scale=0.25]{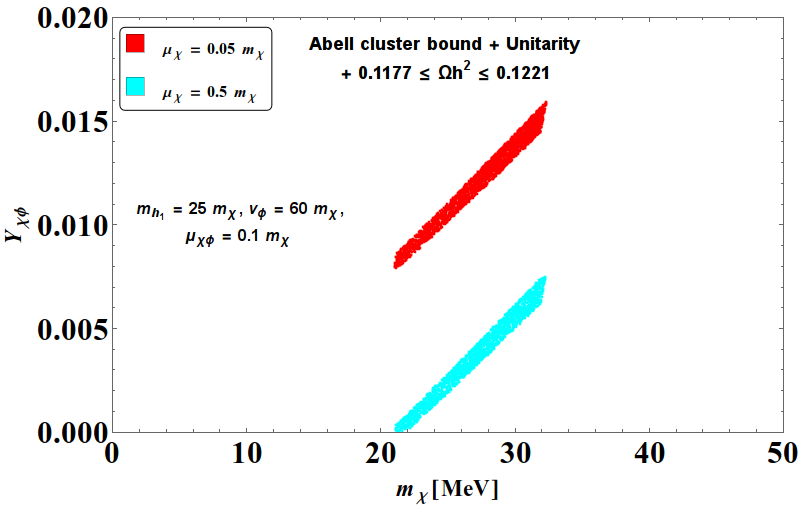}
$$
 \caption{Allowed parameter space in $m_{\chi}-Y_{\chi\phi}$ plane for different choices of $\mu_{\chi}$ from relic density, unitarity and self scattering cross-section coming from [Left Panel] Bullet Cluster, [Right Panel] Abell Cluster constraints.}
 \label{fig:mu-chi}
\end{figure}

In left panel of Fig.~\ref{fig:bullet}, we put together relic density, unitarity bound and self scattering constraint arising from Bullet cluster together 
in $Y_{\chi\phi}-m_\chi$ plane. The right panel figure shows the available parameter space after all these constraints. 
There are two important conclusions that we obtain from here: 
(i) the mass range of the DM is now limited to $\sim$ 200 MeV, while the coupling is restricted to a very small, 
$Y_{\chi\phi}\sim \{0.02 \to 0.08\}$ value. This is obtained with $\lambda_\chi=1$, chosen for this particular scan. We will show later that 
changing $\lambda_\chi$ to $\sim 0.1$ will not change the order of $Y_{\chi\phi}$ significantly.
A similar scan is presented in Fig.~\ref{fig:abell}, but with self scattering cross-section limit dictated by Abell cluster data. The available 
parameter space is further restricted for this case to remain within $\sim$ 40 GeV (right panel of Fig.~\ref{fig:abell}). 

In Fig.~\ref{fig:v-phi}, we show how the allowed parameter space changes due to different choices of $v_{\phi}$. 
Smaller $v_{\phi}$ requires larger $Y_{\chi\phi}$ to keep the annihilation cross-section at right ball park. Similarly in Fig.~\ref{fig:mu-chi}, 
we show how the available parameter space changes due to different choices of $\mu_\chi$ which also serves as an important parameter of the model. 
The behaviour is similar to $v_\phi$. With larger $\mu_\chi$, the coupling $Y_{\chi\phi}$ 
requires to be smaller to adjust right annihilation cross-section. We would also like to point out that in the right panel of 
Fig.~\ref{fig:mu-chi}, the bound from Abell cluster data do not yield a viable parameter 
space for the choice of $\mu_\chi=3m_\chi$, while keeping $Y_{\chi\phi}$ positive. 

\begin{figure}[htb!]
$$ 
\includegraphics[scale=0.25]{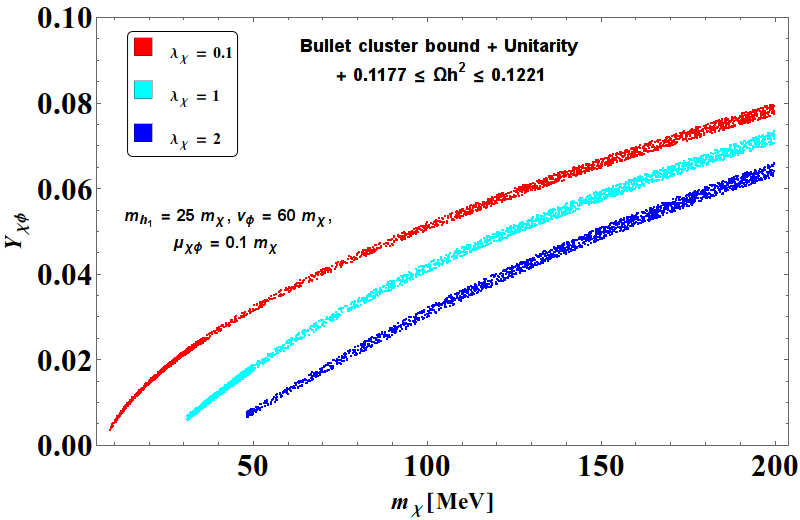}
~~
\includegraphics[scale=0.25]{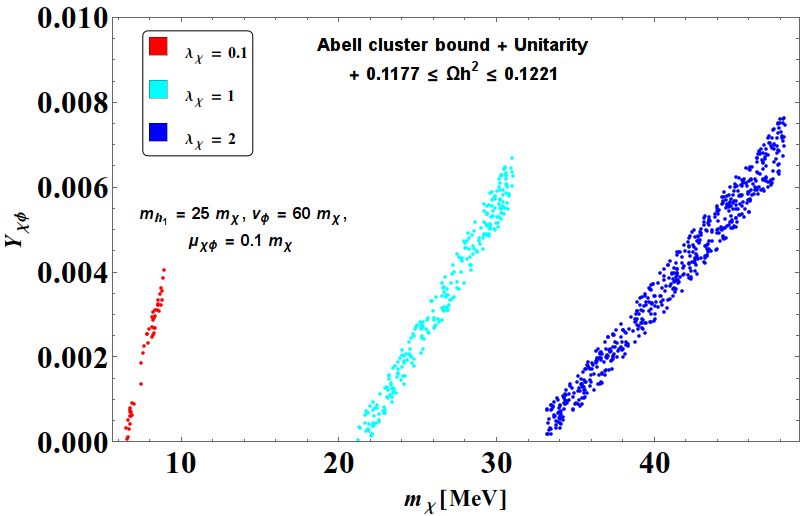}
$$
 \caption{Allowed parameter space in $m_{\chi}-Y_{\chi\phi}$ plane for different choices of $\lambda_{\chi}$ from relic density, unitarity and self scattering cross-section. }
 \label{fig:lam-chi}
\end{figure}

\begin{figure}[htb!]
$$ 
\includegraphics[scale=0.275]{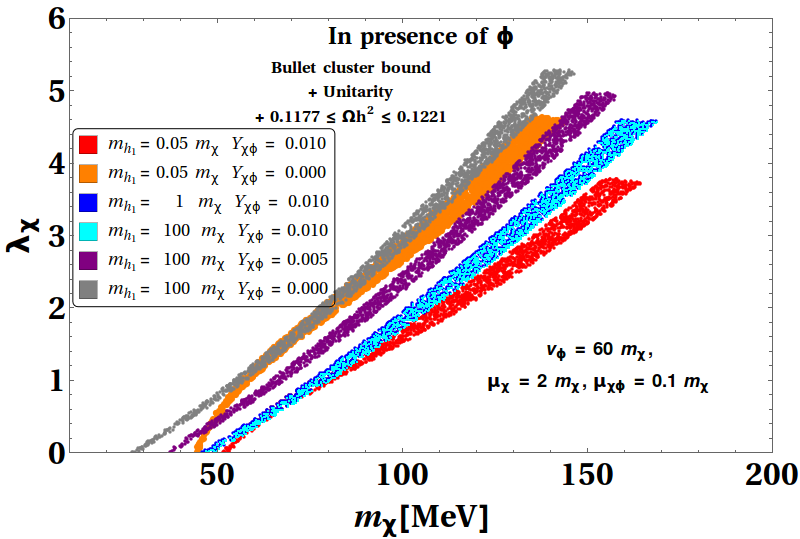}~
\includegraphics[scale=0.275]{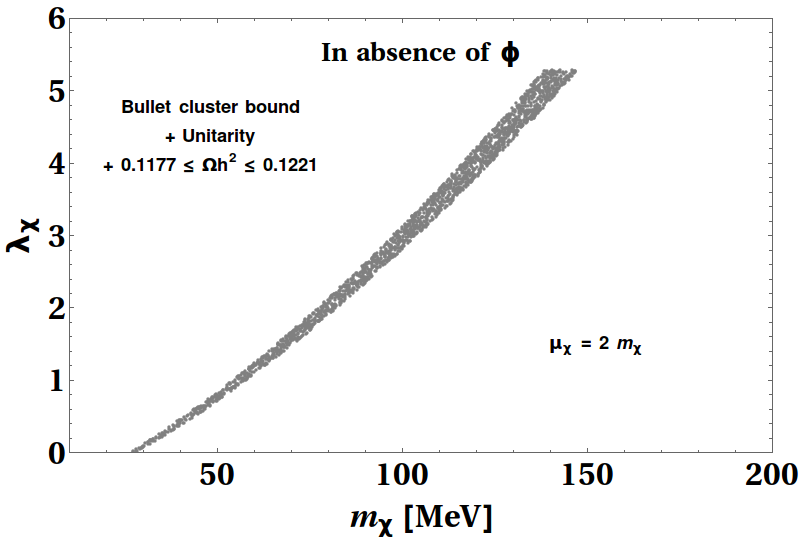}
$$
 \caption{Allowed parameter space in $m_{\chi}-\lambda_{\chi}$ plane from relic density, unitarity and self scattering cross-section [Left Panel] our model, [Right Panel] model in absence of $\phi$.}
 \label{fig:lam-chi12}
\end{figure}

Next we choose to illustrate the importance of $\lambda_\chi$ parameter of the model. 
In Fig.~\ref{fig:lam-chi}, we show the available parameter space in $m_{\chi}-Y_{\chi\phi}$ plane for different choices of 
$\lambda_{\chi}$. Interestingly, we see that a common parameter space $3_{\rm DM} \to 2_{\rm DM}$ available even after choosing $\lambda_\chi=0.1$. 
Finally, we demonstrate the effect of additional scalar ($\phi$) in our model to yield a larger parameter space viable from all the constraints 
in $m_\chi-\lambda_\chi$ plane, shown in Fig.~\ref{fig:lam-chi12}. In the left plot we scan our model and in the right panel 
the case in absence of $\phi$ is presented. It is easily understood that the allowed parameter space is dependent on the choice of $m_{h_1}$ as a light mediator of DM number density depletion processes and 
$Y_{\chi \phi}$, DM-mediator coupling. When $Y_{\chi \phi}\to 0$ and $m_{h_1}>m_\chi$, the model naturally reduces to the case when there is no additional scalar (here $\phi$) present in the set up; 
compare grey bands on left and right panel figures. As we increase $Y_{\chi \phi}$ to a sizeable value within self interaction and unitarity bound (see Fig. \ref{fig:bullet}), 
with the freedom of choosing $m_{h_1}$ as light as $0.05 m_\chi$, the allowed parameter space spans from grey to red region (left panel).
As a result, we see that in our model, we can allow for a larger range of self coupling $\lambda_\chi$ with allowed DM mass ranging between $30-180$ MeV due to the presence of additional 
light scalar. 
\subsection{What keeps the DM in equilibrium in SIMP realisation~?}
\label{sec:equilibrium}
As we have argued before, that SIMP realisation of this model crucially depends on the fact that $2_{\rm DM} \to 2_{\rm SM}~(\chi\chi^{*}\to~f\bar{f})$ annihilation to SM is negligible and that 
has been ensured by vanishingly small $\lambda_{\chi h}$ and $\lambda_{\chi \phi}$ in our model so that thermal freeze-out is governed by 
$3_{\rm DM} \to 2_{\rm DM}~(\chi\chi\chi\to~\chi\chi)$ annihilation in dark sector. Then the question is what keeps the 
DM in equilibrium in the early universe or what ensures the inequality described in Eqn.~\ref{eq:condition}. 
Here we demonstrate that the rate of DM~SM$\to$ DM~SM $(\chi f\to~\chi f)$ scattering is still large enough compared to $2_{\rm DM} \to 2_{\rm SM}$ and $3_{\rm DM} \to 2_{\rm DM}$ annihilations 
even with small $\lambda_{\chi h}$ and small $\lambda_{\chi \phi}$ to keep DM in equilibrium at the early universe and produce a SIMP like freeze-out. 
To show this, we estimate the ratios of the rate of scattering to annihilations in $2_{\rm DM}\to2_{\rm SM}$ and $3_{\rm DM}\to2_{\rm DM}$ which read: 
\bea
\label{ratio}
\frac{\Gamma^{kin}_{{\rm DM ~SM \to DM~SM}}}{\Gamma^{ann}_{2_{\rm DM}\to2_{\rm SM}}}=\frac{\sum\limits_{f}n_{\rm f}\langle \sigma v\rangle_{\chi f\to \chi f}}{n_{\chi}\sum\limits_{f}\langle \sigma v\rangle_{\chi \chi \to f \bar{f}}}~~~{;} ~~
\frac{\Gamma^{kin}_{{\rm DM ~SM \to DM~SM}}}{\Gamma^{ann}_{3_{\rm DM}\to2_{\rm DM}}}=\frac{\sum\limits_{f}n_{\rm f}\langle \sigma v \rangle_{\chi f\to \chi f}}{n_{\chi}^2\langle \sigma v^2 \rangle_{\chi \chi \chi\to \chi \chi}}~.
\eea
In above equations, $f$ denotes SM fermions. Scattering rate is governed by two factors, scattering cross-section $(\langle \sigma v\rangle_{\chi f\to \chi f})$ and number density of SM species ($n_f$). 
The number density of the SM particles is given by,
\bea
n_{f}&=&\frac{3}{4}\frac{g_f \zeta(3)}{\pi^{2}}T^{3}~~~~~\rm{(relativistic)}\nonumber\\
&=& g_{f}\bigg(\frac{m_{f}~T}{2\pi}\bigg)^{3/2}e^{-m_{f}/T}~~~~\rm{(non~relativistic);}
\eea
%
where $g_f$ denotes degrees of freedom and non-relativistic approximation is applied to heavy top quark. DM number density ($n_{\chi}$) can be evaluated by solving the following BEQ as already discussed,
\bea
\frac{dY}{dx}&=&-0.116~g_*^\frac{3}{2}~M_{pl}~\frac{m_{\chi}^4}{x^{5}}~\Big[ \langle \sigma v^2\rangle_{\chi \chi \chi \to \chi\chi} \Big(Y^{3}-Y^2~Y_{eq}\Big)\Big]~ \nonumber \\
&=&-0.116~g_*^\frac{3}{2}~M_{pl}~\frac{m_{\chi}^4}{x^{5}}~\Big[ \Big(\langle \sigma v^2\rangle_{\chi \chi \chi \to \chi\chi^*} + \langle \sigma v^2\rangle_{\chi \chi^* \chi^* \to \chi\chi}\Big) \Big(Y^{3}-Y^2~Y_{eq}\Big)\Big]~;
\eea
where, $Y(x)=n_{\chi}/s$ is the co moving number density.
The analytical form of $\langle \sigma v\rangle_{\chi \chi^{*}\to f \bar{f}}$ and $\langle \sigma v\rangle_{\chi f\to \chi f}$ with corresponding Feynmann diagrams are given in 
Appendix~\ref{annDMSM} and Appendix~\ref{scattDMSM} respectively. The analytical form $3_{\rm DM} \to 2_{\rm DM}$ annihilation processes 
($\langle \sigma v^2\rangle_{\chi \chi \chi \to \chi\chi^*}$ and $\langle \sigma v^2\rangle_{\chi \chi^* \chi^* \to \chi\chi}$) are also discussed in Appendix~\ref{feyn3to2dm1}. 

To verify SIMP conditions described in Eq.~\ref{eq:condition}, we choose DM mass, $m_\chi = 50$ MeV, 
while others parameters are considered as follows:
\[
 \{m_{h_1}=25 m_\chi,~v_\phi=60m_\chi,~\mu_\chi=0.5 m_\chi,~\mu_{\chi\phi}=0.1m_\chi,~Y_{\chi\phi}=0.018, \]
 \[ ~\lambda_{\chi}=1,~\sin\theta=0.999,~ \lambda_{\chi h}=0.001,~\lambda_{\chi\phi}=0.001~\};
\]
consistent with correct relic density and other constraints as obtained in the scans (for example in Fig.~\ref{fig:lam-chi12}).
Now for above choices of parameters at $x=18$ (just before freeze-out, $x_f\simeq$ 19.5, as can be obtained numerically from the solution of BEQ, 
as elaborated in Appendix~\ref{sec:freeze-out-temp}, and can also be verified from analytical solution provided in Eqn.~\ref{analyticformxf}), 
the ratios in Eqn.~\ref{ratio} are obtained as:
\bea
\label{eq:ratiomag}
\frac{\Gamma^{kin}_{{\rm DM ~SM \to DM~SM}}}{\Gamma^{ann}_{2_{\rm DM}\to2_{\rm SM}}}=\frac{\sum\limits_{f}n_{\rm f}\langle \sigma v\rangle_{\chi f\to \chi f}}{n_{\chi}\sum\limits_{f}\langle \sigma v\rangle_{\chi \chi \to f \bar{f}}} && \sim \mathcal{O}(10^{10}), \nonumber \\
\frac{\Gamma^{kin}_{{\rm DM ~SM \to DM~SM}}}{\Gamma^{ann}_{3_{\rm DM}\to2_{\rm DM}}} = \frac{\sum\limits_{f}n_{\rm f}\langle \sigma v \rangle_{\chi f\to \chi f}}{n_{\chi}^2\langle \sigma v^2 \rangle_{\chi \chi \chi\to \chi \chi}}&& \sim \mathcal{O}(10^{3}).
\eea
We clearly see that it satisfies SIMP conditions (as mentioned in Eqn. \ref{eq:condition}) and stops dark sector from heating up.
We can understand the magnitude of the ratios above with some numerical insight; the scattering rate is 
$ \Gamma_{{\rm{DM~SM\to DM~SM}}}^{kin}= 4.07635\times 10^{10}$ GeV, $\sum\limits_{f}\langle \sigma v\rangle_{\chi \chi \to f \bar{f}}=1.5063\times 10^{-15}$ GeV$^{-2}$, 
$\langle \sigma v^2 \rangle_{\chi \chi \chi\to \chi \chi}=1.07191\times 10^{7}$ GeV$^{-5}$, and $n_{\chi}=1.6458\times10^{-15}$.
It is straightforward to check that SIMP condition is satisfied for all the allowed parameter space of the model. Moreover, for $\textrm{DM+SM} \to \textrm{DM+SM}$ to keep the
DM in equilibrium, the interaction rate should dominate over expansion rate of the universe, $\mathcal{H}$ , i.e. $\Gamma_{kin} > \mathcal{H}$. 
We estimate the ratio of $\Gamma_{kin}$ to $\mathcal{H}$ for above choices of parameters at $x=18$ to yield: 
\bea
\frac{\Gamma^{kin}_{{\rm DM ~SM \to DM~SM}}}{\mathcal{H}} &=& \frac{\sum\limits_{f}n_{\rm f}\langle \sigma v\rangle_{\chi f\to \chi f}}{1.66 \sqrt{g_*} \frac{1}{M_{Pl}}T^2} \sim \mathcal{O}(10^4) .
\eea

Further, we would also like to point out that $3_{\rm DM} \to 2_{\rm SM}$ annihilations (to SM) is also non negligible. When two of these processes $3_{\rm DM} \to 2_{\rm DM}$ 
within dark sector and $3_{\rm DM} \to 2_{\rm SM}$ (in SM) contribute together, the BEQ takes the following form:
\bea
\frac{dY}{dx}=& -0.116~g_{*}^{3/2}~M_{pl}~ \frac{m_{\chi}^{4}}{x^{5}}~\bigg[(\langle\sigma v^{2}\rangle_{\chi\chi\chi\to\chi\chi^{*}}+\langle\sigma v^{2}\rangle_{\chi\chi^{*}\chi^{*}\to\chi\chi})(Y^{3}-Y^{2}Y_{eq})\nonumber \\
&\hspace{0.7cm}+\langle\sigma v^{2}\rangle_{\chi\chi\chi \to f\bar{f}}(Y^{3}-Y_{eq}^{3})+\langle\sigma v^{2}\rangle_{\chi\chi^{*} f \to \chi^{*} f}(Y^{2}Y_{eq}-Y Y_{eq}^{2})\bigg].
\label{eq:3to2DMSM}
\eea

\begin{figure}[htb!] 
$$
\includegraphics[scale=0.225]{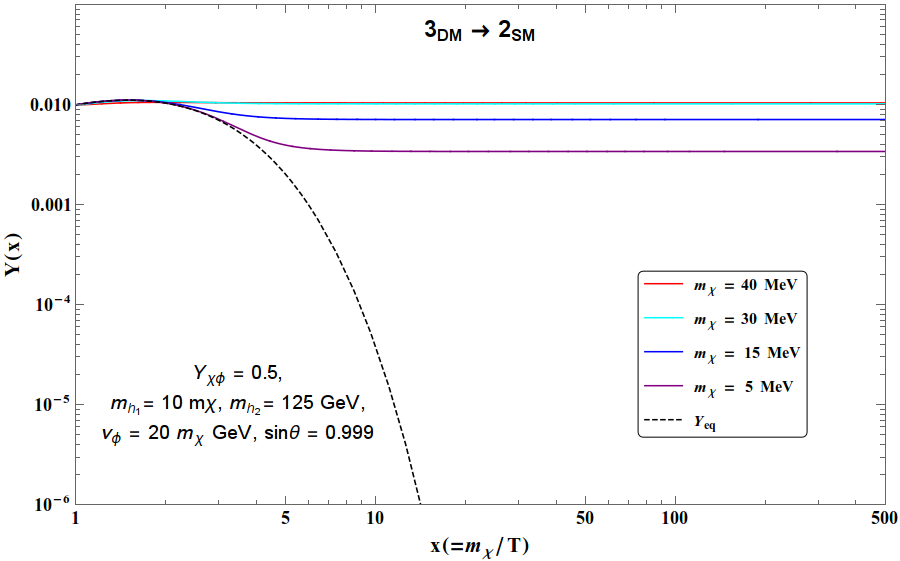}
$$
$$
\includegraphics[scale=0.225]{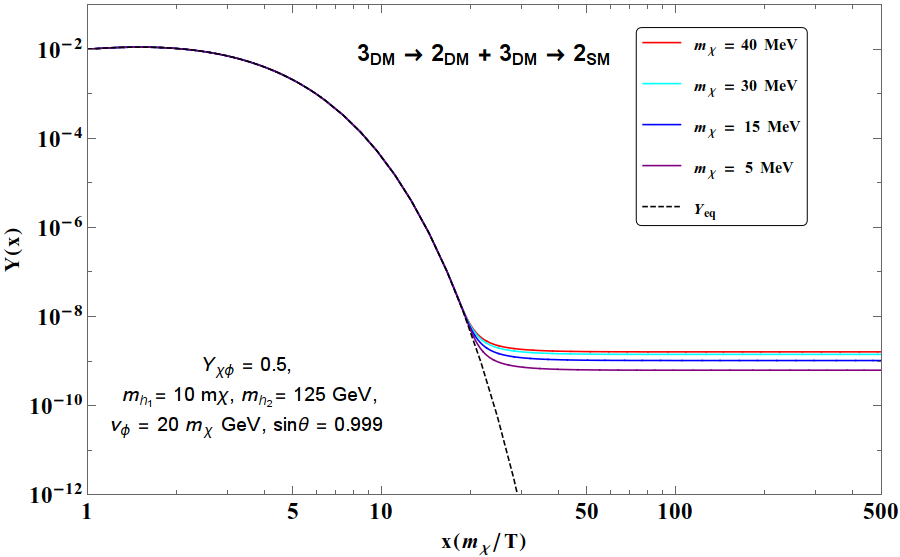}~~
\includegraphics[scale=0.225]{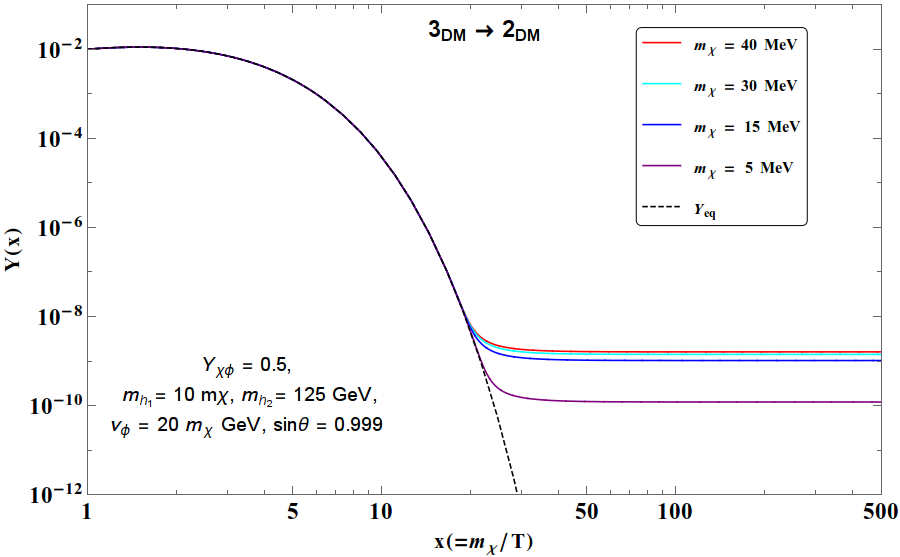}
$$
\caption{Freeze-out of DM $\chi$ from thermal equilibrium in $Y(x)-x$ plane in presence of $3_{\rm{DM}} \to 2_{\rm{SM}}$ i.e. annihilation to SM only [top], $3_{\rm{DM}} \to 2_{\rm{DM,SM}}$ i.e. annihilation to DM and SM [bottom left] 
and $3_{\rm{DM}} \to 2_{\rm{DM}}$ i.e. annihilation to DM only [bottom right].}
\label{modelfreezeout}
\end{figure}

In Fig.~\ref{modelfreezeout}, we demonstrate the freeze-out in such a case. In the top panel, we show the case when DM freeze-out through 
$3_{\rm{DM}} \to 2_{\rm{SM}}$ only (when only the second term is considered in BEQ \ref{eq:3to2DMSM}). The solution shows that 
$3_{\rm{DM}} \to 2_{\rm{SM}}$ interaction is good enough to keep the DM follow equilibrium distribution at low $x$ and yields a typical but early freeze-out. 
On the bottom left panel, when we include additionally the annihilation through $3_{\rm{DM}} \to 2_{\rm{DM}}$ in the dark sector, 
due to enhanced self coupling (as chosen for the SIMP like case), the number changing process in the dark sector dominates over 
$3_{\rm{DM}} \to 2_{\rm{SM}}$ and yields a freeze-out that corresponds to correct relic. This is validated by taking $3_{\rm{DM}} \to 2_{\rm{DM}}$ 
annihilation in the dark sector {\it only} (as we have done for the analysis) in the bottom right 
panel to show that the freeze-out mimics the case of taking both contributions together (as in bottom left Fig.~\ref{modelfreezeout}) and justifies our analysis.


\subsection{$4_{\rm DM} \to 2_{\rm DM}$ SIMP scenario}
\begin{figure}[htb]
$$
\includegraphics[scale=0.40]{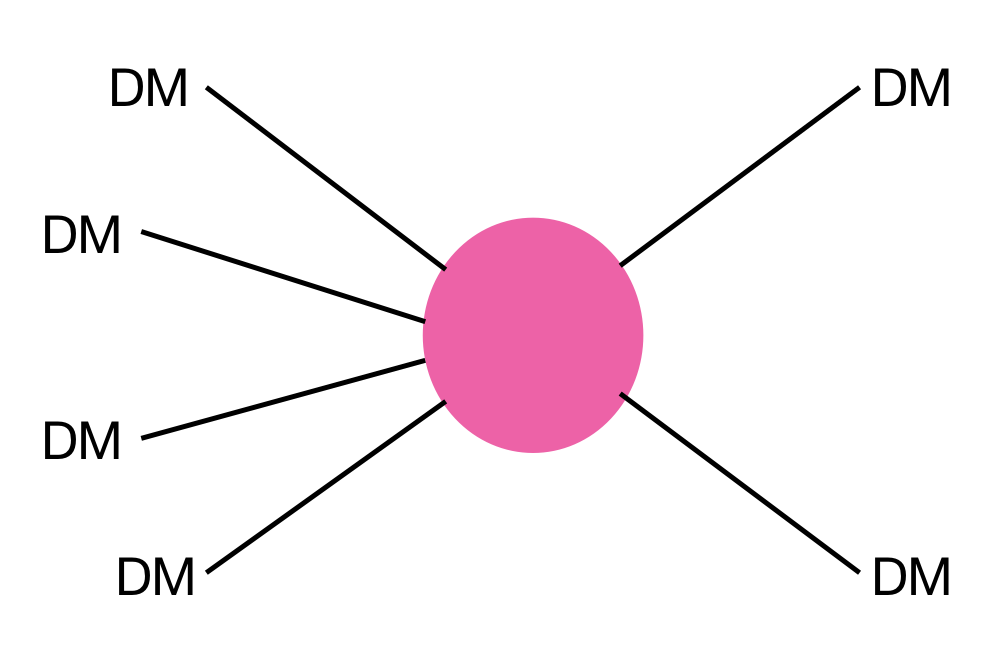}~~
\includegraphics[scale=0.28]{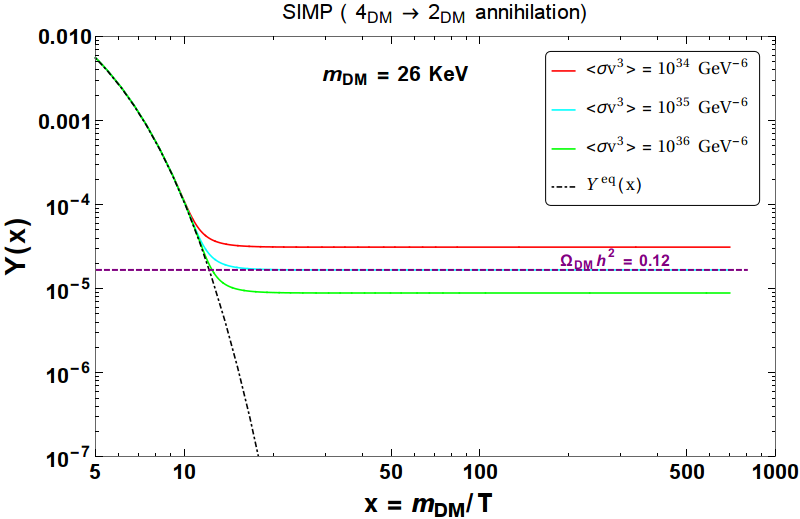}
$$
\caption{[Left] A cartoon of $4_{\rm DM} \to 2_{\rm DM}$ annihilation process in SIMP; [Right] Freeze-out of $4_{\rm DM} \to 2_{\rm DM}$ SIMP DM from $Y_{eq}$ (black dashed line) 
in $Y(x)-x$ plane for DM mass $m_{\rm DM}=26$ KeV in a model independent way for three different choices of $\langle{\sigma v^3}\rangle$ cross-section. 
 Horizontal purple dashed line corresponds to correct relic density. }
\label{BEQ4_MI}
\end{figure}

SIMP like framework can also be realised when the dominant depletion in DM number density occurs through $4_{\rm DM} \to 2_{\rm DM}$ process as shown in the left hand side of 
Fig.~\ref{BEQ4_MI}. The BEQ for such a $4_{\rm DM} \to 2_{\rm DM}$ process is given by:
\bea
\label{eq:BEQ4to2}
\frac{d Y}{d x} = -0.0508~\frac{g_{*s}^{3}}{\sqrt{g_{*}}}~M_{Pl}~\frac{{m_{\rm DM}}^7}{x^8}~{\langle{\sigma v^3}\rangle}_{4_{\rm DM} \to 2_{\rm DM}} ~\Big(Y^{4}-Y^{2} ~Y_{eq}^{2}\Big),
\eea
where $g_{*s} = 3.91$ and $g_*=3.36$ for KeV order DM. The freeze-out solution of BEQ in Eqn.\ref{eq:BEQ4to2} in terms of 
$Y (= \frac{n}{s})$ is shown in RHS of Fig.\ref{BEQ4_MI} with DM mass $26$ KeV for three different choices of 
$\langle{\sigma v^3}\rangle$ of $4_{\rm DM} \to 2_{\rm DM}$ cross-sections. We can see that correct relic density ($\Omega h^2 = 0.12$) can be 
achieved when $\langle{\sigma v^3}\rangle \sim 10^{35}~ {\rm GeV^{-8}}$ for $m_{DM} = 26 ~{\rm KeV}$ in a model independent way. \\~\\
In our model, $4_{\rm DM} \to 2_{ \rm DM}$ processes occur through $\chi \chi^* \chi \chi^* \to \chi \chi$ and $\chi \chi \chi \chi \to \chi \chi^*$ mediated by $
\chi, h_1$ and $h_2$. The amplitude for such process therefore turns out to be:
\bea
|\mathcal{M}_{4_{\rm DM}\to 2_{\rm DM}}|^2 = 2 \Big|\mathcal{M}_{\chi \chi^* \chi \chi^* \to \chi \chi} + \mathcal{M}_{\chi \chi \chi \chi \to \chi \chi^*}\Big|^2,
\eea
where the factor of 2 comes from the corresponding conjugate processes. The thermal average of total cross section for $4_{\rm DM} \to 2_{\rm DM}$ processes is given by:
\bea
{\langle{\sigma v^3}\rangle}_{4_{\rm DM} \to 2_{\rm DM}}= \frac{\sqrt{3}}{256 \pi m_{\chi}^4} |\mathcal{M}_{4_{\rm DM}\to 2_{\rm DM}}|^2 .
\eea
The calculation of $\langle{\sigma v^3}\rangle$ for ${4_{\rm DM} \to 2_{\rm DM}}$ process is described in Appendix \ref{4to2dm1}. We however refrain from elaborating all the Feynman graphs that contribute to 
$\chi \chi^* \chi \chi^* \to \chi \chi$ and $\chi \chi \chi \chi \to \chi \chi^*$ in this model due to the large number of diagrams present. 
\begin{figure}[htb!]
\label{fig:BEQ_MD}
\centering
\includegraphics[scale=0.25]{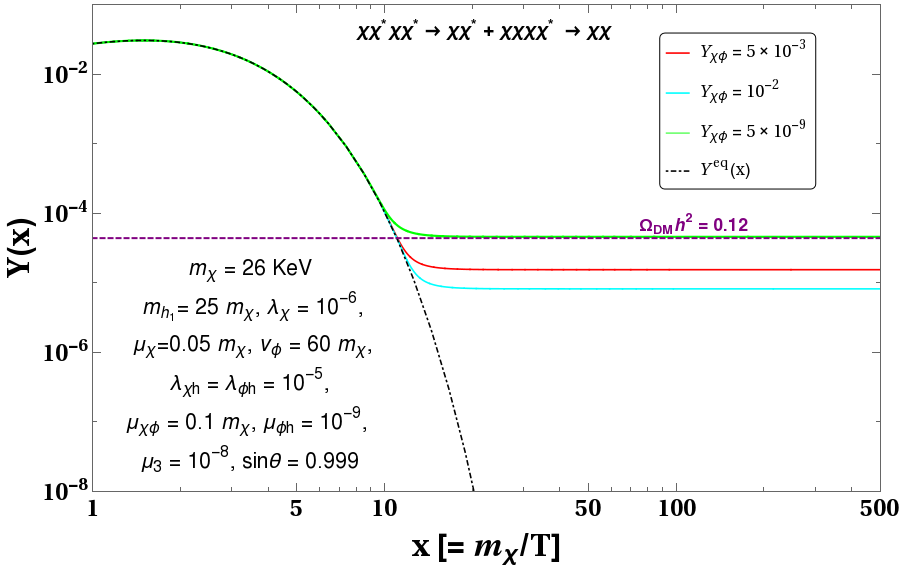}
\caption{Freeze out of $\chi$ through ${4_{\rm DM} \to 2_{\rm DM}}$ process in $Y-x$ plane for three sets of values of $Y_{\chi\phi}$. Other parameters kept 
fixed are mentioned in figure inset and the correct relic density is shown by purple dashed line.}
\label{BEQ4_MD}
\end{figure}
We demonstrate freeze-out of $\chi$ through ${4_{\rm DM} \to 2_{\rm DM}}$ process in Fig.~\ref{BEQ4_MD} in $Y-x$ plane for $m_\chi=26$ KeV in our model. We choose three values 
of $Y_{\chi\phi}$ for demonstration. The one corresponds to correct relic is given by $Y_{\chi\phi}=5 \times 10^{-9}$, with other parameters kept fixed and mentioned 
in figure inset. It is clear that the correct density obtained by a DM mass so light ($\sim$ $\mathcal{O}$(KeV) ), already has compensated for the phase space suppression and therefore 
do not require a coupling in strong limit. With larger DM mass, the coupling gets larger. However, the required couplings to satisfy correct relic density for KeV order DM are 
much smaller compared to MeV order SIMP mass. Therefore, automatically due to the choices of parameters made above, $3_{\rm DM} \to 2_{\rm DM}$ number changing processes are suppressed and the 
freeze-out is governed by $4_{\rm DM} \to 2_{\rm DM}$.
\section{WIMP realisation of the model}
\label{sec4}
\begin{figure}[htb!]
$$\hspace{-0.75cm}{\includegraphics[scale=0.255]{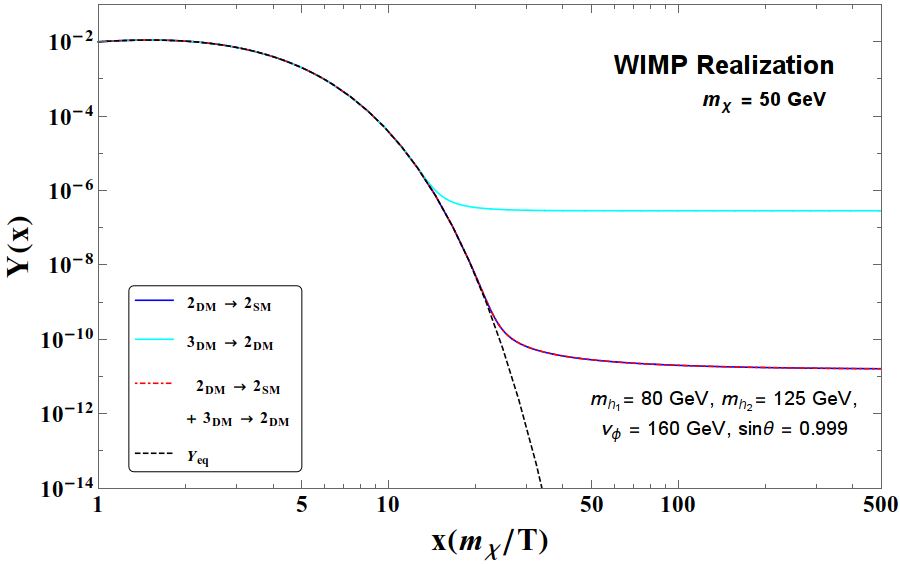}}
$$
\caption{DM freeze out in WIMP scenario following the BEQ given in \eqref{eq:totalbeq} with three choices of DM annihilation: (i) $2_{\rm DM}\to 2_{\rm SM}$ (blue line), 
(ii) $3_{\rm DM}\to 2_{\rm DM}$ (cyan line), (iii)  $2_{\rm DM}\to 2_{\rm SM}$ and $3_{\rm DM}\to 2_{\rm DM}$ together (red dashed). The cases of (i) and (iii) superimpose 
on each other. We choose DM mass of $\sim 50$ GeV, and DM-SM couplings of the order of $\lambda_{\chi \phi}=\lambda_{\chi h}\sim 0.1$. }
\label{fig:simpwimpfo}
\end{figure}

Finally for comparison, we demonstrate the WIMP realisation of the same model that we have studied in this paper. 
The BEQ in WIMP scenario is given by: 
\bea
\frac{dY}{dx} &=&  - 0.264*g_{*s}^{1/2}~ M_{pl}~ \frac{m_{\chi}}{x^{2}}\langle \sigma v \rangle _{2_{\rm DM}\to 2_{\rm SM}}\Big(Y^{2}-Y_{eq}^{2}\Big)\nonumber \\
& &-0.115*g_{*s}^{3/2}~M_{pl}~\frac{m_{\chi}^{4}}{x^{5}}~\langle \sigma v^{2}\rangle _{3_{\rm DM}\to 2_{\rm DM}}\Big(Y^{3}-Y^{2}Y_{eq}\Big).
\label{eq:totalbeq}
\eea
In the above Eqn.~\ref{eq:totalbeq}, we have considered the DM annihilation to SM through $2_{\rm DM} \to 2_{\rm SM}$ and also the one used for SIMP condition, namely $3_{\rm DM} \to 2_{\rm DM}$ process. 
DM freeze-out is shown in Fig.~\ref{fig:simpwimpfo} for three cases: (i) considering only $2_{\rm DM}\to 2_{\rm SM}$ (blue line), 
(ii) only $3_{\rm DM}\to 2_{\rm DM}$ (cyan line), (iii) the actual situation $2_{\rm DM}\to 2_{\rm SM}$ and $3_{\rm DM}\to 2_{\rm DM}$ together (red dashed) 
following Eqn.~\ref{eq:totalbeq}. We clearly see here that $3_{\rm DM}\to 2_{\rm DM}$ annihilation has a very small contribution as the lone process 
of such kind will yield an early freeze-out, whereas when considered together with $2_{\rm DM}\to 2_{\rm SM}$, can not be distinguished from the case 
(iii) where $2_{\rm DM}\to 2_{\rm SM}$ and $3_{\rm DM}\to 2_{\rm DM}$ are addressed together.  Therefore, it is quite justified to neglect the second term in 
BEQ \ref{eq:totalbeq} for WIMP solution.

\begin{figure}[htb!]
$$\hspace{-1cm}{\includegraphics[scale=0.295]{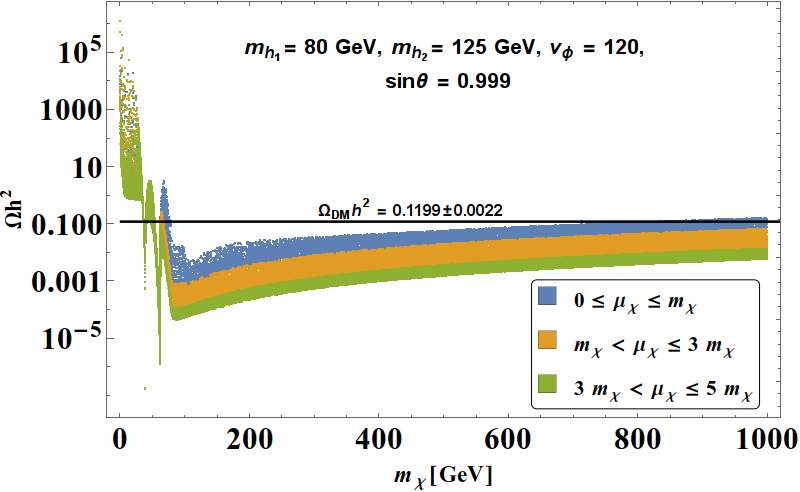}}
~~
{\includegraphics[scale=0.295]{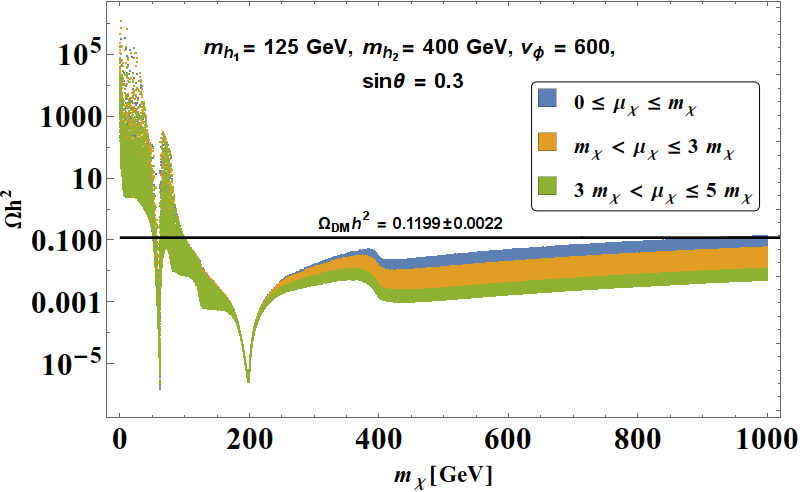}}$$
\caption{Relic density in WIMP condition for DM $\chi$ as a function of DM mass, with the variation in $\mu_{\chi}$. All the parameters 
kept constant are mentioned in figure inset. Notably we have chosen larger $\lambda_{\chi \phi}=\lambda_{\chi h} \sim 0.1$. 
}
\label{fig:muxvaria}
\end{figure}

As has already been mentioned that SIMP realisation of this model was possible by choosing 
the coupling to SM very feeble, namely keeping $\lambda_{\chi \phi}=\lambda_{\chi h}\sim 0.001$, 
altering which the $2_{\rm DM}\to2_{\rm SM}$ annihilation to SM dominates over the $3_{\rm DM}\to2_{\rm DM}$ in dark sector and governs the freeze-out
to reveal WIMP paradigm of the model. 
We show next the variation in relic density with DM mass in Fig.~\ref{fig:muxvaria} for WIMP realisation of the model. 
We choose to illustrate two different values of the additional scalar boson mass: a light scalar mass of 80 GeV for the left plot and a heavy scalar of 400 GeV in the right plot. 
To compute relic density and direct search cross section for the model we have used micrOmegas~\cite{Belanger:2014hqa}.  
We see that two resonance drops at $m_{h_{1,2}}/2$ are clearly observed  for s-channel mediation of $h_{1,2}$ in $2_{\rm DM} \to 2_{\rm SM}$ annihilation 
process. We also point out the variation in $\mu_\chi$ for illustration, the larger the $\mu_\chi$, the larger is the annihilation 
cross-section and therefore smaller is the relic density. There exist a semi-annihilation effect $\chi\chi \to h \chi$ for the WIMP DM here that helps disentangling 
the relic density to direct search; but, to drop below the direct search constraints require a large $\mu_\chi$, that lies in tension with vacuum stability. 
\begin{figure}[htb!]
$${\includegraphics[scale=0.3]{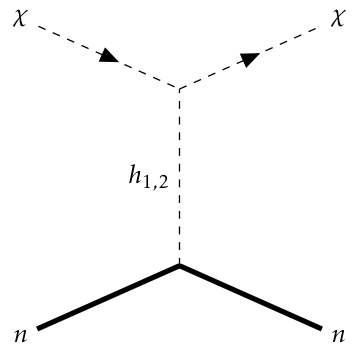}}
$$
\caption{Feynman graph for Direct Search interaction of DM ($\chi$) with nucleon ($n$) through $h_{1,2}$ mediation in WIMP scenario.}
\label{fig:sigmadd}
\end{figure}

\begin{figure}[htb!]
$$
\hspace{-1cm}
{\includegraphics[scale=0.295]{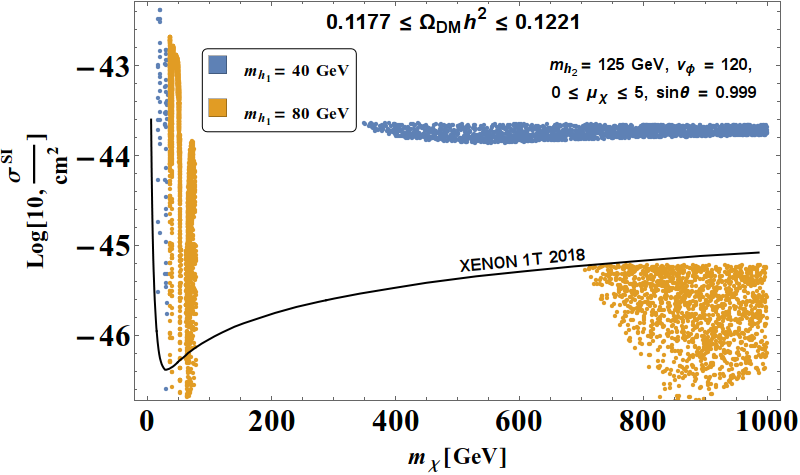}}~~
\includegraphics[scale=0.295]{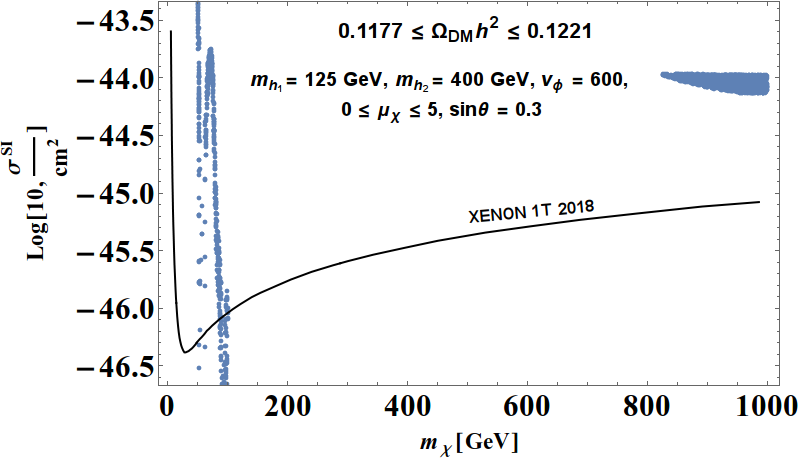}$$
\caption{Direct detection bound (XENON1T) on relic density allowed parameter space of the WIMP DM $\chi$. 
We scan low Higgs mass ($m_{h_1}=40$ GeV [blue] and $=80$ GeV [golden yellow]) on the Left panel, and heavy Higgs mass ($m_{h_2}=400$ GeV) on the right panel. Other parameters are kept steady as mentioned in figure inset.}
\label{fig:DD_mchi}
\end{figure}

We next analyse the constraint coming from direct search to the relic density allowed parameter space of the WIMP scenario of the model. 
The Feynman graph for direct search interaction is shown in Fig.~\ref{fig:sigmadd} through t-channel $h_{1,2}$ mediation. 
The scan for relic density allowed parameter space of the model in spin independent direct search cross section versus DM mass plane 
is shown in Fig.~\ref{fig:DD_mchi}. We have chosen two different possible phenomenological situations for illustration: 
light additional scalar ($m_{h_1}=40$ GeV in blue and $m_{h_1}=80$ GeV in golden yellow) on the left panel and heavy scalar ($m_{h_2}=400$ GeV) on the right panel. 
The main outcome of this analysis is to see that immaterial to the additional scalar mass resonance regions are allowed by direct search. 
Interestingly, when the additional scalar mass is not too far from the SM Higgs, as is the case for $m_{h_1}=80$ GeV as shown by golden yellow points in the left panel, 
there is a large region of heavy DM mass ($\sim 800\to1000$ GeV), which becomes allowed by direct search constraint. This can be explained by realizing that since the spin independent direct search cross-section follows~\cite{Ghosh:2017fmr}:

\begin{equation}
\sigma_{\rm DD}^{\rm SI} =\frac{1}{4\pi} \bigg(\frac{f_{n} \mu_{n}}{m_{\chi}}\bigg)^{2}\bigg(\frac{m_{n}}{v_{h}}\bigg)^{2}\bigg[\frac{\lambda_{a1} \cos \theta}{m_{h_{1}}^{2}}+\frac{\lambda_{b1} \sin \theta}{m_{h_{2}^{2}}}\bigg]^{2}~,
\label{eq:sigmadd}
\end{equation}
where $\lambda_{a1}$ and $\lambda_{b1}$ are DM-Higgs coupling, $f_{n}$ is the form factor, $\mu_{n}=m_{n}m_{\chi}/(m_{n}+m_{\chi})$ is the reduced mass. The cross-section
yields a destructive interference due to opposite sign of $\lambda_{a1}$ and $\lambda_{b1}$ (look at the Table \ref{table:vert} of vertices in Appendix \ref{vert}) when the two scalar masses are close. 
\section{Summary and Conclusion}
\label{sec5}
We have presented a model where both SIMP and WIMP realization of a scalar DM is possible. This is achieved by assuming 
a complex scalar field $\chi$ which transforms under unbroken $\mathcal{Z}_3$. When the portal coupling is small, it provides a SIMP solution and 
when the portal coupling is large, it provides a WIMP like solution. In principle, this bit of model construct is good enough to realise the correct 
relic density in SIMP scenario and perhaps serves as the simplest SIMP DM, where the number changing process within the dark sector 
is solely governed by DM self coupling. However, we add to that another scalar field 
$\phi$ that is even under $\mathcal{Z}_3$, acquires a vev, mixes with SM Higgs and serves as a light scalar mediator to aid DM self scattering to yield a large 
parameter space available to the model. We also see that due to the presence of this additional field, the self coupling to achieve a successful SIMP DM 
paradigm enjoys a larger freedom. The allowed parameter space gets further 
restricted from the self scattering constraints and unitarity bound; for Bullet cluster 
the bound turns out to be within $\sim 200$ MeV, while for Abell cluster data, the bound is more restrictive and remains within $\sim 50$ MeV.

The model can also serve a successful freeze-out through $4_{\rm DM} \to 2_{\rm DM}$ number changing processes, and achieve correct relic density for 
DM mass $\sim$ $\mathcal{O}({\rm KeV})$, where the couplings required are much smaller than that of $3_{\rm DM} \to 2_{\rm DM}$ case, automatically justifying 
the suppression of $3_{\rm DM} \to 2_{\rm DM}$ processes in such circumstances.

The condition to keep the DM in thermal equilibrium at early universe and not heating up through the number changing processes within the dark sector, 
have been verified for points satisfying correct relic density. Additionally, we have verified the kinetic interaction of DM with SM remains larger than the 
Hubble expansion rate before freeze-out.

We also analyse the WIMP limit of the DM for the sake of comparison. Interestingly the direct search allowed parameter space 
for such a framework predict that the additional Higgs mass should be close to the SM Higgs due to a destructive interference in the 
direct search cross-section. On the other hand SIMP realisation is aided when the additional scalar is light of the order of sub-GeV. It is important to 
remind that such a scalar is quite likely to evade the collider search bound due to its singlet nature. 

Thermal freeze out of the DM in SIMP condition for $3_{\rm DM} \to 2_{\rm DM}$ number changing process is performed in details and we advocate an approximate analytical 
solution for relic density which yields agreement to the numerical solution for a certain range of DM mass. 
We also calculate all the cross-sections for freeze out and self scattering in details, 
so that the draft serves as a useful reference for performing phenomenological analysis in any SIMP framework.     

\section*{Acknowledgement} 
SV acknowledges to the BTP project at the department of Physics in IIT Guwahati, where the project was initiated. SB and SV also acknowledges 
DST-INSPIRE Faculty grant IFA-13 PH-57. PG would like to thank MHRD, Government of India for research fellowship.
\clearpage
\appendix
\section*{Appendix}
\section{Vertices and Couplings of the model}
\label{vert}
Here we list all the vertices that appear in the cross-sections for annihilation and scattering processes in this model. We also introduce a shorthand
notation for each vertex that will be used further in computing the amplitudes. 

\begin{table}[htb!]
\centering
\begin{tabular}{|c|c|c|}
\hline
\textbf{ Vertices } & \textbf{ Vertex factor } &\textbf{ Notation}\hspace{0.35cm}\\
\hline 
$\chi^{*}\chi\chi^{*}\chi$ \,\, & $-(2!2!)\:\lambda_{\chi}=\:-4\lambda_{\chi}$ \,\,&$-\lambda_{4} $ \hspace{0.02cm}\\[0.1cm]\hline
$\chi\chi\chi$\,\,& $-\frac{(\mu_{\chi}+Y_{\chi\phi}v_{\phi})}{6}3!=-(\mu_{\chi}+Y_{\chi\phi}v_{\phi})$ \,\,& $-\lambda_{3}$ \hspace{0.05cm}\\[0.1cm]\hline
$\chi^{*}\chi h_{1}$\,\,& $-(\lambda_{\chi h}v_{h}\cos\theta-(\lambda_{\chi\phi}v_{\phi}+\mu_{\chi\phi})\sin\theta)$ \,\,& $-\lambda_{a1}$ \hspace{0.05cm}\\
[0.1cm]\hline
$\chi^{*}\chi h_{2}$\,\,& $-(\lambda_{\chi h}v_{h}\sin\theta+(\lambda_{\chi\phi}v_{\phi}+\mu_{\chi\phi})\cos\theta)$ \,\,& $-\lambda_{b1}$\hspace{0.05cm}\\
[0.1cm]\hline
$\chi\chi\chi h_{1}$\,\,& $-(-\sin\theta Y_{\chi\phi})$ \,\,& $-\lambda_{a2}$\hspace{0.05cm}\\
[0.1cm]\hline
$\chi\chi\chi h_{2}$\,\,& $-(-\cos\theta Y_{\chi\phi})$ \,\,& $-\lambda_{b2}$\hspace{0.05cm}\\
[0.1cm]	\hline 
$f \bar{f} h_{1}$\,\,& $-\frac{m_{f}}{v}\cos\theta$ \,\,& $-\lambda_{f1}$\hspace{0.05cm}\\
[0.1cm]\hline 
$f \bar{f} h_{2}$\,\,& $-\frac{m_{f}}{v}\sin\theta$ \,\,& $-\lambda_{f2}$\\
[0.1cm]\hline 
$h_{1}h_{1}h_{1}$\,\,& $3\cos\theta\sin\theta[\cos\theta(v_{\phi}\lambda_{\phi h}+\mu_{\phi h})-\sin\theta(\lambda_{\phi h}v_{h})]$ & $-\lambda_{H1}$\\&$+6\sin^{3}\theta v_{\phi}\lambda_{\phi}-6\cos^{3}\theta v_{h}\lambda_{h}-2\sin\theta\mu_{3}$ \,\,&\\
[0.1cm]\hline 
$\chi \chi^{*} h_{1}h_{1}$\,\,& $-\cos^{2}\theta  \lambda_{\chi h}-\sin^{2}\theta	\lambda_{\chi \phi}$ \,\,& $-\lambda_{\chi H1}$\\
\hline
\end{tabular}
\caption{Couplings (in terms of the model parameters, see Eq.~\ref{eq:potential}) that appear in the model and is required for 
computing all the processes considered in this analysis. Shorthand notations are introduced.}
\label{table:vert}
\end{table}

\section{Annihilation cross-section for $3_{{\textrm{DM}}} \to 2_{{\textrm{DM}}}$ process}
\label{feyn3to2dm1}
We first note that the dominant contribution in absence of $2_{\rm DM} \to 2_{\rm SM}$ annihilations to SM are $3_{{\textrm{DM}}} \to 2_{{\textrm{DM}}}$
that yields the required freeze out. Apart from $\chi$ mediation, the two other mediators for such diagrams are the two Higgses, which are 
mentioned by the following notation in the matrix element : 
\bea
\nonumber
a\Rightarrow h_{1}\: \textrm{mediation}~,
\\
\nonumber
b\Rightarrow h_{2}\: \textrm{mediation} ~.
\eea
There are two major processes in the model which contribute to such case: $\chi \chi \chi \to \chi \chi^{*}$ and 
$\chi \chi^{*} \chi^{*} \to \chi \chi$ and their conjugates. We will analyse them systematically below.

\newpage
\hspace{6.75cm}\underline{\Large\textbf{$\chi \chi \chi \to \chi \chi^{*}$}}
\subsection*{Feynman Diagrams}
$$
\includegraphics[scale=0.32]{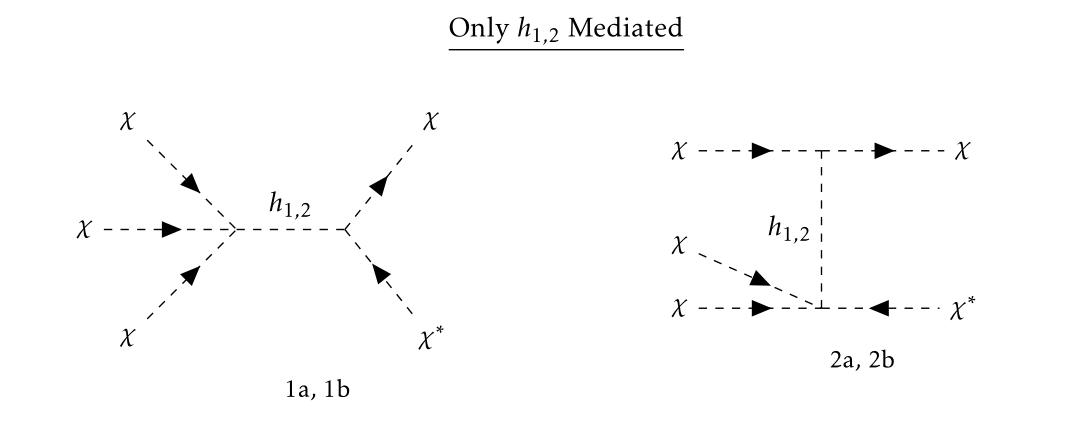}
$$
$$
\includegraphics[scale=0.32]{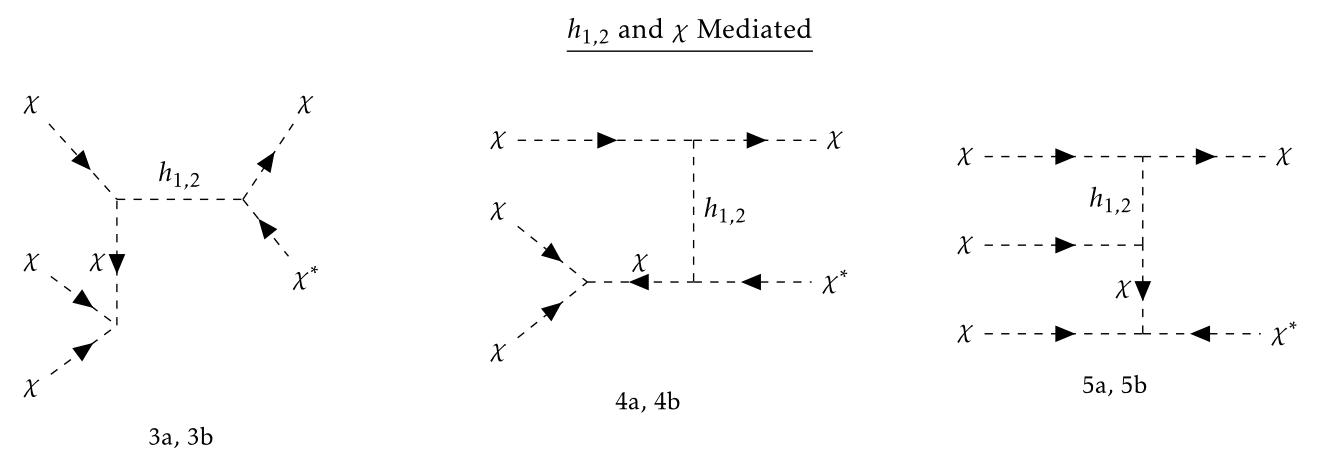}
$$
$$
\includegraphics[scale=0.32]{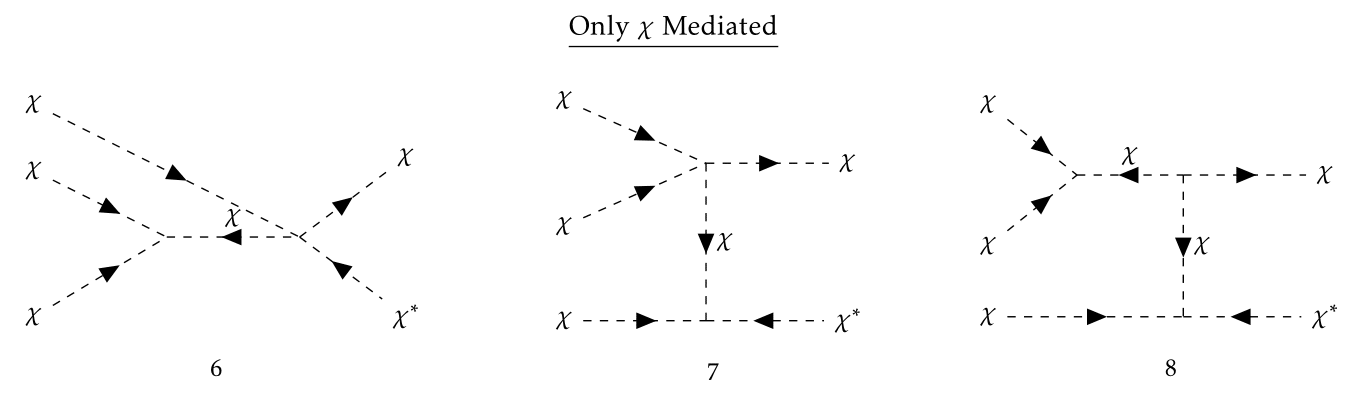}
$$
\vspace{1cm}

\subsection*{Matrix Amplitude}
\begin{varwidth}[t]{.75\textwidth}
\hspace{0.75cm}\underline{Only $h_{1,2}$ mediated}\begin{itemize}
\item $\mathcal{M}_{1a}=\frac{(-\lambda_{a1})(-\lambda_{a2})}{s-m_{h_{1}}^{2}}$
\item $\mathcal{M}_{2a}=\frac{(-\lambda_{a1})(-\lambda_{a2})}{t-m_{h_{1}}^{2}}$
\item $\mathcal{M}_{1b}=\frac{(-\lambda_{b1})(-\lambda_{b2})}{s-m_{h_{2}}^{2}}$
\item $\mathcal{M}_{2b}=\frac{(-\lambda_{b1})(-\lambda_{b2})}{t-m_{h_{2}}^{2}}$
\end{itemize}
\end{varwidth}
\begin{varwidth}[t]{.75\textwidth}
\hspace{0.75cm}\underline{$h_{1,2}$ and $\chi$ mediated}
\begin{itemize}
\item $\mathcal{M}_{3a}=\frac{(-\lambda_{a1})^{2}(-\lambda_{3})}{(s-m_{h_{1}}^{2})(t-m_{\chi}^{2})}$
\item $\mathcal{M}_{4a}=\frac{(-\lambda_{a1})^{2}(-\lambda_{3})}{(t-m_{h_{1}}^{2})(s-m_{\chi}^{2})}$
\item $\mathcal{M}_{5a}=\frac{(-\lambda_{a1})^{2}(-\lambda_{3})}{(t-m_{h_{1}}^{2})(t-m_{\chi}^{2})}$
\item $\mathcal{M}_{3b}=\frac{(-\lambda_{b1})^{2}(-\lambda_{3})}{(s-m_{h_{2}}^{2})(t-m_{\chi}^{2})}$
\item $\mathcal{M}_{4b}=\frac{(-\lambda_{b1})^{2}(-\lambda_{3})}{(t-m_{h_{2}}^{2})(s-m_{\chi}^{2})}$
\item $\mathcal{M}_{5b}=\frac{(-\lambda_{b1})^{2}(-\lambda_{3})}{(t-m_{h_{2}}^{2})(t-m_{\chi}^{2})}$
\end{itemize}
\end{varwidth}
\begin{varwidth}[t]{0.75\textwidth}
\hspace{0.75cm}\underline{Only $\chi$ mediated}
\begin{itemize}
\item $\mathcal{M}_{6}= \frac{(-\lambda_{3})(-\lambda_{4})}{s-m_{\chi}^{2}}$
\item $\mathcal{M}_{7}= \frac{(-\lambda_{3})(-\lambda_{4})}{t-m_{\chi}^{2}}$
\item $\mathcal{M}_{8}= \frac{(-\lambda_{3})^{3}}{(s-m_{\chi}^{2})(t-m_{\chi}^{2})}$
\end{itemize}
\end{varwidth}

\bea
\nonumber
\mathcal{M}_{Net}&=&(\mathcal{M}_{1a}+\mathcal{M}_{2a}+\mathcal{M}_{3a}+\mathcal{M}_{4a}+\mathcal{M}_{5a})+(\mathcal{M}_{1b}+\mathcal{M}_{2b}+\mathcal{M}_{3b}+\mathcal{M}_{4b}+\mathcal{M}_{5b})\\ \nonumber
&&+\mathcal{M}_{6}+\mathcal{M}_{7}+\mathcal{M}_{8}
\eea

Matrix amplitude squared is then 
$$
\Rightarrow |\mathcal{M}_{\chi\chi\chi\to\chi\chi^{*}}|^{2}=|\mathcal{M}_{Net}|^{2} .
$$

The complex conjugate of $\chi\chi\chi\to\chi\chi^{*}$ $i.e.$ $\chi^{*}\chi^{*}\chi^{*}\to\chi^{*}\chi$ also contributes to the total matrix amplitude and has same expression as $\chi\chi\chi\to\chi\chi^{*}$,
$$
|\mathcal{M}_{\chi\chi\chi\to\chi\chi^{*}}|^{2}=|\mathcal{M}_{\chi^{*}\chi^{*}\chi^{*}\to\chi^{*}\chi}|^{2} .
$$
Therefore the thermal average cross-section reads:
\bea
\langle\sigma_{\chi\chi\chi\to\chi\chi^{*}} v^{2}\rangle &=& \frac{\sqrt{5}}{192\pi m_{\chi}^{3}}\bigg[|\mathcal{M}_{\chi\chi\chi\to\chi\chi^{*}}|^{2}+|\mathcal{M}_{\chi^{*}\chi^{*}\chi^{*}\to\chi^{*}\chi}|^{2}\bigg] \nonumber
\\
&=&\frac{\sqrt{5}}{192\pi m_{\chi}^{3}}\bigg[2 \times |\mathcal{M}_{\chi\chi\chi\to\chi\chi^{*}}|^{2}\bigg] .
\eea
We will derive the last expression in a moment.
\clearpage
\hspace{6.5cm}\underline{\Large\textbf{$\chi \chi^{*}\chi^{*} \to \chi \chi$}}
\subsection*{Feynman Diagrams}
\label{feyn3to2dm2}
$$
\hspace{-0.5cm}\includegraphics[scale=0.2]{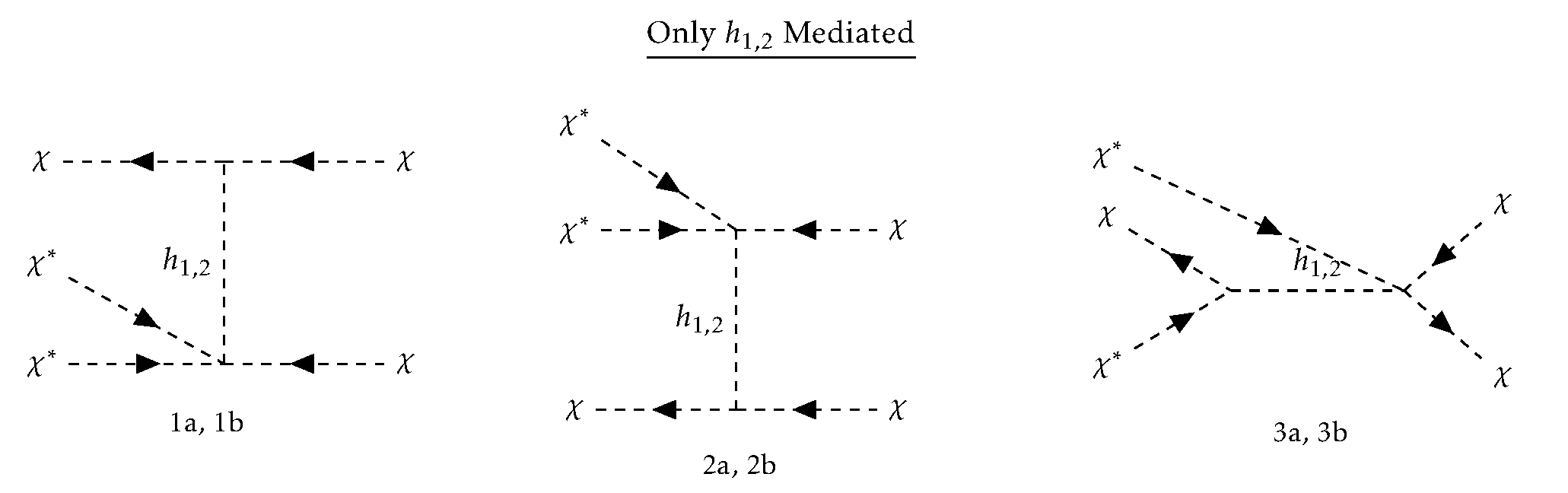}
$$
$$
\hspace{-0.5cm}\includegraphics[scale=0.23]{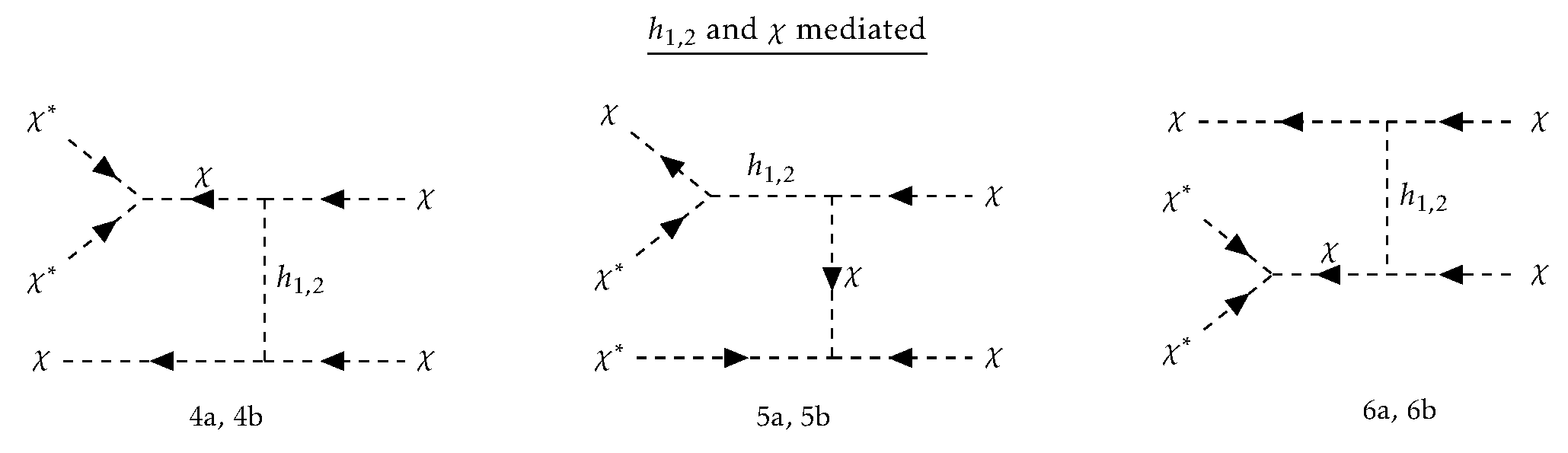}
$$
$$
\hspace{-1cm}\includegraphics[scale=0.23]{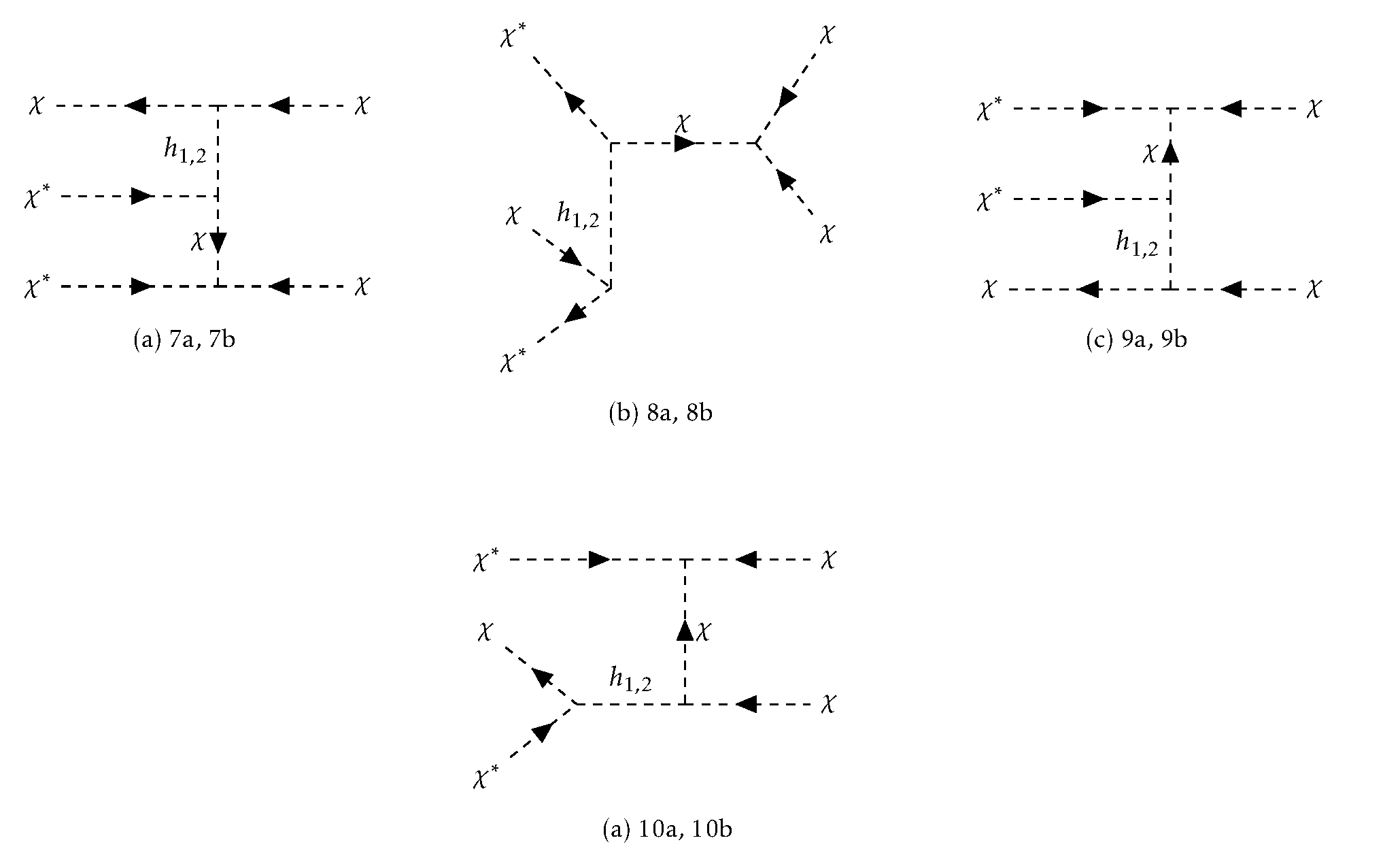}
$$
$$
\hspace{-0.5cm}\includegraphics[scale=0.23]{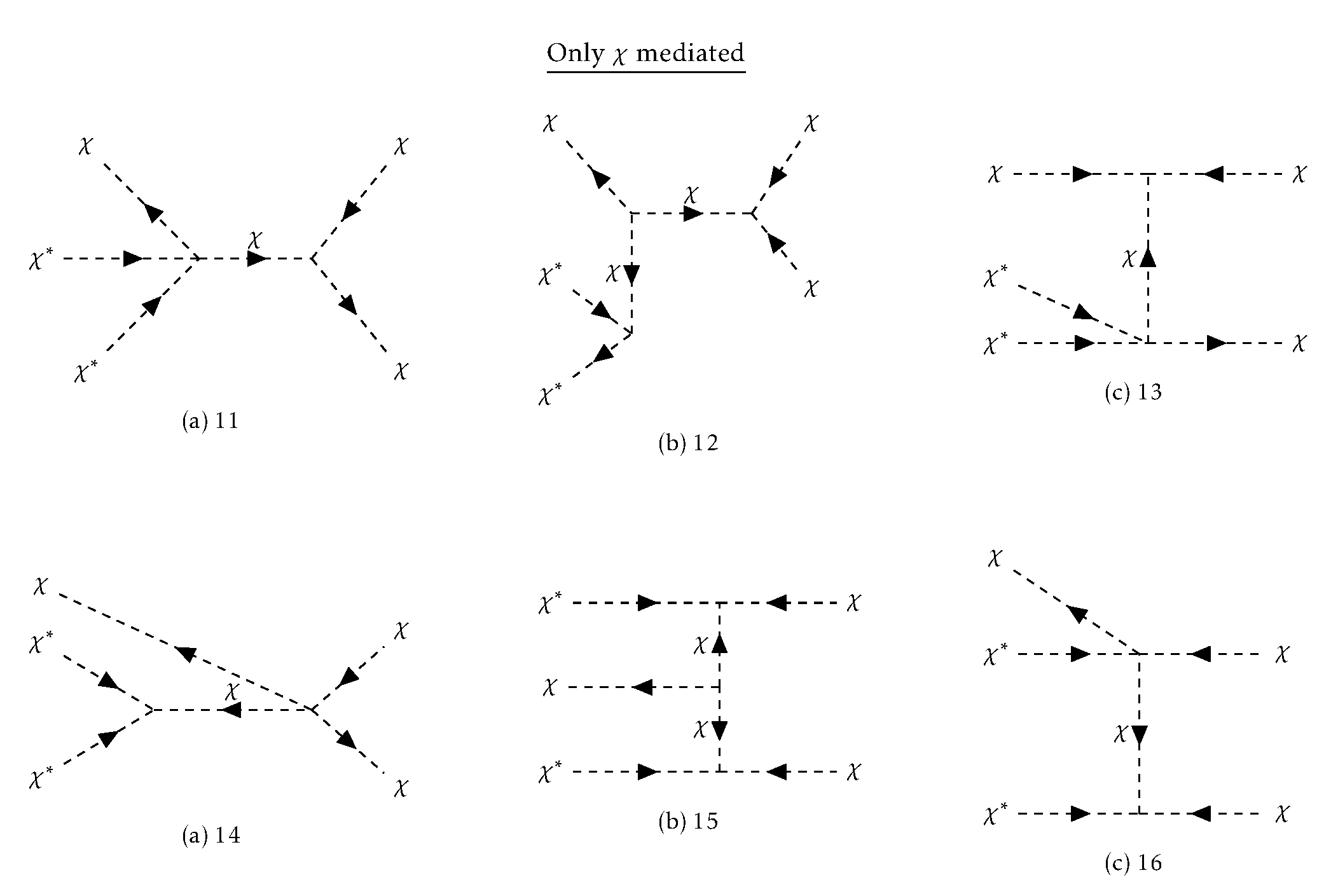}
$$
Note here that we have not shown $u$-channel graphs, which will also contribute to the cross-section.

\subsection*{Matrix Amplitude}
\vspace{-1cm}
\begin{varwidth}[t]{.6\textwidth}
\hspace{0.75cm}\underline{Only $\chi$ mediated}
\begin{itemize}
\item $\mathcal{M}_{11}=\frac{(-\lambda_{3})(-\lambda_{4})}{s-m_{\chi}^{2}}$
\item $\mathcal{M}_{12}= \frac{(-\lambda_{3})^{3}}{(s-m_{\chi}^{2})(t-m_{\chi}^{2})}$
\item $\mathcal{M}_{13}=\frac{(-\lambda_{3})(-\lambda_{4})}{t-m_{\chi}^{2}}$
\item $\mathcal{M}_{14}=\frac{(-\lambda_{3})(-\lambda_{4})}{s-m_{\chi}^{2}}$
\end{itemize}
\end{varwidth}
\begin{varwidth}[t]{.6\textwidth}
\vspace{2cm}
\begin{itemize}
\item $\mathcal{M}_{15t}= \frac{(-\lambda_{3})^{3}}{(t-m_{\chi}^{2})(t-m_{\chi}^{2})}$
\item $\mathcal{M}_{15u}= \frac{(-\lambda_{3})^{3}}{(u-m_{\chi}^{2})(t-m_{\chi}^{2})}$
\item $\mathcal{M}_{16t}=\frac{(-\lambda_{3})(-\lambda_{4})}{t-m_{\chi}^{2}}$
\item $\mathcal{M}_{16u}=\frac{(-\lambda_{3})(-\lambda_{4})}{u-m_{\chi}^{2}}$
\end{itemize}
\end{varwidth}

\begin{varwidth}[t]{.5\textwidth}
\hspace{0.75cm}\underline{Only $h_{1,2}$ mediated},
\begin{itemize}
\item $\mathcal{M}_{1at}=\frac{(-\lambda_{a1})(-\lambda_{a2})}{t-m_{h_{1}}^{2}}$
\item $\mathcal{M}_{1au}=\frac{(-\lambda_{a1})(-\lambda_{a2})}{u-m_{h_{1}}^{2}}$
\item $\mathcal{M}_{2at}=\frac{(-\lambda_{a1})(-\lambda_{a2})}{t-m_{h_{1}}^{2}}$
\item $\mathcal{M}_{2au}=\frac{(-\lambda_{a1})(-\lambda_{a2})}{u-m_{h_{1}}^{2}}$
\item $\mathcal{M}_{3a}=\frac{(-\lambda_{a1})(-\lambda_{a2})}{t-m_{h_{1}}^{2}}$
\item $\mathcal{M}_{1bt}=\frac{(-\lambda_{a1})(-\lambda_{a2})}{t-m_{h_{2}}^{2}}$
\item $\mathcal{M}_{1bu}=\frac{(-\lambda_{a1})(-\lambda_{a2})}{u-m_{h_{2}}^{2}}$
\item $\mathcal{M}_{2bt}=\frac{(-\lambda_{a1})(-\lambda_{a2})}{t-m_{h_{2}}^{2}}$
\item $\mathcal{M}_{2bu}=\frac{(-\lambda_{a1})(-\lambda_{a2})}{u-m_{h_{2}}^{2}}$
\item $\mathcal{M}_{3b}=\frac{(-\lambda_{a1})(-\lambda_{a2})}{t-m_{h_{2}}^{2}}$
\end{itemize}
\end{varwidth}
\begin{varwidth}[t]{0.5\textwidth}
\hspace{0.75cm}\underline{$h_{1}$ and $\chi$ mediated}
\begin{itemize}
\item $\mathcal{M}_{4at}=\frac{(-\lambda_{a1})^{2}(-\lambda_{3})}{(t-m_{h_{1}}^{2})(s-m_{\chi}^{2})}$
\item $\mathcal{M}_{4au}=\frac{(-\lambda_{a1})^{2}(-\lambda_{3})}{(u-m_{h_{1}}^{2})(s-m_{\chi}^{2})}$
\item $\mathcal{M}_{5at}=\frac{(-\lambda_{a1})^{2}(-\lambda_{3})}{(s-m_{h_{1}}^{2})(t-m_{\chi}^{2})}$
\item $\mathcal{M}_{5au}=\frac{(-\lambda_{a1})^{2}(-\lambda_{3})}{(s-m_{h_{1}}^{2})(-m_{\chi}^{2})}$
\item $\mathcal{M}_{6at}=\frac{(-\lambda_{a1})^{2}(-\lambda_{3})}{(t-m_{h_{1}}^{2})(s-m_{\chi}^{2})}$
\item $\mathcal{M}_{6au}=\frac{(-\lambda_{a1})^{2}(-\lambda_{3})}{(u-m_{h_{1}}^{2})(s-m_{\chi}^{2})}$
\item $\mathcal{M}_{7at}=\frac{(-\lambda_{a1})^{2}(-\lambda_{3})}{(t-m_{h_{1}}^{2})(t-m_{\chi}^{2})}$
\item $\mathcal{M}_{7au}=\frac{(-\lambda_{a1})^{2}(-\lambda_{3})}{(s-m_{h_{1}}^{2})(u-m_{\chi}^{2})}$
\item $\mathcal{M}_{8a}=\frac{(-\lambda_{a1})^{2}(-\lambda_{3})}{(t-m_{h_{1}}^{2})(s-m_{\chi}^{2})}$
\item $\mathcal{M}_{9at}=\frac{(-\lambda_{a1})^{2}(-\lambda_{3})}{(t-m_{h_{1}}^{2})(t-m_{\chi}^{2})}$
\item $\mathcal{M}_{9au}=\frac{(-\lambda_{a1})^{2}(-\lambda_{3})}{(s-m_{h_{1}}^{2})(u-m_{\chi}^{2})}$
\item $\mathcal{M}_{10at}=\frac{(-\lambda_{a1})^{2}(-\lambda_{3})}{(s-m_{h_{1}}^{2})(t-m_{\chi}^{2})}$
\item $\mathcal{M}_{10au}=\frac{(-\lambda_{a1})^{2}(-\lambda_{3})}{(s-m_{h_{1}}^{2})(u-m_{\chi}^{2})}$
\end{itemize}
\end{varwidth}
\begin{varwidth}[t]{0.5\textwidth}
\hspace{0.75cm}\underline{$h_{2}$ and $\chi$ mediated}
\begin{itemize}
\item $\mathcal{M}_{4bt}=\frac{(-\lambda_{b1})^{2}(-\lambda_{3})}{(t-m_{h_{1}}^{2})(s-m_{\chi}^{2})}$
\item $\mathcal{M}_{4bu}=\frac{(-\lambda_{b1})^{2}(-\lambda_{3})}{(u-m_{h_{1}}^{2})(s-m_{\chi}^{2})}$
\item $\mathcal{M}_{5bt}=\frac{(-\lambda_{b1})^{2}(-\lambda_{3})}{(s-m_{h_{1}}^{2})(t-m_{\chi}^{2})}$
\item $\mathcal{M}_{5b}=\frac{(-\lambda_{b1})^{2}(-\lambda_{3})}{(s-m_{h_{1}}^{2})(u-m_{\chi}^{2})}$
\item $\mathcal{M}_{6bt}=\frac{(-\lambda_{b1})^{2}(-\lambda_{3})}{(t-m_{h_{1}}^{2})(s-m_{\chi}^{2})}$
\item $\mathcal{M}_{6bu}=\frac{(-\lambda_{b1})^{2}(-\lambda_{3})}{(u-m_{h_{1}}^{2})(s-m_{\chi}^{2})}$
\item $\mathcal{M}_{7bt}=\frac{(-\lambda_{b1})^{2}(-\lambda_{3})}{(t-m_{h_{1}}^{2})(t-m_{\chi}^{2})}$
\item $\mathcal{M}_{7bu}=\frac{(-\lambda_{b1})^{2}(-\lambda_{3})}{(u-m_{h_{1}}^{2})(t-m_{\chi}^{2})}$
\item $\mathcal{M}_{8b}=\frac{(-\lambda_{b1})^{2}(-\lambda_{3})}{(t-m_{h_{2}}^{2})(s-m_{\chi}^{2})}$
\item $\mathcal{M}_{9bt}=\frac{(-\lambda_{b1})^{2}(-\lambda_{3})}{(t-m_{h_{2}}^{2})(t-m_{\chi}^{2})}$
\item $\mathcal{M}_{9bu}=\frac{(-\lambda_{b1})^{2}(-\lambda_{3})}{(u-m_{h_{2}}^{2})(t-m_{\chi}^{2})}$
\item $\mathcal{M}_{10bt}=\frac{(-\lambda_{b1})^{2}(-\lambda_{3})}{(s-m_{h_{2}}^{2})(t-m_{\chi}^{2})}$
\item $\mathcal{M}_{10bt}=\frac{(-\lambda_{b1})^{2}(-\lambda_{3})}{(s-m_{h_{2}}^{2})(u-m_{\chi}^{2})}$
\end{itemize}
\end{varwidth}

\begin{align*}
\mathcal{M}_{net}=&(\mathcal{M}_{1at}+\mathcal{M}_{1au}+\mathcal{M}_{2at}\mathcal{M}_{2au}+\mathcal{M}_{3a}+\mathcal{M}_{4at}+\mathcal{M}_{4au}+\mathcal{M}_{5at}+\mathcal{M}_{5au}\\&+\mathcal{M}_{6at}+\mathcal{M}_{6au}
+\mathcal{M}_{7at}+\mathcal{M}_{7au}+\mathcal{M}_{8a}+\mathcal{M}_{9at}+\mathcal{M}_{9au}+\mathcal{M}_{10at}+\mathcal{M}_{10au})
\\&
+(\mathcal{M}_{1bt}+\mathcal{M}_{1bu}+\mathcal{M}_{2bt}+\mathcal{M}_{2bu}+\mathcal{M}_{3b}+\mathcal{M}_{4bt}+\mathcal{M}_{4bu}+\mathcal{M}_{5bt}+\mathcal{M}_{5bu}\\&+\mathcal{M}_{6bt}+\mathcal{M}_{6bu}+\mathcal{M}_{7bt}+\mathcal{M}_{7bu}+\mathcal{M}_{8b}+\mathcal{M}_{9bt}+\mathcal{M}_{9bu}+\mathcal{M}_{10bt}+\mathcal{M}_{10bu})
\\&
+(\mathcal{M}_{11}+\mathcal{M}_{12}+\mathcal{M}_{13}+\mathcal{M}_{14}+\mathcal{M}_{15t}+\mathcal{M}_{15u}+\mathcal{M}_{16t}+\mathcal{M}_{16u})
\end{align*}
Note above that we have written the u-channel contribution also, which exists corresponding to each t-channel graph as the final state particles here are identical.
Squared matrix amplitude is given as,
$$
\Rightarrow |\mathcal{M}_{\chi\chi^{*}\chi^{*}\to\chi\chi}|^{2}=\frac{1}{2}|\mathcal{M}_{net}|^{2} .
$$

The complex conjugate of $\chi\chi^{*}\chi^{*}\to\chi\chi$ $i.e.$ $\chi^{*}\chi\chi\to\chi^{*}\chi^{*}$ 
also contributes to the total matrix amplitude and has same expression as $\chi\chi^{*}\chi^{*}\to\chi\chi$,
$$
|\mathcal{M}_{\chi\chi^{*}\chi^{*}\to\chi\chi}|^{2}=|\mathcal{M}_{\chi^{*}\chi\chi\to\chi^{*}\chi^{*}}|^{2} .
$$

The thermal average cross-section reads:
\bea 
\langle\sigma_{\chi\chi^{*}\chi^{*}\to\chi\chi} v^{2}\rangle&=&\frac{\sqrt{5}}{192\pi m_{\chi}^{3}}\bigg[|\mathcal{M}_{\chi\chi^{*}\chi^{*}\to\chi\chi}|^{2}+|\mathcal{M}_{\chi^{*}\chi\chi\to\chi^{*}\chi^{*}}|^{2}\bigg] \nonumber \\
&=&\frac{\sqrt{5}}{192\pi m_{\chi}^{3}}\bigg[2\times|\mathcal{M}_{\chi\chi^{*}\chi^{*}\to\chi\chi}|^{2}\bigg]  ~.
\eea

Therefore, the total thermal average cross section for $3_{{\textrm{DM}}} \to 2_{{\textrm{DM}}}$ process turn out to be:

\bea
\langle\sigma_{3_{{\textrm{DM}}} \to 2_{{\textrm{DM}}}}v^{2} \rangle &=&\langle\sigma_{\chi\chi\chi\to\chi\chi^{*}} v^{2}\rangle+\langle\sigma_{\chi\chi^{*}\chi^{*}\to\chi\chi} v^{2}\rangle \nonumber
\\
&=&\frac{\sqrt{5}}{192\pi m_{\chi}^{3}}\bigg[2 \times \bigg(|\mathcal{M}_{\chi\chi\chi\to\chi\chi^{*}}|^{2}+|\mathcal{M}_{\chi\chi^{*}\chi^{*}\to\chi\chi}|^{2}\bigg)\bigg] .
\eea
\subsection*{General expression for $3_{\textrm{DM}} \to 2_{\textrm{DM}}$ annihilation cross-section }
Let us derive the $3_{{\textrm{DM}}} \to 2_{{\textrm{DM}}}$ annihilation cross-section in a model independent way as a 
function of the amplitude. We consider a process like: 
\label{3to2dm}
\bea
\nonumber
\chi(p_{1})~\chi(p_{2})~\chi(p_{3})~\to~\chi(p_{4})~\chi(p_{5})~~.
\eea
In non-relativistic limit, 
\bea
\label{eq:incomingE}
\nonumber
E_{1}=E_{2}=E_{3}&=&m_{\chi}\\
\Rightarrow E_{1}+E_{2}+E_{3}&=&3m_{\chi}.
\eea
$P_{i=1-5}$ are the three-momentum of incoming and outgoing particles. Now, one can express $(\sigma v^{2})_{3_{DM}\to 2_{DM}}$ as~\citep{crossection}:
\bea
\label{eq:bfavgcross}
\nonumber
(\sigma v^{2}) _{3_{\rm DM}\to 2_{\rm DM}}&=&\frac{1}{(2E_{1})(2E_{2})(2E_{3})} \int \frac{d^{3}P_{4}}{(2\pi)^{3}2E_{4}}\frac{d^{3}P_{5}}{(2\pi)^{3}2E_{5}}(2\pi)^{4}\delta^{4}(p_{1}+p_{2}+p_{3}-p_{4}-p_{5})|\mathcal{M}|_{3\to 2}^{2}\\ 
&=&\frac{1}{(2E_{1})(2E_{2})(2E_{3})} \frac{|\mathcal{M}|_{3\to 2}^{2}}{(2\pi)^{6}} \int \frac{d^{3}P_{4}}{2E_{4}}\frac{d^{3}P_{5}}{2E_{5}}(2\pi)^{4}\delta(E_{1}+E_{2}+E_{3}-E_{4}-E_{5}) \nonumber\\
&&\hspace{7.5cm}\delta^{3}(P_{1}+P_{2}+P_{3}-P_{4}-P_{5}),
\eea
assuming that the matrix amplitude is independent of the final outgoing particles. Now, in the centre of mass frame $P_{1}+P_{2}+P_{3}=0$, leads to:
\bea
\nonumber
(\sigma v^{2}) _{3_{\rm DM}\to 2_{\rm DM}} &=&\frac{1}{(2E_{1})(2E_{2})(2E_{3})} \frac{|\mathcal{M}|_{3\to 2}^{2}}{(2\pi)^{2}} \int \frac{d^{3}P_{4}}{2E_{4}}\frac{d^{3}P_{5}}{2E_{5}}\delta(E_{1}+E_{2}+E_{3}-E_{4}-E_{5})\delta^{3}(P_{4}+P_{5}).
\eea
Using Eq.\eqref{eq:incomingE} and the delta function gives us: $P_{4}=-P_{5}$. We also know that $E_{5}=\sqrt{P_{5}^{2}+m_{\chi}^{2}}$. So integrating over $P_{5}$ we get :
\bea
\nonumber
(\sigma v^{2}) _{3_{\rm DM}\to 2_{\rm DM}} &=&\frac{1}{8m_{\chi}^{3}} \frac{|\mathcal{M}|_{3\to 2}^{2}}{(2\pi)^{2}} \int \frac{d^{3}P_{4}}{2E_{4}}\frac{1}{2\sqrt{P_{4}^{2}+m_{\chi}^{2}}}\delta(3m_{\chi}-2\sqrt{P_{4}^{2}+m_{\chi}^{2}})\\ \nonumber
&=&\frac{1}{8m_{\chi}^{3}} \frac{|\mathcal{M}|_{3\to 2}^{2}}{(2\pi)^{2}} \int \frac{P_{4}^{2}d P_{4}d\Omega}{4(P_{4}^{2}+m_{\chi}^{2})}\delta(3m_{\chi}-2\sqrt{P_{4}^{2}+m_{\chi}^{2}})\\
\nonumber
&=&\frac{1}{8m_{\chi}^{3}} \frac{|\mathcal{M}|_{3\to 2}^{2}}{(2\pi)^{2}}\frac{4\pi}{4} \int \frac{P_{4}^{2}d P_{4}}{(P_{4}^{2}+m_{\chi}^{2})}\delta(3m_{\chi}-2\sqrt{P_{4}^{2}+m_{\chi}^{2}})\\\nonumber
&=&\frac{1}{8m_{\chi}^{3}} \frac{|\mathcal{M}|_{3\to 2}^{2}}{4\pi}\int \frac{P_{4}^{2}d P_{4}}{(P_{4}^{2}+m_{\chi}^{2})}\delta(3m_{\chi}-2\sqrt{P_{4}^{2}+m_{\chi}^{2}})\\\nonumber
&=&\frac{1}{2\times 32 \pi m_{\chi}^{3}} |\mathcal{M}|_{3\to 2}^{2}\int \frac{P_{4}^{2}d P_{4}}{(P_{4}^{2}+m_{\chi}^{2})}\delta(\frac{3}{2}m_{\chi}-\sqrt{P_{4}^{2}+m_{\chi}^{2}}).
\eea
Finally integrating over $P_{4}$ we get,
\bea
(\sigma v^{2}) _{3_{\rm DM}\to 2_{\rm DM}}&=&\frac{\sqrt{5}}{3}\times\frac{1}{64\pi m_{\chi}^{3}} |\mathcal{M}|_{3\to 2}^{2}~.
\eea
The thermal averaged cross section under the conditions mentioned above can be written as,
\bea
\label{eq:thermalavg}
\langle\sigma v^{2} \rangle_{3_{\rm DM}\to 2_{\rm DM}}&=&\frac{1}{n_{1}^{eq}~n_{2}^{eq}~n_{3}^{eq}}\int
\frac{g_{DM}~d^{3}P_{1}}{(2\pi)^{3}2E_{1}}~
\frac{g_{DM}~d^{3}P_{2}}{(2\pi)^{3}2E_{2}}~\frac{g_{DM}~d^{3}P_{3}}{(2\pi)^{3}2E_{3}}~\frac{g_{DM}~d^{3}P_{4}}{(2\pi)^{3}2E_{4}}~\frac{g_{DM}~d^{3}P_{5}}{(2\pi)^{3}2E_{5}}\nonumber \\
&&~~~~~~~~~(2\pi)^{4}\delta^{4}(p_{1}+p_{2}+p_{3}-p_{4}-p_{5})
\times |\mathcal{M}_{3\to 2}|^2~f_{1}^{eq}f_{2}^{eq}f_{3}^{eq} ~~.
\eea
Using Eq.\eqref{eq:bfavgcross} and Eq.\eqref{eq:thermalavg}, we can write,
\bea
\langle\sigma v^{2} \rangle_{3_{\rm DM}\to 2_{\rm DM}}&=&\frac{1}{n_{1}^{eq}~n_{2}^{eq}~n_{3}^{eq}}\int\frac{g_{DM} ~d^{3}P_{1}}{(2\pi)^{3}}~\frac{g_{DM}\:d^{3}P_{2}}{(2\pi)^{3}}~\frac{g_{DM}~d^{3}P_{3}}{(2\pi)^{3}}~f_{1}^{eq}f_{2}^{eq}f_{3}^{eq}~(\sigma v^{2})_{3_{DM}\to 2_{DM}}~, \nonumber \\
\eea
where $n_{i}^{eq}$ can be expressed in terms of modified Bessel's function as~\cite{Feng:2014vea},
\bea
n_{i}^{eq}= \frac{g_{DM}}{(2\pi)^{3}} \int  d^{3}P_i~f^{eq}(E_{i},T) ~.
\eea 
Since,
\bea
\nonumber
d^{3}P_{i}~f^{eq}(E_{i},T)&=& 4\pi m_{\chi}^{3}\bigg( \frac{E_{i}}{m_{\chi}}\bigg)~\bigg(\sqrt{\bigg(\frac{E_{i}}{m_{\chi}}\bigg)^{2}-1}\bigg)~e^{-(\frac{E_{i}}{m_{\chi}})(\frac{m_{\chi}}{T})}d\bigg(\frac{E_{i}}{m_{\chi}}\bigg),
\\ \nonumber
\implies \int d^{3}P_{i}~f^{eq}(E_{i},T)&=& 4\pi m_{\chi}^{3}\int \bigg( \frac{E_{i}}{m_{\chi}}\bigg)~\bigg(\sqrt{\bigg(\frac{E_{i}}{m_{\chi}}\bigg)^{2}-1}\bigg)~e^{-(\frac{E_{i}}{m_{\chi}})(\frac{m_{\chi}}{T})}d\bigg(\frac{E_{i}}{m_{\chi}}\bigg)\\
&=&4\pi m_{\chi}^{3}  \frac{K_{2}(m_{\chi}/T)}{m_{\chi}/T}=4\pi m_{\chi}^{2} T K_{2}(m_{\chi}/T)~.
\eea
Therefore, 
\bea
n_{i}^{eq}= \frac{g_{DM}}{(2\pi)^{3}}~4\pi m_{\chi}^{2} T K_{2}(m_{\chi}/T)~.
\eea
Now one can write the $\langle\sigma v^{2} \rangle_{3_{\rm DM}\to 2_{\rm DM}}$ as follows:
\bea 
\nonumber
\Rightarrow \langle \sigma v^{2}\rangle _{3_{\rm DM}\to 2_{\rm DM}} &=& \frac{1}{(4\pi m_{\chi}^{2} T K_{2}(m_{\chi}/T))^{3}} \int (\sigma v^{2}) _{3_{\rm DM}\to 2_{\rm DM}}~f_{1}^{eq}~f_{2}^{eq}~f_{3}^{eq}d^{3}P_{1}~d^{3}P_{2}~d^{3}P_{3} ~~.
\eea 

\bea 
\nonumber
\Rightarrow \langle \sigma v^{2}\rangle _{3_{\rm DM}\to 2_{\rm DM}} &=& \frac{1}{(4\pi m_{\chi}^{2} T K_{2}(m_{\chi}/T))^{3}} \int \frac{\sqrt{5}}{192 \pi E_{1}~E_{2}~E_{3}}|\mathcal{M}|^{2}_{3 \to 2}\\
\nonumber
&&\bigg[4\pi m_{\chi}^{3}\bigg( \frac{E_{1}}{m_{\chi}}\bigg)~\bigg(\sqrt{\bigg(\frac{E_{1}}{m_{\chi}}\bigg)^{2}-1}\bigg)~e^{-(\frac{E_{1}}{m_{\chi}})(\frac{m_{\chi}}{T})}d\bigg(\frac{E_{1}}{m_{\chi}}\bigg)\bigg]\\\nonumber
&&\bigg[4\pi m_{\chi}^{3} \bigg( \frac{E_{2}}{m_{\chi}}\bigg)~\bigg(\sqrt{\bigg(\frac{E_{2}}{m_{\chi}}\bigg)^{2}-1}\bigg)~e^{-(\frac{E_{2}}{m_{\chi}})(\frac{m_{\chi}}{T})}d\bigg(\frac{E_{2}}{m_{\chi}}\bigg)\bigg]\\
&&\bigg[
4\pi m_{\chi}^{3} \bigg( \frac{E_{3}}{m_{\chi}}\bigg)~\bigg(\sqrt{\bigg(\frac{E_{3}}{m_{\chi}}\bigg)^{2}-1}\bigg)~e^{-(\frac{E_{3}}{m_{\chi}})(\frac{m_{\chi}}{T})}d\bigg(\frac{E_{3}}{m_{\chi}}\bigg)\bigg]  \nonumber \\
 &=& \frac{1}{(4\pi m_{\chi}^{2} T K_{2}(m_{\chi}/T))^{3}}~(4\pi m_{\chi}^{2})^{3}~\frac{\sqrt{5}}{192 \pi}|\mathcal{M}|^{2}_{3 \to 2} \nonumber \\
&&\int \bigg[~\bigg(\sqrt{\bigg(\frac{E_{1}}{m_{\chi}}\bigg)^{2}-1}\bigg)~e^{-(\frac{E_{1}}{m_{\chi}})(\frac{m_{\chi}}{T})}d\bigg(\frac{E_{1}}{m{\chi}}\bigg)\bigg] \nonumber \\
&&\bigg[~\bigg(\sqrt{\bigg(\frac{E_{2}}{m_{\chi}}\bigg)^{2}-1}\bigg)~e^{-(\frac{E_{2}}{m_{\chi}})(\frac{m_{\chi}}{T})}d\bigg(\frac{E_{2}}{m_{\chi}}\bigg)\bigg] \nonumber \\
&&\bigg[
~\bigg(\sqrt{\bigg(\frac{E_{3}}{m_{\chi}}\bigg)^{2}-1}\bigg)~e^{-(\frac{E_{3}}{m_{\chi}})(\frac{m_{\chi}}{T})}d\bigg(\frac{E_{3}}{m_{\chi}}\bigg)\bigg] \nonumber \\
&=& \frac{1}{(4\pi m_{\chi}^{2} T K_{2}(m_{\chi}/T))^{3}}~(4\pi m_{\chi}^{2})^{3}~\frac{\sqrt{5}}{192 \pi}|\mathcal{M}|^{2}_{3 \to 2} 
\bigg(\frac{K_{1}(m_{\chi}/T)}{m_{\chi}/T}\bigg)^{3} \nonumber \\
&=& 
\bigg( \frac{K_{1}(m_{\chi}/T)}{K_{2}(m_{\chi}/T)}\bigg)^{3} \frac{\sqrt{5}}{192 \pi m_{\chi}^{3} }|\mathcal{M}|^{2}_{3 \to 2} 
\\ &\approx & \frac{\sqrt{5}}{192 \pi m_{\chi}^{3} }|\mathcal{M}|^{2}_{3 \to 2} ~~.
\eea

\section{General expression for $4_{\textrm{DM}} \to 2_{\textrm{DM}}$ annihilation cross-section}
\label{4to2dm1}
One can also derive the $4_{{\textrm{DM}}} \to 2_{{\textrm{DM}}}$ annihilation cross-section similar like $3_{{\textrm{DM}}} \to 2_{{\textrm{DM}}}$. Let us consider a process like: 
\bea
\nonumber
\chi(p_{1})~\chi(p_{2})~\chi(p_{3})~\chi(p_{4})~\to~\chi(p_{5})~\chi(p_{6})
\eea
In non-relativistic limit, 
\bea
\label{eq:4incomingE1}
\nonumber
E_{1}=E_{2}=E_{3}=E_{4}&=&m_{\chi}\\
\Rightarrow E_{1}+E_{2}+E_{3}+E_{4}&=&4m_{\chi}~.
\eea
Now, one can express the $(\sigma v^{3})_{4_{\rm DM}\to 2_{\rm DM}}$ as,
\bea
\label{eq:4bfavgcross1}
\nonumber
(\sigma v^{3}) _{4_{\rm DM}\to 2_{\rm DM}}&=&\frac{1}{(2E_{1})(2E_{2})(2E_{3})(2E_{4})} \frac{|\mathcal{M}|_{4\to 2}^{2}}{(2\pi)^{6}} \int \frac{d^{3}P_{5}}{2E_{5}}\frac{d^{3}P_{6}}{2E_{6}}(2\pi)^{4}\delta(E_{1}+E_{2}+E_{3}+E_{4}-E_{5}-E_{6})
\\
&&\hspace{7.0cm}\delta^{3}(P_{1}+P_{2}+P_{3}+P_{4}-P_{5}-P_{6}) .
\eea
Here we have considered that the matrix amplitude is independent of the final outgoing particle momentum. Now, in the center-of-mass frame: $P_{1}+P_{2}+P_{3}+P_{4}=0$; the annihilation cross-section 
$(\sigma v^{3}) _{4_{\rm DM}\to 2_{\rm DM}}$ becomes: 
\bea
\nonumber
(\sigma v^{3}) _{4_{\rm DM}\to 2_{\rm DM}} &=&\frac{1}{(2E_{1})(2E_{2})(2E_{3})(2E_{4})} \frac{|\mathcal{M}|_{4\to 2}^{2}}{(2\pi)^{2}} \int \frac{d^{3}P_{5}}{2E_{5}}\frac{d^{3}P_{6}}{2E_{6}}\delta(E_{1}+E_{2}+E_{3}+E_{4}-E_{5}-E_{6}) \\
&& \hspace{8.0cm} \delta^{3}(P_{5}+P_{6}) ~.
\eea
Integrating over $P_{6}$ we get,
\bea
\nonumber
(\sigma v^{3}) _{4_{\rm DM}\to 2_{\rm DM}} &=&\frac{1}{16m_{\chi}^{4}} \frac{|\mathcal{M}|_{4\to 2}^{2}}{(2\pi)^{2}} \int \frac{d^{3}P_{5}}{2E_{5}}\frac{1}{2\sqrt{P_{5}^{2}+m_{\chi}^{2}}}\delta(4m_{\chi}-2\sqrt{P_{5}^{2}+m_{\chi}^{2}})\\ \nonumber
&=&\frac{1}{2\times 64 \pi m_{\chi}^{4}} |\mathcal{M}|_{4\to 2}^{2}\int \frac{P_{5}^{2}d P_{5}}{(P_{5}^{2}+m_{\chi}^{2})}\delta(2m_{\chi}-\sqrt{P_{5}^{2}+m_{\chi}^{2}})~.
\eea
Finally integrating over $P_{5}$ we get,
\bea
\nonumber
\Rightarrow (\sigma v^{3}) _{4_{\rm DM}\to 2_{\rm DM}}&=&\frac{\sqrt{3}}{256~\pi ~m_{\chi}^{4}} |\mathcal{M}|_{4\to 2}^{2} ~,
\eea
where $|\mathcal{M}|_{4\to 2}$ is the matrix amplitude for $4_{\rm DM} \to 2_{\rm DM}$ processes.
The thermal averaged cross section for $4_{\rm DM} \to 2_{\rm DM}$ process can be written as,
\bea
\label{eq:4thermalavg1}
\langle\sigma v^{3} \rangle_{4_{\rm DM} \to 2_{\rm DM}}&=&\frac{1}{n_{1}^{eq}~n_{2}^{eq}~n_{3}^{eq}~n_{4}^{eq}}\int
\frac{g_{DM}~d^{3}P_{1}}{(2\pi)^{3}2E_{1}}~\frac{g_{DM}~d^{3}P_{2}}{(2\pi)^{3}2E_{2}}~\frac{g_{DM}~d^{3}P_{3}}{(2\pi)^{3}2E_{3}}~\frac{g_{DM}~d^{3}P_{4}}{(2\pi)^{3}2E_{4}}\nonumber \\
&&\hspace{3cm}\frac{g_{DM}~d^{3}P_{5}}{(2\pi)^{3}2E_{5}}~\frac{g_{DM}~d^{3}P_{6}}{(2\pi)^{3}2E_{6}}~(2\pi)^{4}\delta^{4}(P_{1}+P_{2}+P_{3}+P_{4}-P_{5}-P_{6})\nonumber \\
&&\hspace{7cm}|\mathcal{M}_{4\to 2}|^2~f_{1}^{eq}f_{2}^{eq}f_{3}^{eq}~f_{4}^{eq} .
\eea
Using Eq.\eqref{eq:4bfavgcross1} and Eq.\eqref{eq:4thermalavg1}, we can write,
\bea
\langle\sigma v^{3} \rangle_{4_{\rm DM} \to 2_{\rm DM}}&=&\frac{1}{n_{1}^{eq}~n_{2}^{eq}~n_{3}^{eq}~n_{4}^{eq}}
\int\frac{g_{DM} ~d^{3}P_{1}}{(2\pi)^{3}}~\frac{g_{DM}~d^{3}P_{2}}{(2\pi)^{3}}~\frac{g_{DM}~d^{3}P_{3}}{(2\pi)^{3}}~\frac{g_{DM}~d^{3}P_{4}}{(2\pi)^{3}}\nonumber\\
&&\hspace{3cm}f_{1}^{eq}f_{2}^{eq}f_{3}^{eq}f_{4}^{eq}~(\sigma v^{3})_{4_{\rm DM} \to 2_{\rm DM}}. \nonumber 
\eea
\bea 
\nonumber
\Rightarrow \langle \sigma v^{3}\rangle _{4_{\rm DM} \to 2_{\rm DM}} &=& \frac{1}{(4\pi m_{\chi}^{2} T K_{2}(m_{\chi}/T))^{4}} \int (\sigma v^{3})_{4\to 2}~f_{1}^{eq}~f_{2}^{eq}~f_{3}^{eq}~f_{4}^{eq}\nonumber\\
&&\hspace{4.5cm}d^{3}P_{1}~d^{3}P_{2}~d^{3}P_{3}~d^{3}P_{4} ~. \nonumber
\eea 
Similarly like $3_{\rm DM} \to 2_{\rm DM}$, we can finally derive
\bea 
\Rightarrow \langle \sigma v^{3}\rangle _{4_{\rm DM} \to 2_{\rm DM}} &=& 
\bigg( \frac{K_{1}(m_{\chi}/T)}{K_{2}(m_{\chi}/T)}\bigg)^{4} \frac{\sqrt{3}}{256 ~\pi ~m_{\chi}^{4} }|\mathcal{M}|^{2}_{4 \to 2} 
\\  &\approx & \frac{\sqrt{3}}{256 \pi m_{\chi}^{4} }|\mathcal{M}|^{2}_{4 \to 2} ~~.
\eea

\section{Self Scattering cross-section of DM}
\label{selfDMann}
We consider here all the processes that yield self scattering. There are two processes in the model essentially: $\chi\chi \to \chi\chi$ and $\chi\chi^* \to \chi\chi^*$ and their conjugates.

\vspace{0.5cm}
\hspace{6.5cm}\underline{\Large\textbf{$\chi \chi \to \chi \chi$}}

\subsection*{Feynman Diagrams}
\includegraphics[scale=0.5]{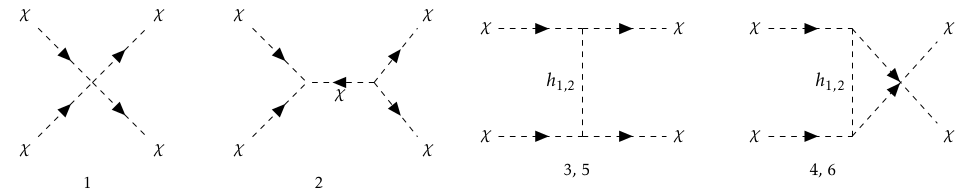}
\subsection*{Matrix Amplitude}
$$\mathcal{M}_{1}=-4\lambda_{\chi}$$
$$\mathcal{M}_{2}=\frac{[-(\mu_{\chi}+Y_{\chi\phi} v_{\phi})]^{2}}{s-m_{\chi}^{2}}$$
$$\mathcal{M}_{3}=\frac{[-(\lambda_{\chi h}v_{h}\cos\theta-(\lambda_{\chi\phi}v_{\phi}+\mu_{\chi\phi})\sin\theta)]^{2}}{t-m_{h_{1}}^{2}}$$
$$\mathcal{M}_{4}=\frac{[-(\lambda_{\chi h}v_{h}\cos\theta-(\lambda_{\chi\phi}v_{\phi}+\mu_{\chi\phi})\sin\theta)]^{2}}{u-m_{h_{1}}^{2}}$$
$$\mathcal{M}_{5}=\frac{[-(\lambda_{\chi h}v_{h}\sin\theta+(\lambda_{\chi\phi}v_{\phi}+\mu_{\chi\phi})\cos\theta)]^{2}}{t-m_{h_{2}}^{2}}$$
$$\mathcal{M}_{6}=\frac{[-(\lambda_{\chi h}v_{h}\sin\theta+(\lambda_{\chi\phi}v_{\phi}+\mu_{\chi\phi})\cos\theta)]^{2}}{u-m_{h_{2}}^{2}}$$

Net matrix amplitude for $\chi\chi\to\chi\chi$ is 
$$
\mathcal{M}_{Net}=\mathcal{M}_{1}+\mathcal{M}_{2}+\mathcal{M}_{3}+\mathcal{M}_{4}+\mathcal{M}_{5}+\mathcal{M}_{6}~~.
$$

So the squared matrix amplitude is given by 
$$
\Rightarrow |\mathcal{M}_{\chi\chi\to\chi\chi}|^{2}=\frac{1}{2}|\mathcal{M}_{Net}|^{2} ~~.
$$

The complex conjugate of $\chi\chi\to\chi\chi$ $i.e.$ $\chi^{*}\chi^{*}\to\chi^{*}\chi^{*}$ also contributes to the total matrix amplitude and has same expression as $\chi\chi\to\chi\chi$,
$$
|\mathcal{M}_{\chi\chi\to\chi\chi}|^{2}=|\mathcal{M}_{\chi^{*}\chi^{*}\to\chi^{*}\chi^{*}}|^{2} ~.
$$
The cross section turns out to be
\bea
\sigma_{\chi\chi\to\chi\chi}&=&\frac{1}{64\pi m_{\chi}^{2}}\bigg[|\mathcal{M}_{\chi\chi\to\chi\chi}|^{2}+|\mathcal{M}_{\chi^{*}\chi^{*}\to\chi^{*}\chi^{*}}|^{2}\bigg] \nonumber
\\
&=&\frac{1}{64\pi m_{\chi}^{2}}\bigg[2* |\mathcal{M}_{\chi\chi\to\chi\chi}|^{2}\bigg].
\eea

\vspace{0.5cm}
\hspace{6.5cm}\underline{\Large\textbf{$\chi \chi^{*} \to \chi \chi^{*}$}}

\subsection*{Feynman diagrams}
\includegraphics[scale=0.325]{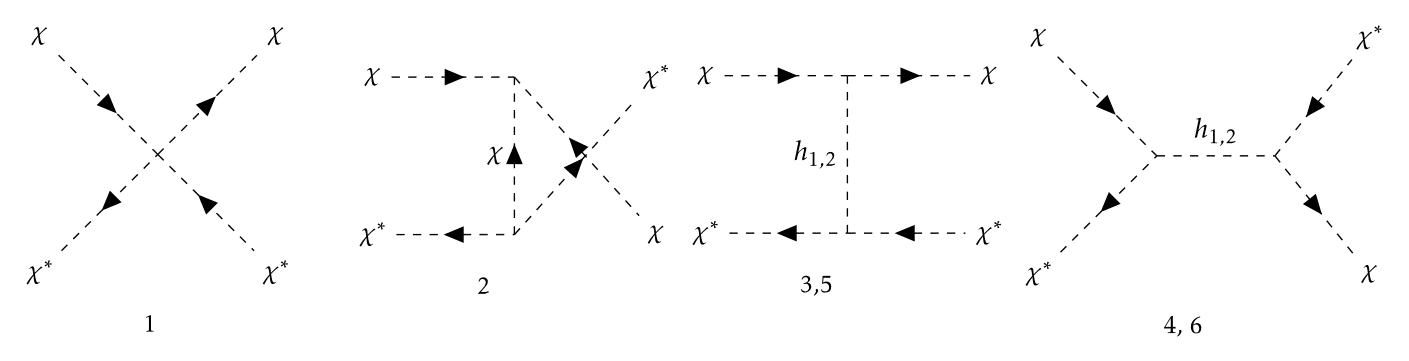}

\subsection*{Matrix Amplitude}
$$\mathcal{M}_{1}=-4\lambda_{\chi}$$
$$\mathcal{M}_{2}=\frac{[-(\mu_{\chi}+Y_{\chi\phi}v_{\phi})]^{2}}{u-m_{\chi}^{2}}$$
$$\mathcal{M}_{3}=\frac{[-(\lambda_{\chi h}v_{h}Cos\theta-(\lambda_{\chi\phi}v_{\phi}+\mu_{\chi\phi})Sin\theta)]^{2}}{t-m_{h_{1}}^{2}}$$
$$\mathcal{M}_{4}=\frac{[-(\lambda_{\chi h}v_{h}Cos\theta-(\lambda_{\chi\phi}v_{\phi}+\mu_{\chi\phi})Sin\theta)]^{2}}{s-m_{h_{1}}^{2}}$$
$$\mathcal{M}_{5}=\frac{[-(\lambda_{\chi h}v_{h}Sin\theta+(\lambda_{\chi\phi}v_{\phi}+\mu_{\chi\phi})Cos\theta)]^{2}}{t-m_{h_{2}}^{2}}$$
$$\mathcal{M}_{6}=\frac{[-(\lambda_{\chi h}v_{h}Sin\theta+(\lambda_{\chi\phi}v_{\phi}+\mu_{\chi\phi})Cos\theta)]^{2}}{s-m_{h_{2}}^{2}}$$

Net Matrix amplitude for $\chi\chi^{*}\to\chi\chi^{*}$ is written as,
$$
\mathcal{M}_{Net}=\mathcal{M}_{1}+\mathcal{M}_{2}+\mathcal{M}_{3}+\mathcal{M}_{4}+\mathcal{M}_{5}+\mathcal{M}_{6} ~.
$$

Squared matrix amplitude is given as,
$$
\Rightarrow |\mathcal{M}_{\chi\chi^{*}\to\chi\chi^{*}}|^{2}=|\mathcal{M}_{Net}|^{2}~~.
$$

The complex conjugate of $\chi\chi^{*}\to\chi\chi^{*}$ $i.e.$ $\chi^{*}\chi\to\chi^{*}\chi$ also contributes to the total matrix amplitude and has same expression as $\chi\chi^{*}\to\chi\chi^{*}$,
$$
|\mathcal{M}_{\chi\chi^{*}\to\chi\chi^{*}}|^{2}=|\mathcal{M}_{\chi^{*}\chi\to\chi^{*}\chi}|^{2}  ~.
$$
The cross section for this process then turns out to be
\begin{align*}
\sigma_{\chi\chi^{*}\to\chi\chi^{*}}=\frac{1}{64\pi m_{\chi}^{2}}\bigg[|\mathcal{M}_{\chi\chi^{*}\to\chi\chi^{*}}|^{2}\bigg] ~.
\end{align*}
Finally, adding both contributions, the total scattering cross-section is obtained as
\begin{align}
\sigma_{self}&=2 \times (\sigma_{\chi\chi\to\chi\chi}+\sigma_{\chi\chi^{*}\to\chi\chi^{*}}) \nonumber \\
&=\frac{1}{64\pi m_{\chi}^{2}}\bigg[2 \times \bigg(|\mathcal{M}_{\chi\chi\to\chi\chi}|^{2}+|\mathcal{M}_{\chi\chi^{*}\to\chi\chi^{*}}|^{2}\bigg)\bigg] ~.
\end{align}

\section{$3_{\rm DM} \to 2_{\rm SM}$ cross-section}

We have focused on two types of annihilations here: $3_{\rm DM} \to 2_{\rm SM}$ and $2_{\rm DM} \to 2_{\rm SM}$. We will first analyse the 
processes that contribute to $3_{\rm DM} \to 2_{\rm SM}$ annihilation in this model and also compute the generic form of such cross-section.
\\

\hspace{4.5cm}\underline{\Large\textbf{$\chi(p_{1}) \chi(p_{2})\chi(p_{3}) \to f(k_{1}) \bar{f}(k_{2}) $}}

\subsection*{Feynman Diagrams}
\begin{figure}[htb!]
\hspace{2.5cm}
\includegraphics[scale=0.45]{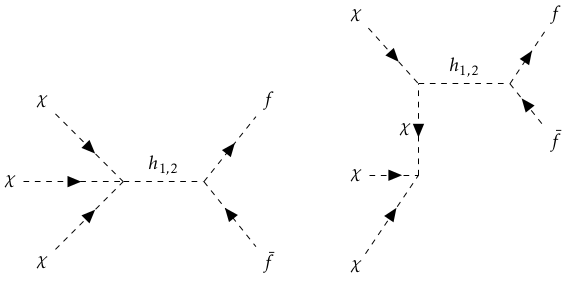}
\end{figure}

\subsection*{Matrix Amplitude}
\begin{align*}
|\mathcal{M}_{\chi\chi\chi\to f\bar{f}}|^{2}&=2(s-4m_{f}^{2})\bigg[\frac{\lambda_{a2}\lambda_{f1}}{s-m_{h_{1}}^{2}}+\frac{\lambda_{b2}\lambda_{f2}}{s-m_{h_{2}}^{2}}+\frac{\lambda_{3}\lambda_{a1}\lambda_{f1}}{(s-m_{h_{1}}^{2})(t-m_{\chi}^{2})}+\frac{\lambda_{3}\lambda_{b1}\lambda_{f2}}{(s-m_{h_{2}}^{2})(t-m_{\chi}^{2})}\bigg]^{2}~.
\end{align*}
The complex conjugate of $\chi\chi\chi\to f\bar{f}$ $i.e.$ $\chi^{*}\chi^{*}\chi^{*}\to \bar{f} f$also contributes to the total matrix amplitude and has same expression as $\chi\chi\chi\to f\bar{f}$,
$$
|\mathcal{M}_{\chi\chi\chi\to f\bar{f}}|^{2}=|\mathcal{M}_{\chi^{*}\chi^{*}\chi^{*}\to \bar{f} f}|^{2} ~.
$$
Therefore the cross-section for $3_{\rm DM} \to 2_{\rm SM}$ is :
\bea
\langle\sigma v^2\rangle_{{\chi\chi\chi\to f\bar{f}}}&=&\frac{1}{64\pi m_{\chi}^{3}}\bigg(1-\frac{4m_{f}^{2}}{9m_{\chi}^{2}}\bigg)^{1/2}\bigg[|\mathcal{M}_{\chi\chi\chi\to f\bar{f}}|^{2}+|\mathcal{M}_{\chi^{*}\chi^{*}\chi^{*}\to \bar{f} f}|^{2}\bigg] \nonumber \\
&=&\frac{1}{64\pi m_{\chi}^{3}}\bigg(1-\frac{4m_{f}^{2}}{9m_{\chi}^{2}}\bigg)^{1/2}\bigg[2* |\mathcal{M}_{\chi\chi\chi\to f \bar{f} }|^{2}\bigg] ~.
\eea
\subsection*{General expression for $3_{\textrm{DM}} \to 2_{\textrm{SM}}$ annihilation cross-section }
Let us quickly derive the $3_{{\textrm{DM}}} \to 2_{{\textrm{SM}}}$ annihilation cross-section in a model independent way as a 
function of the amplitude. We consider a process like: 
\label{3to2dmsm}
\bea
\nonumber
\chi(p_{1})~\chi(p_{2})~\chi(p_{3})~\to~ f(p_{4})~f(p_{5})~~.
\eea
Following a similar procedure that we adopted for $3_{\rm DM}\to 2_{\rm DM}$  annihilation crossection we can derive an expression for $3_{\rm DM}\to 2_{\rm SM}$ as follows,
\bea
\nonumber
(\sigma v^{2})_{3_{\rm DM}\to 2_{\rm SM}} &=&\frac{1}{8m_{\chi}^{3}} \frac{|\mathcal{M}|_{3\to 2}^{2}}{(2\pi)^{2}} \int \frac{d^{3}P_{4}}{2E_{4}}\frac{1}{2\sqrt{P_{4}^{2}+m_{f}^{2}}}\delta(3m_{\chi}-2\sqrt{P_{4}^{2}+m_{f}^{2}})\\ \nonumber
&=&\frac{1}{8m_{\chi}^{3}} \frac{|\mathcal{M}|_{3\to 2}^{2}}{(2\pi)^{2}} \int \frac{P_{4}^{2}d P_{4}d\Omega}{4(P_{4}^{2}+m_{f}^{2})}\delta(3m_{\chi}-2\sqrt{P_{4}^{2}+m_{f}^{2}})\\
\nonumber	
&=&\frac{1}{8m_{\chi}^{3}} \frac{|\mathcal{M}|_{3\to 2}^{2}}{(2\pi)^{2}}\frac{4\pi}{4} \int \frac{P_{4}^{2}d P_{4}}{(P_{4}^{2}+m_{f}^{2})}\delta(3m_{\chi}-2\sqrt{P_{4}^{2}+m_{f}^{2}})\\\nonumber
&=&\frac{1}{8m_{\chi}^{3}} \frac{|\mathcal{M}|_{3\to 2}^{2}}{4\pi}\int \frac{P_{4}^{2}d P_{4}}{(P_{4}^{2}+m_{f}^{2})}\delta(3m_{\chi}-2\sqrt{P_{4}^{2}+m_{f}^{2}})\\\nonumber
&=&\frac{1}{2\times 32 \pi m_{\chi}^{3}} |\mathcal{M}|_{3\to 2}^{2}\int \frac{P_{4}^{2}d P_{4}}{(P_{4}^{2}+m_{f}^{2})}\delta(\frac{3}{2}m_{\chi}-\sqrt{P_{4}^{2}+m_{f}^{2}}) ~.
\eea
Now, integrating over $P_{4}$ we get,
\bea
(\sigma v^{2}) _{3_{\rm DM}\to 2_{\rm SM}} &=&\frac{1}{64\pi m_{\chi}^{3}}~\bigg(1-\frac{4m_{f}^{2}}{9m_{\chi}^{2}}\bigg)^{1/2}~|\mathcal{M}|_{3_{\rm DM}\to 2_{\rm SM}}^{2} ~.
\eea
We can write the thermally averaged crossection for $3_{\rm DM}\to 2_{\rm SM}$ just like we did for $3_{\rm DM}\to 2_{\rm DM}$ in \ref{eq:thermalavg}. So, we can write the thermally averaged $3_{\rm DM}\to 2_{\rm SM}$ cross-section as,
\bea
\langle\sigma v^{2}\rangle _{3_{\rm DM}\to 2_{\rm SM}}&=&\frac{1}{64\pi m_{\chi}^{3}}~\bigg(1-\frac{4m_{f}^{2}}{9m_{\chi}^{2}}\bigg)^{1/2}~|\mathcal{M}|_{3_{\rm DM}\to 2_{\rm SM}}^{2}.
\eea

\section{$2_{\rm DM} \to 2_{\rm SM}$ cross-section}
\label{annDMSM}
Calculation of such $2_{\rm DM} \to 2_{\rm SM}$ processes are well known. We only demonstrate the one ($\chi(p_{1}) \chi^{*}(p_{2}) \to f(k_{1}) \bar{f}(k_{2})$) 
which helps us to achieve the SIMP inequality Eq.~\ref{eq:condition} in this model.

\begin{figure}[htb!]
\centering
\includegraphics[scale=0.4]{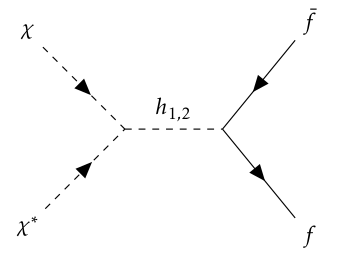}
\caption{Feynman diagram for annihilation of scalar DM to fermion pairs in this model.}
\label{FD:2to2SM}
\end{figure}
The Feynman graphs for DM annihilation to fermion pairs (relevant for DM mass $\sim$ MeV) is shown in Fig.~\ref{FD:2to2SM}.
Corresponding matrix elements from the graphs are:
$$\mathcal{M}_{1}=\lambda_{a_{1}}\frac{1}{s-m_{h_{1}}^{2}}\bar{u}(k_{1})\lambda_{f_{1}} v(k_{2})$$

$$\mathcal{M}_{2}=\lambda_{b_{1}}\frac{1}{s-m_{h_{2}}^{2}}\bar{u}(k_{1})\lambda_{f_{2}} v(k_{2}) .$$ 

Net Matrix amplitude for $\chi\chi^{*}\to f\bar{f}$ is,
$$
\mathcal{M}_{Net}=\mathcal{M}_{1}+\mathcal{M}_{2} .
$$

Squared matrix amplitude is given as,
\bea
|\mathcal{M}_{\chi\chi^{*}\to f\bar{f}}|^{2}&=&|\mathcal{M}_{Net}|^{2} \nonumber \\
&=&  2(s-4m_{f}^{2})\bigg(\frac{\lambda_{a1}\lambda_{f_{1}}}{(s-m_{h_{1}}^{2})}+\frac{\lambda_{b1}\lambda_{f_{2}}}{(s-m_{h_{2}}^{2})}\bigg)^{2}~.
\eea
The complex conjugate of $\chi\chi^{*}\to f\bar{f}$ $i.e.$ $\chi^{*}\chi\to \bar{f} f$also contributes to the total matrix amplitude and has same expression as $\chi\chi^{*}\to f\bar{f}$,
$$
|\mathcal{M}_{\chi\chi^{*}\to f\bar{f}}|^{2}=|\mathcal{M}_{\chi^{*}\chi\to \bar{f} f}|^{2} ~.
$$
Therefore, the total cross-section can be written as
\bea
(\sigma v_{\chi \chi^{*}\to f\bar{f}})&=&\frac{1}{8\pi s\sqrt{s}}\sqrt{s-4m_{f}^{2}}\bigg[|\mathcal{M}_{\chi\chi^{*}\to f\bar{f}}|^{2}+|\mathcal{M}_{\chi^{*}\chi\to \bar{f} f}|^{2}\bigg]\nonumber
\\
&=&\frac{1}{8\pi s\sqrt{s}}\sqrt{s-4m_{f}^{2}}\bigg[2 \times |\mathcal{M}_{\chi\chi^{*}\to f\bar{f}}|^{2}\bigg]~.
\eea
The thermal average cross-section is followed as
\bea
\langle\sigma v\rangle_{\chi \chi^{*}\to f\bar{f}}&=& \frac{x}{16~T~m_{\chi}^{4}~(K_{2}(x))^{2}}\int_{4m_{\chi}^{2}}^{\infty} (\sigma v_{\chi \chi^{*}\to f\bar{f}})~ K_{1}\bigg(\frac{\sqrt{s}}{T}\bigg)~s\sqrt{s-4m_{\chi}^{2}}~ds    ~.
\eea
\section{Scattering cross-section of DM with SM}
\label{scattDMSM}
We compute the scattering cross-section for the DM with SM fermions. This is required for analysing 
the kinetic equilibrium of the DM in early universe as well as for the direct search prospects of the DM. 


\begin{figure}[htb!]
\centering
\includegraphics[scale=0.55]{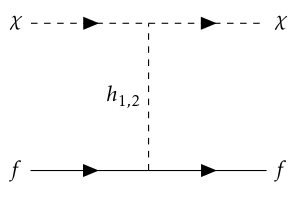}
\caption{DM-SM scattering in our model.}
\label{FD:scattering}
\end{figure}
DM-SM scattering in our model is governed by the interactions shown in Fig.~\ref{FD:scattering}. The matrix elements for the processes are given by

$$\mathcal{M}_{1}=\lambda_{a_{1}}\frac{1}{t-m_{h_{1}}^{2}}\bar{u}(k_{1})\lambda_{f_{1}} v(k_{2})$$

$$\mathcal{M}_{2}=\lambda_{b_{1}}\frac{1}{t-m_{h_{2}}^{2}}\bar{u}(k_{1})\lambda_{f_{2}}v(k_{2})$$

Net Matrix amplitude for $\chi f \to \chi f$ is,
$$
\mathcal{M}_{Net}=\mathcal{M}_{1}+\mathcal{M}_{2}
$$

Squared matrix amplitude is given as,
\bea
|\mathcal{M}_{\chi f\to \chi f}|^{2}= (-2)(t-4m_{f}^{2})\bigg(\frac{\lambda_{a1}\lambda_{f_{1}}}{(t-m_{h_{1}}^{2})}+\frac{\lambda_{b1}\lambda_{f_{2}}}{(t-m_{h_{2}}^{2})}\bigg)^{2}~.
\eea
The complex conjugate of $\chi f \to \chi f$  also contributes to the total matrix amplitude and has same expression as $\chi f \to \chi f$.
Therefore the cross-section for $2_{{\rm DM}~+~{\rm SM}} \to 2_{{\rm DM}~+~{\rm SM}} $ scattering turns out to be:
\bea
(\sigma v_{\chi f\to\chi f})
&=&\frac{1}{4\pi s\sqrt{s}}\frac{1}{2\sqrt{s}}\sqrt{(s-(m_{\chi}+m_{f})^{2})(s-(m_{\chi}-m_{f})^{2})}\bigg[2 \times  |\mathcal{M}_{\chi f \to \chi f}|^{2}\bigg]~.
\eea

and the thermal average scattering cross-section is followed as
\bea
\langle\sigma v\rangle_{\chi f\to\chi f}&=& \frac{x}{16~T~m_{\chi}^{2}~m_{f}^{2}~K_{2}(m_{\chi}/T)~K_{2}(m_{f}/T)}\int_{(m_{f}+m_{\chi})^{2}}^{\infty} (\sigma v_{\chi f\to\chi f})~ K_{1}\bigg(\frac{\sqrt{s}}{T}\bigg)~s\sqrt{s-4m_{\chi}^{2}}~ds~~.
\eea
\section{Freeze-out temperature of MeV order SIMP DM in our model}
\label{sec:freeze-out-temp}
\begin{figure}[htb!]
$$
\hspace{-1cm}
{\includegraphics[scale=0.25]{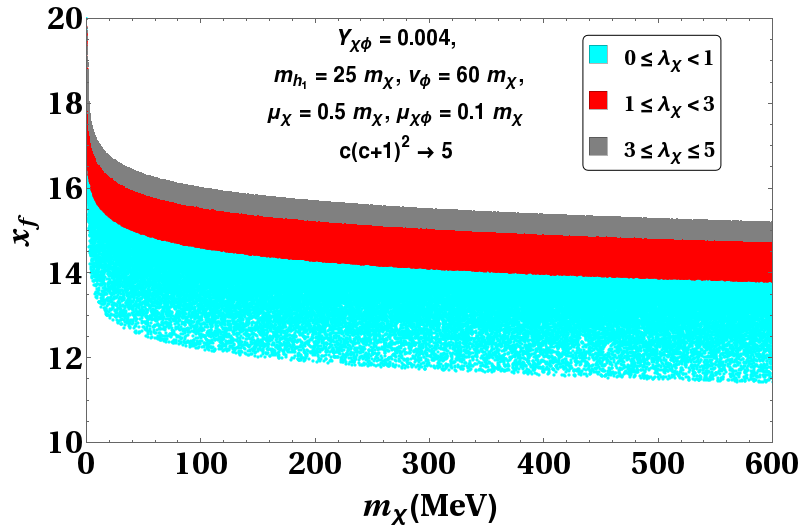}}~~
\includegraphics[scale=0.25]{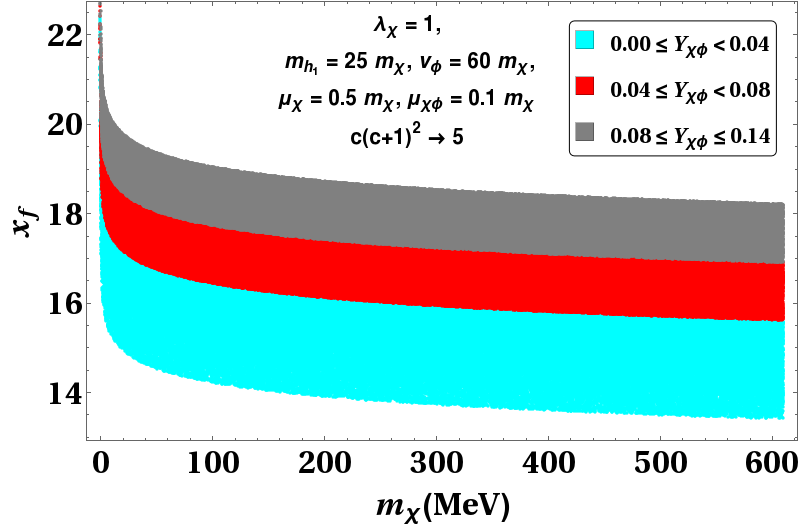}$$
\caption{Variation of $x_f=m_{\chi}/T_{f}$ with DM mass $m_\chi$ for different ranges of $\lambda_\chi$ (left panel) and $Y_{\chi \phi}$ (right panel). Other parameters kept fixed are mentioned in each figure inset.}
\label{fig:xf}
\end{figure}
SIMP type DM satisfy correct relic density for light mass of the order of MeV or below. Question then arises whether SIMP type DM is relativistic or non-relativistic. Relativistic and non-relativistic nature of thermally 
produced DM depends on freeze-out $x_f =m_{\chi}/T_{f}$ \cite{Kolb}:
\begin{itemize}
\item Relativistic:~$x_f < 3$
\item Non-Relativistic :~$x_f > 3$.
\end{itemize}
Therefore, evaluating freeze-out point is good enough to test above credential. Here, we have plotted the freeze-out temperature in terms of $x_f$  with DM mass $m_{\chi}$ (obtained using the Eqn. \ref{analyticformxf}) keeping other parameters fixed in Fig.~\ref{fig:xf}. The range 
of parameter space scanned certainly encapsulate the relic density allowed points as obtained in this model framework. 
It is clearly seen that $x_f \gtrsim 12$, which indicates non relativistic behaviour of SIMP type DM in our model as assumed.

\bibliographystyle{JHEP}
\bibliography{references}
\end{document}